\documentclass{article}
\usepackage{jheppub}
\usepackage[T1]{fontenc}
\usepackage{microtype} 
\usepackage{mathtools} 
\usepackage{amsmath,amsfonts,amsthm,amssymb}
\usepackage{mathrsfs}
\usepackage{tikz-cd}
\DeclareFontFamily{T1}{calligra}{}
\DeclareFontShape{T1}{calligra}{m}{n}{<->s*[1.44]callig15}{}
\DeclareMathAlphabet\mathxg      {U}{rsfso}{m}{n}

\newcommand{\cG}{\mathcal{G}}
\newcommand{\xG}{\mathscr{G}}

\def\beq{\begin{equation}}
\def\eeq{\end{equation}}
\def\bsp#1\esp{\begin{split}#1\end{split}}

\newcommand{\eps}{\epsilon}
\newcommand{\cN}{\mathcal{N}}
\newcommand{\cK}{\mathcal{K}}
\newcommand{\cF}{\mathcal{F}}
\newcommand{\cX}{\mathcal{X}}
\newcommand{\cC}{\mathcal{C}}
\newcommand{\fa}{\mathfrak{a}}
\newcommand{\fb}{\mathfrak{b}}
\newcommand{\fc}{\mathfrak{c}}
\newcommand{\fd}{\mathfrak{d}}
\renewcommand{\ln}{\log}
\newcommand{\zb}{\overline{z}}

\newcommand{\omegab}{\overline{\omega}}
\newcommand{\Ti}{\text{Ti}}

\preprint{CP3-17-14, CERN-TH-2017-113}

\title{The analytic structure and the transcendental weight of the BFKL ladder at NLL accuracy}

\author[a,1]{Vittorio Del Duca}
\note{On leave from INFN, Laboratori Nazionali di Frascati, Italy.}
\author[b,c]{Claude Duhr}
\author[c]{Robin Marzucca}
\author[c]{Bram Verbeek}
\affiliation[a]{Institute for Theoretical Physics, ETH Z\"urich, 8093 Z\"urich, Switzerland.}
\affiliation[b]{Theoretical Physics Department, CERN, CH-1211 Geneva 23, Switzerland.}
\affiliation[c]{Center for Cosmology, Particle Physics and Phenomenology (CP3), Universit\'e catholique de Louvain,
Chemin du Cyclotron 2, 1348 Louvain-La-Neuve, Belgium.}

\emailAdd{delducav@itp.phys.ethz.ch}
\emailAdd{claude.duhr@cern.ch}
\emailAdd{robin.marzucca@uclouvain.be}
\emailAdd{bram.verbeek@uclouvain.be}

\abstract{We study some analytic properties of the BFKL ladder at next-to-leading logarithmic accuracy (NLLA).
We use a procedure by Chirilli and Kovchegov to construct the NLO eigenfunctions, and we show that the BFKL ladder can be evaluated order by order in the coupling in terms of certain generalised single-valued multiple polylogarithms recently introduced by Schnetz. We develop techniques to evaluate the BFKL ladder at any loop order, and we present explicit results up to five loops.
Using the freedom in defining the matter content of the NLO BFKL eigenvalue, we obtain conditions for the BFKL ladder in momentum space at NLLA to have maximal transcendental weight. We observe that, unlike in moment space, the result in momentum space in $\cN=4$ SYM is not identical to the maximal weight part of QCD, and moreover that there is no gauge theory with this property. We classify the theories for which the BFKL ladder at NLLA has maximal weight in terms of their field content, and we find that these theories are highly constrained: there are precisely four classes of theories with this property involving only fundamental and adjoint matter, all of which have a vanishing one-loop beta function and a matter content that fits into supersymmetric multiplets. Our findings indicate that theories which have maximal weight are highly constrained and point to the possibility that there is a connection between maximal transcendental weight and superconformal symmetry.}


\begin{document}

\maketitle

\section{Introduction}
\label{sec:intro}

In the limit in which the squared center-of-mass energy is much greater than
the momentum transfer, $s \gg |t|$, any QCD scattering process is dominated
by gluon exchange in the $t$ channel. In this limit
the Balitsky-Fadin-Kuraev-Lipatov (BFKL) theory models
strong-interaction processes
by resumming the leading radiative corrections to parton-parton
scattering. This is achieved at leading logarithmic accuracy (LLA), in $\ln(s/|t|)$, through the BFKL 
equation~\cite{Fadin:1975cb,Kuraev:1976ge,Kuraev:1977fs,Balitsky:1978ic}, to which the corrections at
next-to-leading-logarithmic accuracy (NLLA) were computed in refs.~\cite{Fadin:1998py,Ciafaloni:1998gs}.

In this paper, we address three questions related to the next-to-leading-order (NLO) corrections to the {singlet} eigenvalue of the BFKL equation.
Firstly, the NLO corrections to the BFKL eigenvalue were computed by Fadin and Lipatov~\cite{Fadin:1998py} by acting with the NLO BFKL kernel on the leading order eigenfunctions. This procedure is not consistent, and it was already clear to Fadin and Lipatov
that the terms which make the procedure inconsistent are related to the running of the coupling. The consistent NLO eigenfunctions were
constructed by Chirilli and Kovchegov~\cite{Chirilli:2013kca,Chirilli:2014dcb}, who found indeed that the additional pieces which occur at NLO
are proportional to the beta function. We show that the NLO corrections to the eigenfunctions
can be made to vanish by taking the scale of the coupling to be the geometric mean of the transverse momenta at the ends of the BFKL ladder. 

Secondly, in ref.~\cite{DelDuca:2013lma} it was shown that the functions which describe the analytic structure of the BFKL ladder at LLA are single-valued iterated integrals on the moduli space ${\cal M}_{0,4}$ of Riemann spheres with four marked points, which are single-valued
harmonic polylogarithms (SVHPLs)~\cite{BrownSVHPLs}. We extend the results of ref.~\cite{DelDuca:2013lma} and show that the functions which describe the analytic structure of the BFKL ladder at NLLA are a generalisation of  SVHPLs recently introduced by Schnetz~\cite{Schnetz:2016fhy}. We use this insight and develop techniques to evaluate the BFKL ladder in momentum space perturbatively to any loop order, and we provide explicit results through five loops. 

Finally, it has been guessed~\cite{Kotikov:2000pm,Kotikov:2002ab}, and verified at NLO accuracy~\cite{Kotikov:2003fb}, that the anomalous dimensions of the leading-twist operators which control the Bjorken scaling violations in ${\cal N}=4$ SYM have a uniform and maximal transcendental weight in moment space, which matches the maximal weight part of the corresponding anomalous dimensions in QCD. This has been used to derive the anomalous dimensions in ${\cal N}=4$ SYM at NNLO accuracy~\cite{Kotikov:2004er} from the known anomalous dimensions in QCD at NNLO~\cite{Moch:2004pa}. 
We consider the BFKL ladder in a generic gauge theory with arbitrary matter content and we use the explicit results in momentum space through five loops to analyse the transcendental weight properties of the BFKL ladder at NLLA. {It is well known~\cite{Kotikov:2000pm,Kotikov:2002ab} that in moment space the singlet BFKL eigenvalue in $\cN=4$ SYM is given by the maximal weight part of the corresponding QCD result. Our results show that the corresponding statement is not true in momentum space: We find that in momentum space the BFKL ladder in $\cN=4$ SYM is \emph{not} identical the maximal weight part of QCD}, and, moreover that there is no theory with additional scalar or fermionic matter with that property. While so far all considerations were independent of the colour representation of the matter fields, in the case of only adjoint and fundamental matter we derive a set of necessary conditions for the BFKL ladder to have maximal transcendental weight in momentum space to all loop orders. We find that the theories that satisfy these constraints are extremely rare: there are only four classes of theories that satisfy these constraints, and all of them have a vanishing one-loop beta function and a matter content that fits into supersymmetric multiplets. Although our analysis is restricted to the BFKL ladder at NLLA, our findings indicate that the property of maximal transcendentality is a very special property shared by only very few and very special theories.

This paper is organised as follows: In Section~\ref{sec:setup} and~\ref{sec:LLA} we review the BFKL ladder and its analytic structure in perturbation theory at LLA. In Section~\ref{sec:Chirilli} we review the Chirilli-Kovchegov procedure to define the NLO eigenfunctions of the BFKL kernel, and we show
that the NLO corrections to the eigenfunctions can be made to vanish by taking the scale of the coupling to be the geometric mean of the transverse momenta at the ends of the BFKL ladder.
In Section~\ref{sec:nll} we compute the BFKL ladder at NLLA in terms of the generalised single-valued multiple polylogarithms introduced by Schnetz in ref.~\cite{Schnetz:2016fhy}. In Section~\ref{sec:weight} we analyse the transcendental weight properties of the BFKL ladder. The appendices collect technical proofs omitted throughout the main text as well as explicit results for generalised single-valued multiple polylogarithms and the BFKL ladder through five loops.

\section{The BFKL equation}
\label{sec:setup}

The main object of interest in this note is the BFKL ladder, which appears in the total cross section for parton scattering in the high-energy limit,
\beq
\sigma(s) \simeq \int\frac{d^2q_1\,d^2q_2}{(2\pi)^2\,q_1^2\,q_2^2}\,\Phi_A(q_1)\,\Phi_B(q_2)\,f(q_1,q_2,\log(s/s_0))\,,
\eeq
where $s$ is the center-of-mass energy and 
\beq\label{eq:s0}
s_0\equiv \sqrt{q_1^2\,q_2^2}
\eeq
is the geometric mean of the two transverse momenta. $\Phi_{A/B}$ denote the impact factors and $f$ is the BFKL ladder, which is written as,
\beq\label{eq:fdef}
f(q_1,q_2,y) = \int_{C}\frac{d\omega}{2\pi i}\,e^{y\,\omega}\,f_{\omega}(q_1,q_2)\,.
\eeq
The integration contour $C$ is a straight vertical line such that all poles in $\omega$ are to the right of the contour and $f_\omega$ is a solution to the BFKL equation,
\begin{equation}
\omega\, f_\omega(q_1,q_2) = {1\over 2}\,\delta^{(2)}(q_1-q_2) + (\cK\star f_\omega)(q_1,q_2) \,,
\label{bfklop}
\end{equation}
where the convolution is defined by
\beq
(\cK\star f_\omega)(q_1,q_2) \equiv \int d^2k\,K(q_1,k)\,f_{\omega}(k,q_2)\,,
\eeq
with $K(q_1,q_2)$ the BFKL kernel.
The kernel is real and symmetric, $K(q_1,q_2) = K(q_2,q_1)$, and so the integral operator $\cK$ is hermitian and its eigenvalues are real.

The BFKL equation can be solved by finding a suitable set of eigenfunctions of the BFKL integral operator,
\beq
(\cK\star\Phi_{\nu n})(q) = \omega_{\nu n}\,\Phi_{\nu n}(q)\,.
\eeq
The eigenfunctions are labeled by $(\nu, n)$, where $\nu$ is a real number and $n$ an integer, and form a complete and orthonormal set of functions. Hence, they satisfy
\beq
2\int d^2q\,\Phi_{\nu n}(q)\,\Phi_{\nu^\prime n^\prime}^\ast(q) =  \int_0^{\infty} dq^2 \int_0^{2\pi}d\theta\, \Phi_{\nu n}(q)\, \Phi^\ast_{\nu^\prime n^\prime}(q)
= \delta(\nu-\nu^\prime)\, \delta_{nn^\prime} \,,
\label{orthonorm}
\eeq
and
\beq\bsp
\sum_{n=-\infty}^{+\infty} \int_{-\infty}^{+\infty} d\nu\, \Phi_{\nu n}(q)\, \Phi_{\nu  n}^\ast({q^\prime}) 
= {1\over 2}\,\delta^{(2)}(q-q^\prime) = \delta\left(q^2-{q^\prime}^2\right)\, \delta(\theta-\theta^\prime) \,.
\label{complete}
\esp\eeq
In a conformally-invariant theory, the eigenfunctions are fixed by conformal symmetry~\cite{Lipatov:1985uk},
\beq
\label{loeigenf}
\Phi_{\nu n}^{CFT}(q)\equiv\varphi_{\nu n}(q) = {1\over 2\pi}\, (q^2)^{-1/2+i\nu}\, e^{i n \theta}\,.
\eeq
It is easy to check that the functions $\varphi_{\nu n}$ form a complete and orthonormal set of eigenfunctions.
In a non conformally-invariant theory, like QCD, the form of the eigenfunctions may differ.

In terms of the eigenfunctions and eigenvalue, the solution to eq.~\eqref{bfklop} takes the form,
\begin{equation}
f_\omega(q_1,q_2)\, = \sum_{n=-\infty}^{+\infty} \int_{-\infty}^{+\infty} d\nu\, {1\over \omega -\omega_{\nu n}}\,
\Phi_{\nu n}(q_1)\, \Phi_{\nu n}^\ast(q_2)\,.
\label{solb}
\end{equation}
Indeed, we have
\beq\bsp
(\cK\star f_\omega)(q_1,q_2) &\,= \sum_{n=-\infty}^{+\infty} \int_{-\infty}^{+\infty} d\nu\, {1\over \omega -\omega_{\nu n}}\,
(\cK\star\Phi_{\nu n})(q_1)\, \Phi_{\nu n}^\ast(q_2)\\
&\,= \sum_{n=-\infty}^{+\infty} \int_{-\infty}^{+\infty} d\nu\, {\omega_{\nu n}\over \omega -\omega_{\nu n}}\,
\Phi_{\nu n}(q_1)\, \Phi_{\nu n}^\ast(q_2)\\
&\,=-\frac{1}{2}\delta^{(2)}(q_1-q_2) + \omega\,f_{\omega}(q_1,q_2)\,,
\esp\eeq
where in the last step we used the completeness relation satisfied by the eigenfunctions. Finally, inserting eq.~\eqref{solb} into eq.~\eqref{eq:fdef}, we find
\beq\label{eq:GreenGluon}
f(q_1,q_2,y) = \sum_{n=-\infty}^{+\infty} \int_{-\infty}^{+\infty} d\nu\,
\Phi_{\nu n}(q_1)\, \Phi_{\nu n}^\ast(q_2)\,e^{y\,\omega_{\nu n}}\,.
\eeq

In the following we are interested in the perturbative expansion of the BFKL ladder.
The kernel of the integral equation admits the expansion,
\begin{equation}
K(q_1,q_2) = \overline{\alpha}_\mu\sum_{l=0}^\infty\overline{\alpha}_\mu^l\, K^{(l)}(q_1,q_2)\,.
\end{equation}
where $\overline{\alpha}_\mu = N_C\, \alpha_S(\mu^2)/\pi$ is the renormalised strong coupling constant evaluated at an arbitrary scale $\mu^2$. $K^{(0)}$ is the leading order (LO) BFKL kernel~\cite{Fadin:1975cb,Kuraev:1976ge,Kuraev:1977fs,Balitsky:1978ic}, 
which leads to the resummation of the terms of ${\cal O}\left( (\overline{\alpha}_\mu y)^n\right)$, i.e., terms at LLA, and the NLO kernel $K^{(1)}$~\cite{Fadin:1998py,Ciafaloni:1998gs} resums the terms at NLLA, i.e. of ${\cal O}\left(\overline{\alpha}_\mu (\overline{\alpha}_\mu y)^n\right)$, and so forth. The BFKL integral operators $\cK^{(k)}$ are defined in an obvious way. The BFKL eigenvalue and eigenfunctions also admit an expansion in the strong coupling,
\begin{equation}\label{eq:expansion}
\omega_{\nu n} = \overline{\alpha}_\mu\sum_{l=0}^\infty\overline{\alpha}_\mu^l\, \omega_{\nu n}^{(l)} {\rm~~and~~}
\Phi_{\nu n}(q) = \sum_{l=0}^\infty\overline{\alpha}_\mu^l\, \Phi_{\nu n}^{(l)}(q)\,.
\end{equation}
Note that in a conformally-invariant theory the quantum corrections to the eigenfunctions must vanish, and so we expect the quantum corrections to the eigenfunctions to be proportional to the beta function.
The truncated eigenvalue and eigenfunctions,
\beq
\omega_{\nu n}^{N^kLO} = \overline{\alpha}_\mu\sum_{l=0}^k\overline{\alpha}_\mu^l\, \omega_{\nu n}^{(l)} {\rm~~and~~}
\Phi_{\nu n}^{N^kLO}(q) = \sum_{l=0}^k\overline{\alpha}_\mu^l\, \Phi_{\nu n}^{(l)}(q)\,,
\end{equation}
are eigenvalues and eigenfunctions of the truncated BFKL integral operator,
\beq
\left(\cK^{N^kLO}\star \Phi_{\nu n}^{N^kLO}\right)(q) = \omega_{\nu n}^{N^kLO}\,\Phi_{\nu n}^{N^kLO}(q)+\mathcal{O}(\overline{\alpha}_\mu^{k+1})\,,\textrm{ with } \cK^{N^kLO} = \overline{\alpha}_\mu\sum_{l=0}^k\overline{\alpha}_\mu^l\, \cK^{(l)}\,.
\eeq
In the remainder of this note we discuss the first two terms in the expansion of the BFKL ladder,
\beq
f(q_1,q_2,y) = f^{LL}(q_1,q_2,\eta_\mu) + \overline{\alpha}_{\mu}\, f^{NLL}(q_1,q_2,\eta_\mu) + \ldots\,,\qquad \eta_\mu = \overline{\alpha}_\mu\,y\,.
\eeq
The LO term $f^{LL}(q_1,q_2,\eta_\mu)$ is the BFKL ladder at LLA, and the NLO term $f^{NLL}(q_1,q_2,\eta_\mu)$ is the ladder at NLLA.
We start by discussing the LO case in the next section, before extending the discussion to NLO in subsequent sections.

\section{The BFKL ladder at leading logarithmic accuracy}
\label{sec:LLA}

At LO the BFKL kernel is conformally-invariant (independently of the theory under consideration), and thus the LO eigenfunctions are fixed to eq.~\eqref{loeigenf}. The LO eigenvalue is given by~\cite{Kuraev:1977fs,Balitsky:1978ic},
\begin{equation}
\omega_{\nu n}^{(0)}\equiv\chi_{\nu n} =  -2\gamma_E - \psi\left( \frac{|n|+1}{2}+i\nu \right) - \psi\left( \frac{|n|+1}{2}-i\nu \right) \,,
  \label{bfklloeigen}
\end{equation}
where $\gamma_E=-\Gamma'(1)$ is the Euler-Mascheroni constant and $\psi(z) = \frac{d}{dz}\log\Gamma(z)$ is the digamma function. We thus have
\beq
(\cK^{LO}\star \varphi_{\nu n})(q) = \chi_{\nu n}\,\varphi_{\nu n}(q)\,.
\eeq
The LO eigenvalue is symmetric under $\nu\to -\nu$. 
Inserting the LO eigenvalue and eigenfunctions into eq.~\eqref{eq:GreenGluon}, we find the expression for the BFKL ladder at LLA,
\begin{eqnarray}
f^{LL}(q_1,q_2,\eta_\mu)
&=& \sum_{n=-\infty}^{+\infty} \int_{-\infty}^{+\infty} d\nu\, e^{\eta_\mu\,\chi_{\nu n}}\, \varphi_{\nu n}(q_1)\, \varphi^\ast_{\nu n}(q_2) \,.
\label{mellin}
\end{eqnarray}
The dependence of the strong coupling on the renormalisation scale $\mu^2$ in eq.~\eqref{mellin} is immaterial, since the effect of changing the scale is NLLA, i.e., beyond the LL accuracy at which we are working. At LLA, we can expand $f^{LL}$ in powers of $\eta_\mu$,
\beq
f^{LL}(q_1,q_2,\eta_\mu) = \frac{1}{2}\delta^{(2)}(q_1-q_2) + \frac{1}{2\pi\,\sqrt{q_1^2\,q_2^2}}\sum_{k=1}^{\infty}\frac{\eta_\mu^k}{k!}\,f^{LL}_k(z)\,.
\eeq
The coefficients of the expansion depend on a single complex variable $z$ defined by
\beq
z \equiv \frac{\tilde{q}_1}{\tilde{q}_2}\,,\qquad \textrm{with } \tilde{q}_k\equiv q_k^x+iq_k^y\,.
\eeq
The coefficients can then be cast in the form of a Fourier-Mellin transform,
\beq\label{eq:FM_LL}
f^{LL}_k(z) = \cF\left[\chi_{\nu n}^k\right] \equiv \sum_{n=-\infty}^{+\infty}\left(\frac{z}{\bar z}\right)^{n/2}\int_{-\infty}^{+\infty}\frac{d\nu}{2\pi}\,|z|^{2i\nu}\,\chi_{\nu n}^k\,.
\eeq
The inverse transform is given by
\beq\label{eq:inv_FM}
\cF^{-1}\left[f(z)\right] = \int\frac{d^2z}{\pi}\,z^{-1-i\nu-n/2}\,\overline{z}^{-1-i\nu+n/2}\,f(z)\,.
\eeq
In ref.~\cite{Dixon:2012yy} it was shown that the natural space of functions to which Fourier-Mellin transforms of this type evaluate are single-valued harmonic polylogarithms (SVHPLs)~\cite{BrownSVHPLs}, which we review in the following.

Ordinary, i.e., not necessarily single-valued, harmonic polylogarithms (HPLs)~\cite{Remiddi:1999ew} are a special class of multiple polylogarithms (MPLs)\footnote{There is a conventional sign difference in the literature between HPLs and generic MPLs. Throughout this paper, we strictly follow the sign convention of eq.~\eqref{eq:MPL_def}.}~\cite{Goncharov:1998,Goncharov:2001}. The latter are defined as the iterated integrals,
\beq\label{eq:MPL_def}
G(a_1,\ldots,a_n;z) = \int_0^z\frac{dt}{t-a_1}\,G(a_2,\ldots,a_n;t)\,,
\eeq
except if $(a_1,\ldots,a_n) = (0,\ldots,0)$, in which case we define
\beq
G(\underbrace{0,\ldots,0}_{n\textrm{ times}};z) = \frac{1}{n!}\log^nz\,.
\eeq
The case of HPLs is recovered for $a_i\in \{-1,0,1\}$. The number $n$ of integrations is called the \emph{weight} of the MPL. MPLs are endowed with a lot of algebraic structure. In particular, they form a shuffle algebra, which allows one to write the product of two MPLs of weight $n_1$ and $n_2$ as a linear combination of MPLs of weight $n_1+n_2$.

In general, MPLs define multi-valued functions, and the branch cut structure of a scattering amplitude is connected to the concept of unitarity. It is however possible to consider linear combinations of MPLs such that all discontinuities cancel and the resulting function is single-valued. As a simple example, we can consider the linear combination,
\beq\label{eq:SV_example}
\cG(a;z) \equiv G(a;z) + G(\overline{a};\overline{z}) = \log\left(1-\frac{z}{a}\right) + \log\left(1-\frac{\overline{z}}{\overline{a}}\right) = \log\left|1-\frac{z}{a}\right|^2\,.
\eeq
The argument of the logarithm in eq.~\eqref{eq:SV_example} is positive-definite, and thus the function is single-valued. It is possible to generalise this construction to MPLs of higher weight. In particular, in the case where the position of the singularities $a_i$ is independent of the variable $z$ (which covers the case of HPLs), one can show that there is a map ${\bf s}$ which assigns to an MPL $G(\vec a;z)$ its single-valued version $\cG(\vec a;z) \equiv {\bf s}(G(\vec a;z))$. Single-valued multiple polylogarithms (SVMPLs) inherit many of the properties of ordinary MPLs. In particular, SVMPLs form a shuffle algebra and satisfy the same holomorphic differential equations and boundary conditions as their multi-valued analogues. There are several ways to explicitly construct the map ${\bf s}$, based on the Knizhnik-Zamolodchikov equation~\cite{BrownSVHPLs,BrownSVMPLs}, the coproduct and the action of the motivic Galois group on MPLs~\cite{Brown:2013gia,Brown_Notes,DelDuca:2016lad} and the existence of single-valued primitives of MPLs~\cite{Schnetz:2016fhy}.

In ref.~\cite{DelDuca:2013lma} a (conjectural) generating functional of the BFKL ladder was given that allows one to express each coefficient $f^{LL}_k(z)$ as a linear combination of SVHPLs without singularities at $z=-1$. Writing
\beq\label{eq:LL_decomposition}
f^{LL}_k(z) = \frac{|z|}{2\pi\,|1-z|^2}\,F_k(z)\,,
\eeq
we can express the first few coefficients as~\cite{DelDuca:2013lma},
\beq\bsp
F_1(z) &\,= 1\,,\\
F_2(z) &\,= 2\,\cG_1(z) - \cG_0(z)\,,\\
F_3(z) &\,= 6\,\cG_{1,1}(z) -3\,\cG_{0,1}(z)-3\,\cG_{1,0}(z)+\cG_{0,0,0}(z)\,,\\
F_4(z) &\,= 24\,\cG_{1,1,1}(z)+4\,\cG_{0,0,1}(z)+6\, \cG_{0,1,0}(z)-12\, \cG_{0,1,1}(z)+4\,\cG_{1,0,0}(z)\\
&\,-12\, \cG_{1,0,1}(z)-12\, \cG_{1,1,0}(z)-\cG_{0,0,0}(z)+8\,\zeta_3 \,,
\esp\eeq
where we used the shorthand $\cG_{a_1,\ldots,a_n}(z)\equiv\cG(a_1,\ldots,a_n;z)$. The conjecture of ref.~\cite{DelDuca:2013lma} implies that the functions $F_k$ have a particularly simple form: at any loop order, the functions $F_k$ are pure~\cite{ArkaniHamed:2010gh}. More precisely, the $F_k$ are conjectured to be linear combinations of SVHPLs  of uniform weight $(k-1)$ with singularities at most at $z=0$ or $z=1$, and the coefficients of the linear combination are rational numbers (note that we consider SVHPLs to include single-valued multiple zeta values~\cite{Brown:2013gia}). This claim is a consequence of the proof given in Appendix \ref{app:proof}.

The purpose of this note is to extend the results of ref.~\cite{DelDuca:2013lma} and to explore the analytic structure of the BFKL ladder at NLLA. We start by deriving the correct Fourier-Mellin representation at NLLA in terms of the NLO BFKL eigenvalue in Section~\ref{sec:Chirilli}, and we develop techniques to evaluate $f^{NLL}$ perturbatively in Section~\ref{sec:nll}.

\section{The BFKL ladder at next-to-leading logarithmic accuracy}
\label{sec:Chirilli}
\subsection{Beyond the leading order: the Chirilli-Kovchegov procedure}

The NLO corrections to the BFKL kernel in QCD were obtained in ref.~\cite{Fadin:1998py}. The corresponding NLO corrections to the BFKL singlet eigenvalue were computed in ref.~\cite{Fadin:1998py} for $n=0$ and 
in ref.~\cite{Kotikov:2000pm,Kotikov:2002ab} for arbitrary $n$, albeit in the approximation that the NLO eigenfunctions are identical to the LO eigenfunctions given in eq.~\eqref{loeigenf}. In other words, the NLO corrections $\delta_{\nu n}$ to the BFKL eigenvalue of ref.~\cite{Kotikov:2000pm,Kotikov:2002ab} are \emph{defined} by the equation,
\begin{equation}
(\cK^{NLO}\star\varphi_{\nu n})(q) \equiv
 \overline{\alpha}_S(q^2) \left( \chi_{\nu n} + \overline{\alpha}_S(q^2) \frac{\delta_{\nu n}}4 \right)\, \varphi_{\nu n}(q) +\mathcal{O}(\overline{\alpha}_S^3(q^2))\,.
\label{homognlo}
\end{equation}
The NLO corrections to the eigenvalue $\delta_{\nu n}$ in QCD are given in this approximation by~\cite{Fadin:1998py,Kotikov:2000pm,Kotikov:2002ab},
\begin{align}
  \delta_{\nu n} &= 6\zeta_3 - \frac{1}{2}\beta_0\, \chi^2_{\nu n} + 4 \gamma_K^{(2)}\, \chi_{\nu n} + \frac{i}{2}\beta_0\, \partial_\nu \chi_{\nu n}\
+ \partial_\nu^2 \chi_{\nu n}\nonumber \\
  &-2\Phi(n,\gamma) - 2\Phi(n,1-\gamma) - \frac{\Gamma(\frac{1}{2}+i\nu)\Gamma(\frac{1}{2}-i\nu)}{2i\nu}\left[ \psi \left( \frac{1}{2} +i\nu \right) - \psi \left( \frac{1}{2} - i\nu \right) \right] \label{nloeigenv} \\
  &\times\left[ \delta_{n0}\left( 3+ \left( 1 + \frac{N_f}{N^3_c} \right)\frac{2 + 3\gamma (1-\gamma)}{(3-2\gamma)(1+2\gamma)} \right) - \delta_{|n|2}\left(\left( 1 + \frac{N_f}{N^3_c} \right)\frac{\gamma(1-\gamma)}{2(3-2\gamma)(1+2\gamma)} \right) \right],\nonumber
\end{align}
with $\beta_0$ the one-loop beta function and $\gamma_K^{(2)}$ the two-loop cusp anomalous dimension for QCD in the dimensional reduction (DRED) scheme,
\begin{equation}
 \beta_0 = \frac{11}{3} - \frac{2 N_f}{3 N_c}\,, \qquad
 \gamma_K^{(2)} = \frac{1}{4}\left( \frac{64}{9} - \frac{10 N_f}{9 N_c} \right)-\frac{\zeta_2}2\,.
 \label{beta0}
 \end{equation}
In eq.~\eqref{nloeigenv} we use the shorthand $\gamma = 1/2 + i\nu$, with $\Phi(n,\gamma)$ defined as,
\begin{align}
\Phi(n,\gamma)	&= \sum^{\infty}_{k=0} \frac{(-1)^{k+1}}{k + \gamma + |n|/2}\bigg\{ \psi'(k + |n|+1)-\psi'(k+1)+ (-1)^{k+1}[\beta'(k + |n|+1)+\beta'(k+1)] \nonumber\\
&-\frac{1}{k + \gamma + |n|/2}[\psi(k+|n|+1)-\psi(k+1)]\bigg\} \,,
\end{align}
with
\begin{equation}
\beta'(z) = \frac{1}{4}\left[ \psi'\left( \frac{1+z}{2}\right) -\psi'\left( \frac{z}{2}\right) \right].
\end{equation}
Note that for $\cN=4$ SYM the eigenvalue is
\begin{equation}\label{eq:delta_N=4}
 \delta_{\nu n}^{\cN=4} = 6\zeta_3 + 4 \gamma_K^{(2) \cN=4} \chi_{\nu n} 
+ \partial_\nu^2 \chi_{\nu n} -2\Phi(n,\gamma) - 2\Phi(n,1-\gamma)\,,
\end{equation}
with $\chi_{\nu n}$ defined in eq.~\eqref{bfklloeigen}, and
$\gamma_K^{(2) \cN=4}$ the two-loop cusp anomalous dimension in $\cN=4$ SYM,
\begin{equation}\label{eq:cusp_N=4}
\gamma_K^{(2) \cN=4} = -\frac{1}2\,\zeta_2 \,.
\end{equation}
Equations~\eqref{eq:delta_N=4} and~\eqref{eq:cusp_N=4} are valid in DRED which preserves supersymmetry.
As $\cN=4$ SYM is conformally invariant, the eigenfunctions are fixed to all orders by eq.~\eqref{loeigenf},
\beq
\Phi_{\nu n}^{\cN=4}(q) = \varphi_{\nu n}(q)\,.
\eeq
Hence, $\delta_{\nu n}^{\cN=4}$ is the correct NLO BFKL eigenvalue in $\cN=4$ SYM.

While the NLO eigenvalue in eq.~\eqref{nloeigenv} was derived under the assumption that the eigenfunctions are the same at LO and NLO, we have seen in Section~\ref{sec:setup} that the LO eigenfunctions \eqref{loeigenf} may themselves receive higher-order corrections in a non conformally-invariant theory, cf. eq.~\eqref{eq:expansion}.
In fact, the true NLO eigenvalue must be real (as the eigenvalue of a hermitian operator) and independent of $q^2$. $\delta_{\nu n}$ fails to meet either criterion: the right-hand side of eq.~\eqref{homognlo} depends on $q^2$ through the strong coupling constant and eq.~(\ref{nloeigenv}) contains the term $i\beta_0\,\partial_\nu \chi_{\nu n}$, which is imaginary. Note that both of these issues are absent in a conformally-invariant theory, where the strong coupling does not depend on the scale and the beta function vanishes. In particular, the term proportional to the $\beta$ function is absent in $\cN=4$ SYM, cf. eq.\eqref{eq:delta_N=4}, and in that case the LO eigenfunctions are indeed eigenfunctions of the NLO kernel.

In ref.~\cite{Fadin:1998py}, Fadin and Lipatov already hinted that one could get
rid of the undesired properties of $\delta_{\nu n}$ by modifying the LO eigenfunctions through the running-coupling terms. This was made explicit by Chirilli and Kovchegov~\cite{Chirilli:2013kca,Chirilli:2014dcb}. In the remainder of this section we shall review the Chirilli-Kovchegov procedure, and construct 
accordingly the NLO eigenfunctions and the corresponding NLO eigenvalue for any value of $n$.

Our goal is to construct functions $\omega^{(1)}_{\nu n}$ and $\Phi_{\nu n}^{(1)}(q)$ such that 
\beq\bsp\label{eq:KPhi}
\left[\cK^{NLO}\star\left(\varphi_{\nu n}+\overline{\alpha}_\mu\,\Phi_{\nu n}^{(1)}\right)\right](q) &\,= \overline{\alpha}_\mu\,\left(\chi_{\nu n}+\overline{\alpha}_\mu\,\omega^{(1)}_{\nu n}\right)\left[\varphi_{\nu n}(q)+\overline{\alpha}_\mu\,\Phi_{\nu n}^{(1)}(q)\right]+\mathcal{O}(\overline{\alpha}_\mu^3)\,.
\esp\eeq
We parametrise the NLO eigenvalue $\omega^{(1)}_{\nu n}$ in terms of $\delta_{\nu n}$ and an unknown function $c_{\nu n}$ as
\beq\label{eq:chi_nlo}
\omega^{(1)}_{\nu n} = \frac{\delta_{\nu n}}4 + c_{\nu n} = i \frac{\beta_0}8\, \partial_\nu \chi_{\nu n} + \Delta_{\nu n} + c_{\nu n}\,,
\eeq
where $\Delta_{\nu n}$ collects all  the terms in eq.~\eqref{nloeigenv} that are symmetric under $\nu \to -\nu$, and we expect the function $\omega^{(1)}_{\nu n}$ to be symmetric. Inserting the parametrisation in eq.~\eqref{eq:chi_nlo} into eq.~\eqref{eq:KPhi} and using eq.~\eqref{homognlo} and the one-loop running of the strong coupling,
\begin{equation}
\overline{\alpha}_S(q^2) = \frac{\overline{\alpha}_S(\mu^2)}{1+\frac{\beta_0}{4} \,\overline{\alpha}_S(\mu^2) \ln\frac{q^2}{\mu^2}}
=\overline\alpha_\mu\left[1-\overline\alpha_\mu\,\frac{\beta_0}{4}\, \ln\frac{q^2}{\mu^2}+\mathcal{O}(\overline\alpha_\mu^2)\right]\,,
\end{equation}
we obtain
\beq
\left(\cK^{LO}\star\Phi_{\nu n}^{(1)}\right)(q) = \left(c_{\nu n}+\frac{\beta_0}{4}\,\chi_{\nu n}\,\log\frac{q^2}{\mu^2}\right)\,\varphi_{\nu n}(q)+\chi_{\nu n}\,\Phi_{\nu n}^{(1)}(q)\,.
\label{cnunsolv}
\eeq
Following Chirilli and Kovchegov~\cite{Chirilli:2013kca,Chirilli:2014dcb}, since eq.~\eqref{cnunsolv} must be satisfied for arbitrary values of $q$,
we must take $\Phi^{(1)}_{\nu n}$ proportional to $\varphi_{\nu n}$. We therefore make the following ansatz,
\begin{equation}
\Phi^{(1)}_{\nu n}(q) =  \left( a_{0,\nu n} + a_{1,\nu n} \ln\frac{q^2}{\mu^2} + a_{2,\nu n} \ln^2\frac{q^2}{\mu^2} \right) \varphi_{\nu n}(q)\,,
\label{phiguess}
\end{equation}
where $a_{j,\nu n}$ for $j=0, 1, 2$ are arbitrary complex coefficients. Inserting eq.~(\ref{phiguess}) into \eqref{cnunsolv}, one finds
%
%
\begin{equation}\bsp
a_{2,\nu n} &\,= i\, \frac{\beta_0}8\, \frac{\chi_{\nu n}}{\partial_\nu \chi_{\nu n}} \,,\\
c_{\nu n} &\,= -i\, a_{1,\nu n}\, \partial_\nu \chi_{\nu n} + i\, \frac{\beta_0}8\, \frac{\chi_{\nu n}\, \partial_\nu^2 \chi_{\nu n}}{\partial_\nu \chi_{\nu n}} \,.
\esp
\end{equation}
We emphasise that the previous equations are only valid for $\nu\neq0$, because the denominator has a simple pole for $\nu=0$, $\partial_\nu \chi_{\nu n}{}_{|\nu=0}=0$. 
We `regulate' this singularity by interpreting the eigenfunctions as distributions, with a principal value prescription for the pole at $\nu=0$,
\begin{equation}\bsp
a_{2,\nu n} &\,= i\, \frac{\beta_0}8\, P\frac{\chi_{\nu n}}{\partial_\nu \chi_{\nu n}} \,,\\
c_{\nu n} &\,= -i\, a_{1,\nu n}\, \partial_\nu \chi_{\nu n} + i\, \frac{\beta_0}8\, P\frac{\chi_{\nu n}\, \partial_\nu^2 \chi_{\nu n}}{\partial_\nu \chi_{\nu n}} \,,
\label{a2nun}
\esp
\end{equation}
where the principal value is defined by
\begin{equation}
\int_{-\infty}^{+\infty}d\nu\left(P\frac{1}{\nu}\right)f(\nu) \equiv \lim_{\eps\to0}\left(\int_{-\infty}^{-\eps}\frac{d\nu}{\nu}f(\nu)+\int_\eps^{+\infty}\frac{d\nu}{\nu}f(\nu)\right)\,.
\end{equation}
Note that if $g_\nu$ is regular at $\nu=0$, we must have 
\begin{equation}
\label{eq:cla_dist_1}
P(g_\nu/\nu) \equiv g_\nu\,P\frac{1}{\nu} {\rm~~and~~} Pg_{\nu} \equiv g_\nu\,. 
\end{equation}
Hence, it is natural to define the principal value for a function $X_\nu$ with a simple pole at $\nu=0$ to be
\begin{equation}
\label{eq:cla_dist_2}
PX_\nu \equiv \nu\,X_\nu\,P\frac{1}{\nu}\,.
\end{equation}
%
%
Then the eigenfunctions can be written as,
\begin{equation}
\Phi_{\nu n}(q) = \varphi_{\nu n}(q)\,\left[ 1 + \overline{\alpha}_\mu \left( a_{0,\nu n} + a_{1,\nu n} \ln\frac{q^2}{\mu^2} 
+ i\, \frac{\beta_0}8\, P\frac{\chi_{\nu n}}{\partial_\nu \chi_{\nu n}} \ln^2\frac{q^2}{\mu^2} \right) +\mathcal{O}(\overline{\alpha}_\mu^2)\right]  \,,
\label{nloeigenfexpl}
\end{equation}
with the coefficients $a_{0,\nu n}$ and $a_{1,\nu n}$ still to be determined.
The free coefficients can be further constrained by requiring the eigenfunctions in eq.~\eqref{nloeigenfexpl} to form a complete and orthonormal set. 
In particular, through NLO the completeness relation for the eigenfunctions implies that
%
%
\begin{equation}\bsp
{\rm Re}[a_{1,\nu n}] &\,= \frac{\beta_0}8\, \partial_\nu P\frac{\chi_{\nu n}}{\partial_\nu \chi_{\nu n}} \,,\\
2{\rm Re}[a_{0,\nu n}] &\,= \partial_\nu {\rm Im}[a_{1,\nu n}] \,.
\label{rea1}
\esp
\end{equation}
%
Thus, after imposing the completeness relation~\eqref{complete}, the NLO eigenfunction can be written as,
\begin{eqnarray}
\Phi_{\nu n}(q) &=& 
\varphi_{\nu n}(q)\,\Bigg[1 + \overline{\alpha}_\mu \Bigg( 
\frac1{2}\, \partial_\nu {\rm Im}[a_{1,\nu n}] + i\, {\rm Im}[a_{0,\nu n}] + i\, {\rm Im}[a_{1,\nu n}]\, \ln\frac{q^2}{\mu^2} \nonumber\\
&&\quad + \frac{\beta_0}8 \ln\frac{q^2}{\mu^2} \,\partial_\nu P\frac{\chi_{\nu n}}{\partial_\nu \chi_{\nu n}}
+\, i\, \frac{\beta_0}8\, \ln^2\frac{q^2}{\mu^2}\,P\frac{\chi_{\nu n}}{\partial_\nu \chi_{\nu n}}  \Bigg)\Bigg] \,.
\label{nloeigenfexpl2}
\end{eqnarray}
The orthogonality condition in eq.~\eqref{orthonorm}
is now automatically fulfilled through NLO and does not add any new constraint.
Hence, the NLO eigenfunctions are determined up to two unknown real parameters, ${\rm Im}[a_{0,\nu n}]$ and ${\rm Im}[a_{1,\nu n}]$,
which can be absorbed through the freedom of defining the phase of the eigenfunctions and the translation invariance of the $\nu$ integral.
Thus, one finds the following result for the NLO eigenfunctions,
\begin{equation}
\Phi_{\nu n}(q) =
\varphi_{\nu n}(q)\left[1 + \overline{\alpha}_\mu \frac{\beta_0}8\, \ln\frac{q^2}{\mu^2}\, \left(  \partial_\nu P\frac{\chi_{\nu n}}{\partial_\nu \chi_{\nu n}}
+\, i\, \ln\frac{q^2}{\mu^2}\,P\frac{\chi_{\nu n}}{\partial_\nu \chi_{\nu n}} \right) + \mathcal{O}(\overline{\alpha}_\mu^2) \right]\,,
\label{nloeigenfexpl4}
\end{equation}
in agreement with ref.~\cite{Chirilli:2014dcb}. Furthermore, with this choice of eigenfunctions, 
the NLO eigenvalue becomes
\begin{equation}
\omega_{\nu n}^{(1)} = \Delta_{\nu n} =  \frac{\delta_{\nu n}}4 -i \frac{\beta_0}8\, \partial_\nu \chi_{\nu n} \,,
\label{nloeigenvmod3}
\end{equation}
where $\delta_{\nu n}$ is given in eq.~\eqref{nloeigenv}.
Equations~\eqref{nloeigenfexpl4} and~\eqref{nloeigenvmod3} are the correct BFKL eigenvalue and eigenfunction through NLO. Let us make some comments about the result. First, we see that the eigenvalue in eq.~\eqref{nloeigenvmod3} is the real part of $\delta_{\nu n}$. Hence, the eigenvalue $\Delta_{\nu n}$ is real and independent of $q^2$, as expected. Note that the eigenvalue is left unchanged for $\cN=4$ SYM, or more generally any conformally-invariant theory. Furthermore, we see that the quantum corrections to the eigenfunction in eq.~\eqref{nloeigenfexpl4} are proportional to the beta function, and so they vanish in a conformally-invariant theory, in agreement with eq.~\eqref{loeigenf}.


\subsection{The BFKL ladder through NLLA}

We now discuss the BFKL ladder through NLLA of eq.~\eqref{eq:GreenGluon} when we use the true NLO eigenvalue and eigenfunctions of eq.~\eqref{nloeigenfexpl4} and~\eqref{nloeigenvmod3}. 
We start by expanding the product of two eigenfunctions through NLO. We have
\begin{equation}\begin{split}
\Phi_{\nu n}&(q_1)\, \Phi_{\nu n}^{\ast}(q_2)\\
&\,= \varphi_{\nu n}(q_1)\,  \varphi^\ast_{\nu n}(q_2)\left[1 + \overline{\alpha}_\mu \frac{\beta_0}8\, \ln\frac{q_1^2q_2^2}{\mu^4}\, \left(  \partial_\nu P\frac{\chi_{\nu n}}{\partial_\nu \chi_{\nu n}}
+\, i\, \ln\frac{q_1^2}{q_2^2}\,P\frac{\chi_{\nu n}}{\partial_\nu \chi_{\nu n}} \right) +\mathcal{O}(\overline{\alpha}_\mu^2) \right]\\
&\,= \varphi_{\nu n}(q_1)\, \varphi^\ast_{\nu n}(q_2)\left[1 + \overline{\alpha}_\mu \frac{\beta_0}4\, \ln\frac{s_{0}}{\mu^2}\,X_{\nu n}(q_1^2/q_2^2) +\mathcal{O}(\overline{\alpha}_\mu^2) \right]\,,
\end{split}\end{equation}
where we defined
\begin{equation}
X_{\nu n}(x)
= \partial_\nu P\frac{\chi_{\nu n}}{\partial_\nu \chi_{\nu n}} + i\, \ln x\,P\frac{\chi_{\nu n}}{\partial_\nu \chi_{\nu n}} \,,
\end{equation}
and the scale $s_0$ is the geometric mean defined in eq.~\eqref{eq:s0}.
Inserting the previous expression into eq.~\eqref{eq:GreenGluon}, we find
\begin{equation}
f(q_1,q_2,y) =\sum_{n=-\infty}^{+\infty} \int_{-\infty}^{+\infty} d\nu\, e^{y\,\omega_{\nu n}}\, \varphi_{\nu n}(q_1)\,  \varphi^\ast_{\nu n}(q_2)
\left(1 + \overline{\alpha}_\mu \frac{\beta_0}4\, \ln\frac{s_{0}}{\mu^2}\,X_{\nu n}(q_1^2/q_2^2)+\mathcal{O}(\overline{\alpha}_\mu^2)  \right)\,.
\end{equation}
{Upon integration by parts, we have}
\begin{equation}\begin{split}
e^{y\,\omega_{\nu n}}\,&\varphi_{\nu n}(q_1)\,  \varphi^\ast_{\nu n}(q_2)\, \partial_\nu P\frac{\chi_{\nu n}}{\partial_\nu \chi_{\nu n}}\\
&\,= -e^{y\,\omega_{\nu n}}\,\varphi_{\nu n}(q_1)\,  \varphi^\ast_{\nu n}(q_2)\left(y\,\partial_\nu\omega_{\nu n}+i\log\frac{q_1^2}{q_2^2}\right)P\frac{\chi_{\nu n}}{\partial_\nu \chi_{\nu n}}\\
&\,= e^{y\,\omega_{\nu n}}\,\varphi_{\nu n}(q_1)\,  \varphi^\ast_{\nu n}(q_2)\left(-y\,\overline{\alpha}_\mu\,\partial_\nu\chi_{\nu n}-i\log\frac{q_1^2}{q_2^2}+\mathcal{O}(\overline{\alpha}_\mu^2)\right)P\frac{\chi_{\nu n}}{\partial_\nu \chi_{\nu n}}\\
&\,= e^{y\,\omega_{\nu n}}\,\varphi_{\nu n}(q_1)\,  \varphi^\ast_{\nu n}(q_2)\left(-i\log\frac{q_1^2}{q_2^2}P\frac{\chi_{\nu n}}{\partial_\nu \chi_{\nu n}}-y\,\overline{\alpha}_\mu\,\chi_{\nu n}+\mathcal{O}(\overline{\alpha}_\mu^2)\right)\,,
\end{split}\end{equation}
and so
\begin{equation}\begin{split}
e^{y\,\omega_{\nu n}}\,&\varphi_{\nu n}(q_1)\,  \varphi^\ast_{\nu n}(q_2)\,X_{\nu n}(q_1^2/q_2^2) = e^{y\,\omega_{\nu n}}\,\varphi_{\nu n}(q_1)\,  \varphi^\ast_{\nu n}(q_2)\,(-y\,\overline{\alpha}_{\mu}\,\chi_{\nu n}+\mathcal{O}(\overline{\alpha}_\mu^2))\,.
\end{split}\end{equation}
Finally, we find
\begin{equation}
f(q_1,q_2,y) 
=\sum_{n=-\infty}^{+\infty} \int_{-\infty}^{+\infty} d\nu\, e^{y\,\omega_{\nu n}}\, \varphi_{\nu n}(q_1)\,  \varphi^\ast_{\nu n}(q_2)
\left(1 - \overline{\alpha}_\mu^2 \frac{\beta_0}4\, \ln\frac{s_{0}}{\mu^2}\,y\,\chi_{\nu n}+\mathcal{O}(\overline{\alpha}_\mu^3)  \right)\,.
\label{eq:GreenGluon2}
\end{equation}
The previous expression for the BFKL ladder is valid through NLLA, and it agrees with the result of refs.~\cite{Chirilli:2013kca,Chirilli:2014dcb}. 
Through NLLA, the term proportional to the $\beta$ function can be interpreted as resetting the scale used in the strong coupling constant. Indeed, we have
\begin{equation}\begin{split}
 \exp y&\left[\overline{\alpha}_S(s_0)\,\chi_{\nu n} + \overline{\alpha}_S(s_0)^2\,\Delta_{\nu n}+\mathcal{O}(\overline{\alpha}_S^3)\right]\\
&\,= \exp y\left[\overline{\alpha}_\mu\,\left(1-\overline{\alpha}_\mu\frac{\beta_0}{4}\,\log\frac{s_0}{\mu^2}\right)\chi_{\nu n} + \overline{\alpha}_\mu^2\,\Delta_{\nu n}+\mathcal{O}(\overline{\alpha}_\mu^3)\right]\\
&\,=e^{y\,\omega_{\nu n}}\left(1-\overline{\alpha}_\mu^2\frac{\beta_0}{4}\,\log\frac{s_0}{\mu^2}\,y\,\chi_{\nu n}+\mathcal{O}(\overline{\alpha}_\mu^3)\right)\,.
\end{split}\end{equation}
Hence, through NLLA, we can cast eq.~\eqref{eq:GreenGluon2} in the equivalent form,
\begin{equation}\begin{split}
\label{eq:f_NLL}
f(q_1,q_2,y) &\,=\sum_{n=-\infty}^{+\infty} \int_{-\infty}^{+\infty} d\nu\,\varphi_{\nu n}(q_1)\,  \varphi^\ast_{\nu n}(q_2) \,e^{y\,\overline{\alpha}_S(s_0)[\chi_{\nu n} + \overline{\alpha}_S(s_0)\Delta_{\nu n}]}
 +\ldots \,,
\end{split}\end{equation}
where the dots indicate terms that are beyond NLLA.
In other words, if we choose the scale of the strong coupling to be the geometric mean of the transverse momenta, $\mu^2=s_0 = \sqrt{q_1^2q_2^2}$, then we can use the LO eigenfunctions instead of the NLO ones.


\section{Analytic results for the BFKL ladder at NLLA in QCD}
\label{sec:nll}

\subsection{Fourier-Mellin representation of the BFKL ladder at NLLA}
In this section we obtain analytic results for the BFKL ladder at NLLA. The discussion from the previous section implies that it is sufficient to study the case where the renormalisation scale is set to the geometric mean of the two transverse momenta. We define
\beq
f^{NLL}(q_1,q_2,\eta_{s_0}) = \frac{1}{2\pi\,\sqrt{q_1^2q_2^2}}\,\sum_{k=1}^{\infty}\frac{\eta_{s_0}^k}{k!}\,f^{NLL}_{k+1}(z)\,,
\eeq
with $\eta_{s_0} = y\,\overline{\alpha}_S(s_0)$. The perturbative coefficients are given by the Fourier-Mellin transform,
\beq\label{eq:FM_NLL}
f^{NLL}_{k}(z)= \cF\left[\Delta_{\nu n}\,\chi_{\nu n}^{k-2}\right] = \sum_{n=-\infty}^{+\infty}\left(\frac{z}{\bar z}\right)^{n/2}\int_{-\infty}^{+\infty}\frac{d\nu}{2\pi}\,|z|^{2i\nu}\,\Delta_{\nu n}\,\chi_{\nu n}^{k-2}\,.
\eeq
Our goal is to develop a strategy to evaluate the Fourier-Mellin transform in eq.~\eqref{eq:FM_NLL}. It will be useful to split the NLO eigenvalue $\Delta_{\nu n}$ into a sum of terms,
\begin{equation}
	\Delta_{\nu n} = \frac{1}{4}\,\delta^{(1)}_{\nu n} + \frac{1}{4}\,\delta^{(2)}_{\nu n} + \frac{1}{4}\,\delta^{(3)}_{\nu n}  +\frac{3}{2}\,\zeta_3+ \gamma_K^{(2)}\chi_{\nu n} - \frac{1}{8} \beta_0 \chi^2_{\nu n}\,,
\end{equation}
where we singled out terms proportional to powers of the LO eigenvalue, because their Fourier-Mellin transform at any order will evaluate to the coefficients appearing in the expansion of the BFKL ladder at LLA, cf. eq.~\eqref{eq:FM_LL}. In QCD, the remaining terms are given by
\begin{align}\label{eq:DELTAS}
 \delta^{(1)}_{\nu n} =& \,\, \partial^2_\nu \chi_{\nu n}\,, \\
 \label{eq:DELTAS_2}
 \delta^{(2)}_{\nu n} =& -2\Phi(n,\gamma) - 2\Phi(n,1-\gamma)\,,\\
 \label{eq:DELTAS_3}
 \delta^{(3)}_{\nu n} =& - \frac{\Gamma(\frac{1}{2}+i\nu)\Gamma(\frac{1}{2}-i\nu)}{2i\nu}\left[ \psi \left( \frac{1}{2} +i\nu \right) - \psi \left( \frac{1}{2} - i\nu \right) \right]\\ &\times\left[ \delta_{n0}\left( 3+ A\frac{2 + 3\gamma (1-\gamma)}{(3-2\gamma)(1+2\gamma)} \right) - \delta_{|n|2}\left(A\frac{\gamma(1-\gamma)}{2(3-2\gamma)(1+2\gamma)} \right) \right]\,, \nonumber
 \end{align}
with
 \begin{equation}
 A = \left( 1 + \frac{N_f}{N^3_c} \right) \,.
\end{equation}
The coefficients in eq.~\eqref{eq:FM_NLL} can then be written as
\beq
\label{eq:NLL_breakdown}
	f^{{NLL}}_{k}(z) = \frac{1}{4}\,C^{(1)}_{k}(z)+\frac{1}{4}\,C^{(2)}_{k}(z)+\frac{1}{4}\,C^{(3)}_{k}(z)
+ \frac{3}{2}\,\zeta_3\,f^{{LL}}_{k-2}(z)	+ \gamma_K^{(2)}\,f^{{LL}}_{k-1}(z) - \frac{1}{8}\, \beta_0\, f^{{LL}}_{k}(z)\,,	
\eeq
where we set $f_0^{LL}(z) = \cF[1] = \pi\,\delta^{(2)}(1-z)$. The only unknowns in eq.~\eqref{eq:NLL_breakdown} are the functions $C_k^{(i)}$, which are defined by
\beq
C^{(i)}_{k}(z) = \cF\left[\delta_{\nu n}^{(i)}\,\chi_{\nu n}^{k-2}\right]\,,
\eeq
with $k\ge 2$.
In the remainder of this section we discuss the computation of each of these quantities in turn.

\subsection{The contribution from $\delta_{\nu n}^{(3)}$}
\label{sec:C3}
We start by discussing the computation of $C^{(3)}_{k}=C_k^{(3,0)}(z)+C_k^{(3,2)}(z)$, where $C_k^{(3,i)}(z)$ is due to the terms proportional to $\delta_{|n|i}$. Since the dependence of $\delta_{\nu n}^{(3)}$ on $n$ is only through Kronecker deltas, the Fourier-Mellin transform reduces to an ordinary Mellin-type integral,
\beq\bsp\label{eq:C3_int_rep}
C_k^{(3)}(z)&\,=-\int_{-\infty}^{+\infty}\frac{d\nu}{2\pi}\,|z|^{2i\nu}\,\frac{\Gamma(\frac{1}{2}+i\nu)\Gamma(\frac{1}{2}-i\nu)}{2i\nu}\left[ \psi \left( \frac{1}{2} +i\nu \right) - \psi \left( \frac{1}{2} - i\nu \right) \right]\\
&\,\qquad\times\left[\chi_{\nu 0}^{k-2}\,A_0(\nu)
+\left(\frac{z}{\overline{z}}+\frac{\overline{z}}{z}\right)\,\chi_{\nu 2}^{k-2}\,A_2(\nu)\right]\,,
\esp\eeq
with
\beq\bsp
A_0(\nu) &\,= 3+ A\,\frac{2 + 3\gamma (1-\gamma)}{(3-2\gamma)(1+2\gamma)} {\rm~~and~~}A_2(\nu)= - A\,\frac{\gamma(1-\gamma)}{2(3-2\gamma)(1+2\gamma)} \,.
\esp\eeq
The integral in eq.~\eqref{eq:C3_int_rep} can be evaluated by closing the contour in the upper half-plane and summing up the residues at $\nu = i\left(\frac{1}{2}+m\right)$, $m\in\mathbb{N}$. The resulting sum of residues can always be performed using the techniques of ref.~\cite{Vermaseren:1998uu,Moch:2001zr,Weinzierl:2004bn}, and the result can be expressed in terms of MPLs of the type $G(a_1,\ldots,a_n;|z|)$, with $a_k\in\{-i,0,i\}$. One can check that these functions are single-valued functions of the complex variable $z$, because the functions have no branch cut on the positive real axis.

Complex conjugation acts in a simple and natural way on this class of functions: it leaves $|z|$ invariant and exchanges the purely imaginary arguments.
It is then natural to decompose the functions into real and imaginary parts. For example, we can write
\beq\bsp
G(0,i;|z|) \pm G(0,-i;|z|)  = \left\{\begin{array}{ll}
  \frac{1}{2} G(0,-1,|z|^2)\,,\\
 2 i\, \Ti_2(|z|)\,,
 \end{array}\right.
\esp\eeq
where $\Ti_n(z)$ are the inverse tangent integrals,
\beq
\Ti_n(z) = \textrm{Im}\,\textrm{Li}_n(i\,z) = -\textrm{Im}\,G(\vec 0_{n-1},1;i\,z)  = \textrm{Im}\,G(\vec 0_{n-1},i;\,z)\,,
\eeq
with $\vec 0_n = (\underbrace{0,\ldots,0}_{n\textrm{ times}})$. We observe that we can always express the results for $C^{(3)}_{k}$ in terms of HPLs of the form $G(b_1,\ldots,b_n;|z|^2)$, with $b_i \in \lbrace -1,0 \rbrace$, and \emph{generalised inverse tangent integrals},
\begin{equation} \label{eqn:invTan}
\Ti_{m_1,\dots,m_k}(|z|) = \textrm{Im}\,\text{Li}_{m_1,\dots,m_k}(\sigma_1,\ldots,\sigma_{k-1}, i\,\sigma_k\, |z|)\,,\qquad \sigma_j = \textrm{sign}(m_j)\,,
\end{equation}
where $\text{Li}_{m_1,\dots,m_k}$ denotes the sum representation of MPLs,
\beq\bsp
\text{Li}_{m_1,\dots,m_k}(z_1,\dots,z_k) &\,= \sum_{0 < n_1<n_2 < \dots < n_k} \frac{z_1^{n_1} \dots z_k^{n_k}}{n_1^{m_1}\dots n_k^{m_k}} \\
&\,= (-1)^k G\Big(\underbrace{0,\dots,0}_{m_k-1},\frac{1}{z_k},\dots,\underbrace{0,\dots,0}_{m_1-1},\frac{1}{z_1\dots z_k};1\Big)\,.
\esp\eeq

The explicit results for the functions $C_k^{(3)}$ in QCD are rather lengthy and are not shown here, can be found through five loops in Appendix~\ref{app:C3results}. 
We observe that neither of the functions $C^{(3,i)}_{k}$ is uniform in transcendental weight, but both $C^{(3,0)}_{k}$ and $C^{(3,2)}_{k}$ involve functions of weight $0\le w\le k$. 

\subsection{The contribution from $\delta_{\nu n}^{(1)}$}
\label{sec:C1result}
The term $\delta_{\nu n}^{(1)}$ is given by the second derivative of the LO eigenvalue $\chi_{\nu n}$, and so the functions $C_k^{(1)}$ can be computed using the same techniques as for the BFKL ladder in LLA, by closing the contour in the upper half of the complex $\nu$-plane and summing the residues of the poles of the polygamma functions. The resulting double sums can be performed in terms of $S$-sums~\cite{Vermaseren:1998uu,Moch:2001zr}. As a result, we find that $C_k^{(1)}$ can be expressed in terms of SVHPLs with singularities at most for $z=0$ and $z=1$, just like at LLA. In the following we describe an alternative method for computing the functions $C_k^{(1)}$, which can be generalised to more general functions, in particular $C_k^{(2)}$.

The Fourier-Mellin transform maps ordinary products into convolutions,
\beq\label{eq:conv_thm}
\cF\left[A_{\nu n}\,B_{\nu n}\right] = \cF\left[A_{\nu n}\right]\ast\cF\left[B_{\nu n}\right] \,,
\eeq
where the convolution product is defined by
\beq
(f\ast g)(z) = \frac{1}{\pi}\int\frac{d^2w}{|w|^2}f(w)\,g\left(\frac{z}{w}\right)\,.
\eeq
We can use eq.~\eqref{eq:conv_thm} to obtain a recursion in the number of loops for the perturbative coefficients~\cite{DelDuca:2016lad},
\beq\label{eq:C1_loop_recursion}
C_{k+1}^{(1)}(z) = \left(\cX \ast C_k^{(1)}\right)(z)\,,
\eeq
with $k\ge 2$, and where we defined
\beq
\cX(z) \equiv\cF\left[\chi_{\nu n}\right] = f_1^{LL}(z)= \frac{|z|}{2\pi\,|1-z|^2}\,.
\eeq
The starting point of the recursion is the two-loop coefficient. In order to compute it, we start by noting that 
\beq
\cF\left[\partial_\nu A_{\nu n}\right] = -i\,\cF\left[A_{\nu n}\right]\,\log|z|^2= -i\,\cF\left[A_{\nu n}\right]\,\cG_0(z)\,.
\eeq
Hence, we find
\beq\label{eq:C212_result}
C_2^{(1)}(z) = \cF\left[\partial^2_\nu \chi_{\nu n}\right] = -\cX(x)\,\log^2|z|^2 = -\frac{|z|}{\pi\,|1-z|^2}\cG_{0,0}(z)\,.
\eeq

Next, we can increase the loop number by convoluting with $\cX(z)$, and the convolution integral can always be reduced to a sum over residues. Indeed, consider a single-valued function $f(z)$ with isolated singularities at $z=a_i$ and $z=\infty$. Close to any of these singularities, $f$ can be expanded into a series of the form,
\beq\bsp
f(z) &\,= \sum_{k,m,n}\,c^{a_i}_{k,m,n}\,\log^k\left|1-\frac{z}{a_i}\right|^2\,(z-a_i)^m\,(\zb-\bar{a}_i)^n\,, \quad z\to a_i\,,\\
f(z) &\,= \sum_{k,m,n}\,c^{\infty}_{k,m,n}\,\log^k\frac{1}{|z|^2}\,\frac{1}{z^m}\,\frac{1}{\zb^n}\,, \quad z\to \infty\,.
\esp\eeq
The \emph{holomorphic residue} of $f$ at the point $z=a$ is then defined as the coefficient of the simple holomorphic pole without logarithmic singularities,
\beq
\textrm{Res}_{z=a}f(z) \equiv c^{a}_{0,-1,0}\,.
\eeq
Antiholomorphic residues $\overline{\textrm{Res}}_{\bar{z}=\bar{a}}f(z)$ are defined in a similar manner.

In ref.~\cite{Schnetz:2013hqa} it was shown that the integral of $f$ over the whole complex plane, if it exists, can be computed in terms of holomorphic residues. More precisely, if $F$ is an antiholomorphic primitive of $f$, $\bar{\partial}_zF=f$, then
\beq
\int \frac{d^2z}{\pi}\,f(z) = \textrm{Res}_{z=\infty}F(z) - \sum_i\textrm{Res}_{z=a_i}F(z)\,.
\eeq
This result is essentially an application of Stokes' theorem to the punctured complex plane.

Since $\cX$ and $C_2^{(1)}$ only have isolated singularities at $z=0$ and $z=1$, we can perform all convolution integrals in terms of holomorphic residues. 
We have computed the functions $C_k^{(1)}$ for $k\le 5$. The results are presented in Appendix~\ref{app:C1results}.
We observe that, up to an overall algebraic prefactor, the functions $C_k^{(1)}$ in eq.~\eqref{eq:C1} are pure functions of weight $k$. In Appendix~\ref{app:proof} we show that this feature is true at any loop order: $C_k^{(1)}$ can be expressed as a linear combination of uniform weight $k$ of SVHPLs with singularities at most at $z=0$ and $z=1$.

\subsection{The contribution from $\delta_{\nu n}^{(2)}$}
We now apply the convolution-based technique from the previous section to the computation of $C_k^{(2)}$. Unlike $\delta_{\nu n}^{(1)}$, the contribution from $\delta_{\nu n}^{(2)}$ cannot be related to the LO eigenvalue. The start of the loop recursion is the two-loop result, which we compute by closing the contour in the complex $\nu$-plane and summing residues. We find
\begin{equation}\label{eq:C22}
	C_2^{(2)}(z) = \mathcal{F}\left[\delta^{(2)}_{\nu n}\right] =C_2^{(2,1)}(z)+C_2^{(2,2)}(z)\,,
\end{equation}
with
\beq\bsp
\label{eq:C221n2}
C_2^{(2,1)}(z) =&  \frac{|z|\,(z-\zb)}{2\pi\,|1+z|^2|1-z|^2}\left[\cG_{1,0}(z)-\cG_{0,1}(z)\right]\,,\\
C_2^{(2,2)}(z) =& \frac{|z|\,(1-|z|^2)}{2\pi\,|1+z|^2|1-z|^2}\left[\cG_{1,0}(z)+\cG_{0,1}(z)-G_{-1,0}\left(|z|^2\right)-\zeta_2\right] \,.
\esp\eeq
We see that the contribution from $\delta_{\nu n}^{(2)}$ is not only very different from the contribution of $\delta_{\nu n}^{(1)}$ in moment space, but also the analytic structure of the Fourier-Mellin transform is very different. First, we see that unlike $C_2^{(1)}$, $C_2^{(2)}$ is not a pure function (up to an overall rational prefactor), but it is the sum of two pure functions $C_2^{(2,1)}$ and $C_2^{(2,2)}$ appearing with different rational prefactors. Second, we see that $C_2^{(2)}$ has a different analytic structure, with singularities at $z=-1$. While $C_2^{(2,1)}$ is a linear combination of SVHPLs with singularities at most at $z=0$ and $z=1$, $C_2^{(2,2)}$ is expressed in terms of both SVHPLs and ordinary HPLs evaluated at $|z|^2$. We note that $C_2^{(2,2)}$ is still single-valued as a function of the complex variable $z$, because the argument of  $G_{-1,0}\left(|z|^2\right)$ is positive-definite and the function has no branch cut on the positive real axis. 
In the next section we review the generalised SVMPLs introduced by Schnetz, which allow us to extend the technique used for the computation of $C_k^{(1)}(z)$ to the functions $C_k^{(2)}(z)$.

\subsubsection{Generalised SVMPLs}
Since the two-loop result in eq.~\eqref{eq:C221n2} is single-valued, all the convolutions resulting from the loop recursion will also be single-valued at any loop order. Hence, we would like to perform the convolution integrals in terms of residues using Stokes' theorem. The single-valued polylogarithms in eq.~\eqref{eq:C221n2}, however, do not all fall into the class of SVMPLs studied in ref.~\cite{BrownSVMPLs,BrownSVHPLs}, because the holomorphic derivative involves non-holomorphic rational functions, e.g.,
\beq\label{eq:der}
\partial_zG_{-1}\left(|z|^2\right) = \frac{1}{z+1/\zb}\,. 
\eeq
A detailed understanding of the functions and their properties is needed for the computation of the holomorphic residues and the antiholomorphic primitives.

In ref.~\cite{Schnetz:2016fhy} Schnetz defined a more general class of single-valued multiple polylogarithms in one complex variable  with singularities at 
\beq\label{eq:loc_sing}
z=\frac{\alpha\,\zb+\beta}{\gamma\,\zb+\delta}\,,\qquad \alpha,\beta,\gamma,\delta\in\mathbb{C}\,.
\eeq
These functions obviously reduce to the SVMPLs of ref.~\cite{BrownSVMPLs,BrownSVHPLs} in the case where the singularities are at constant locations. Since eq.~\eqref{eq:der} has a singularity at $z=-1/\zb$, we expect that the coefficients $\cC_k^{(2,1)}$ can be expressed in terms of Schnetz' generalised SVMPLs (gSVMPLs). In the following we show that that is indeed the case.

We start by reviewing the definition and the construction of gSVMPLs~\cite{Schnetz:2016fhy}. Consider a set of functions $\xG(a_1,\ldots,a_n;z)$ defined by the following conditions,
\begin{enumerate}
\item The functions $\xG(a_1,\ldots,a_n;z)$ are single-valued.
\item They form a shuffle algebra.
\item They satisfy the holomorphic differential equation,
\beq\label{eq:diff_eq_xG}
\partial_z\xG(a_1,\ldots,a_n;z)= \frac{1}{z-a_1}\,\xG(a_2,\ldots,a_n;z)\,.
\eeq
\item They vanish for $z=0$, except if all $a_i$ are 0, in which case we have 
\beq\label{eq:init_cond}
\xG(\underbrace{0,\ldots,0}_{n\textrm{ times}};z) = \frac{1}{n!}\log^n|z|^2\,.
\eeq
\item The singularities in eq.~\eqref{eq:diff_eq_xG} are antiholomorphic functions of $z$ of the form,
\beq\label{eq:sing_def}
a_ i = \frac{\alpha\,\zb+\beta}{\gamma\,\zb+\delta}\,,\textrm{ for some } \alpha,\beta,\gamma,\delta\in\mathbb{C}\,.
\eeq
\end{enumerate}
In ref.~\cite{Schnetz:2016fhy} it is shown that these conditions uniquely define the functions  $\xG(a_1,\ldots,a_n;z)$. Every linear combination $f$ of such functions with singularities at most for $z=a_i$, with $a_i$ defined in eq.~\eqref{eq:sing_def}, has both a single-valued holomorphic and antiholomorphic primitive~\cite{Schnetz:2016fhy}, i.e., there are single-valued functions $F_1$ and $F_2$ such that
\beq
\partial_zF_1 =f = \overline{\partial}_zF_2\,.
\eeq
If we denote a single-valued holomorphic (antiholomorphic) primitive of $f$ by $\int_{\scriptstyle{SV}}dz\, f$ ($\int_{\scriptstyle{SV}}d{\zb}\, f$), then it agrees with any ordinary (i.e., not necessarily single-valued) holomorphic primitive up to an arbitrary antiholomorphic function, e.g.,
\beq
\int_{\scriptstyle{SV}}dz\, f = \delta(\zb) + \int dz\, f\,.
\eeq
In particular, any two single-valued holomorphic primitives must agree up to an antiholomorphic rational function.

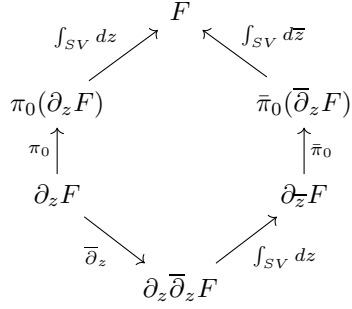
\begin{figure}[!t]
\[
\begin{tikzcd}[row sep = normal, column sep = tiny]
&&F && \\
&{\pi}_0(\partial_{{z}} F) \arrow[ur, "\int_{SV}dz"] & &\bar{\pi}_0(\overline{\partial}_{z} F) \arrow[ul,"\int_{SV}d\zb"']&\\
&\partial_z F \arrow[u, "\pi_0"] \arrow[dr, "\overline{\partial}_{z}"'] & & \partial_{\zb} F \arrow[u, "\bar{\pi}_0"']&\\
&&\partial_z\overline{\partial}_{z}F  \arrow[ur, "\int_{SV}dz"'] &&
\end{tikzcd}
\]
\caption{\label{fig:commutative}Commutative diagrams of ref.~\cite{Schnetz:2016fhy} illustrating the computation of the single-valued primitive.}
\end{figure}

A single-valued primitive $F$ can be computed in an algorithmic way using the commutative diagram of fig.~\ref{fig:commutative}~\cite{Schnetz:2016fhy}, where $\pi_0$ (or $\bar{\pi}_0$) denotes the projection onto the corresponding function with no holomorphic (or antiholomorphic) residues,
\beq\bsp
\pi_0&: f \mapsto f - \sum_{z_0 \in \mathbb{C}} \frac{\text{Res}_{z=z_0}f
}{z-z_0}\,,	\\
\bar{\pi}_0&: f \mapsto f - \sum_{\zb_0 \in \mathbb{C}} \frac{\overline{\text{Res}}_{\zb=\zb_0}f
}{\zb-\zb_0}\,,
\esp\eeq
where the sum runs over all the poles of $f$.

More concretely, assume that we want to compute the single-valued holomorphic primitive $F$ of a single-valued function $f=\partial_zF$ of weight $n$. The commutativity of the diagram in fig.~\ref{fig:commutative} implies that any single-valued holomorphic primitive can  be expressed as a single-valued antiholomorphic primitive. Let us define,
\begin{align}
F_1(z,\zb) = \int dz\, {\pi}_0(\partial_zF(z,\zb)) {\rm~~and~~} \overline{F}_2(z,\zb) = \int d\zb\, \bar{\pi}_0(\overline{\partial}_{z} F(z,\zb))\,.
\end{align}
Note that the diagram in fig.~\ref{fig:commutative} implies that $F_1$ and $\overline{F}_2$ can be computed explicitly if we assume recursively how to compute single-valued primitives of lower weight.
It is clear that these functions must equal the single-valued primitive up to an (anti)holomorphic function,
\begin{align}
F(z,\zb) &= F_1(z,\zb) + \delta_1(\zb)= \overline{F}_2(z,\zb) + \delta_2({z})\,.
\end{align}
From this, one can see that
\begin{equation}
F_1(z,\zb)-\overline{F}_2(z,\zb)=\delta_2(z)-\delta_1(\zb)\,,	
\end{equation}
and so we can determine $\delta_1$ and $\delta_2$ up to a constant from $F_1$ and $\overline{F}_2$. The single-valued primitive of $f$ is then recovered by adding back the residues,
\beq\bsp
F &\,=\delta(\zb)+ \int_{SV}dz\,f\\
&\, = \delta(\zb)+ \int_{SV}dz\,\pi_0(f) + \sum_{z_0 \in \mathbb{C}} \text{Res}_{z=z_0}f\,\int_{SV}\frac{dz}{z-z_0} \\
&\,=\delta(\zb)+ \int_{SV}dz\,\pi_0(f) + \sum_{z_0 \in \mathbb{C}} \text{Res}_{z=z_0}f\,\xG(z_0;z)\,,
\esp\eeq
where $\delta(\zb)$ is any antiholomorphic rational function.

In our case we need to consider gSVMPLs with $a_i\in \{-1,0,1,-1/\zb\}$.
One can use the existence of the single-valued primitive and the commutative diagram of fig.~\ref{fig:commutative} to explicitly construct a basis for the gSVMPLs~\cite{Schnetz:2016fhy}. Assume that we have constructed all gSVMPLs up to weight $n$. Equation~\eqref{eq:diff_eq_xG} implies that gSVMPLs of weight $n+1$ can be obtained by computing single-valued holomorphic primitives. The primitive is only defined up to an arbitrary constant, which can be fixed from eq.~\eqref{eq:init_cond}. Note that the algebra generated by these gSVMPLs contains two natural subalgebras,
\begin{enumerate}
\item If $a_i\neq -1/\zb$, then the gSVMPL reduces to an ordinary SVMPL,
\beq
\xG(a_1,\ldots,a_n;z) = \cG(a_1,\ldots,a_n;z) \,, \qquad \textrm{if }a_i\in\{-1,0,1\}\,.
\eeq
\item If $a_i\neq \pm1$, then the gSVMPL reduces to an ordinary HPL evaluated at $|z|^2$,
\beq
\xG(a_1,\ldots,a_n;z) = G(\zb\,a_1,\ldots,\zb\,a_n;|z|^2) \,, \qquad \textrm{if }a_i\in\{0,-1/\zb\}\,.
\eeq
\end{enumerate}
These subalgebras cover the class of functions encountered in eq.~\eqref{eq:C221n2}. As we will see, at higher loops we obtain functions that cannot be reduced to these two subalgebras.

Let us illustrate the construction of the gSVMPLs on the example of the function $F(z,\zb) = \xG_{-1/\zb,1}(z)$, where we introduce the same shorthand notation as for SVMPLs. Equation~\eqref{eq:diff_eq_xG} implies that $F$ satisfies the holomorphic differential equation,
\beq\label{eq:example1}
\partial_zF(z,\zb) = \partial_z \xG_{-1/\zb,1}(z) = \frac{1}{z+1/\zb}\, \xG_{1}(z) = \frac{1}{z+1/\zb}\, \cG_{1}(z) \equiv f(z,\zb)\,.
\eeq
The initial condition is given by the requirement that $F$ vanishes at the origin. Our goal is thus to compute the holomorphic single-valued primitive of $f$ that vanishes at the origin. Clearly, the denominator in eq.~\eqref{eq:example1} never vanishes. Hence, $f$ is free of poles, and so $\pi_0(f)=f$. We start by computing a non single-valued holomorphic primitive of $f$,
\begin{equation}
F_1(z,\zb) = \int^z_0\frac{dt}{t+1/\zb}\xG_{1}(t)=G_1(\zb)\,G_{-1/\zb}(z)+G_{-1/\zb,1}(z)\,.
\end{equation}
Then there is a antiholomorphic function $\delta_1$ such that
\beq\label{eq:path1}
F(z,\zb) = F_1(z,\zb)+\delta_1(\zb)\,.
\eeq
This completes the evaluation of one branch of the commutative diagram in fig.~\ref{fig:commutative}. The value of $\delta_1$ is at this point undetermined.

Next, let us compute the antiholomorphic derivative of $f$,
\beq
\overline{\partial}_zf(z,\zb) = \frac{1}{(1+|z|^2)^2}\,\cG_{1}(z) + \frac{1}{(\zb-1)\,(z+1/\zb)}\,.
\eeq
All the terms in $\overline{\partial}_zf$ have either lower weight or poles of higher order. Using partial fractioning, we can compute the holomorphic single-valued primitive of $\overline{\partial}_zf$\,
\begin{equation}\bsp
\overline{F}'_2(z,\zb)&\,=\int_{SV}dz\,\overline{\partial}_{z}f(z,\bar{z})\\
&\,=	\frac{1}{\zb+1/z}\,\xG_1(z)-\frac{1}{\zb}\,\xG_{-1/\zb}(z)+\frac{1}{\zb-1}\,\xG_{-1/\zb}(z)+\frac{1}{\zb+1}\,[\xG_{-1/\zb}(z)-\xG_1(z)]\,.
\esp\end{equation}
$\overline{F}'_2$ has antiholomorphic poles at $\zb=0$ and $\zb=\pm1$. The residue at $\zb=0$ vanishes, while
\beq
\overline{\textrm{Res}}_{\zb = \pm1}\overline{F}'_2(z,\zb) = \pm\log2\,.
\eeq
Hence we have
\beq\bsp
\overline{\pi}_0(\overline{F}'_2(z,\zb))&\, = 
\frac{1}{\zb+1/z}\,\xG_1(z)-\frac{1}{\zb}\,\xG_{-1/\zb}(z)+\frac{1}{\zb-1}\,[\xG_{-1/\zb}(z)-\log2]\\
&\,+\frac{1}{\zb+1}\,[\xG_{-1/\zb}(z)-\xG_1(z)+\log 2]\,.
\esp\eeq
An ordinary antiholomorphic primitive of this expression is easily obtained,
\beq\bsp
\overline{F}_2(z,\zb) &\,= \int^{\zb}_0 d\bar{t}\,\overline{\pi}_0(\overline{F}'_2(z,\bar{t}))\\
&\,=G_1(\zb)\,G_{-1/\zb}(z)+G_{-1/\zb,1}(z)-G_{-1,1}(\zb)+\log{2}\,G_{-1}(\zb)-\log{2}\,G_{1}(\zb)\,.
\esp\eeq
Then there is a holomorphic function $\delta_2$ such that
\beq\label{eq:path2}
F(z,\zb) = \overline{F}_2(z,\zb)+\delta_2(z)\,.
\eeq
This completes the evaluation of second branch of the commutative diagram in fig.~\ref{fig:commutative}. The value of $\delta_2$ is at this point undetermined.

The values of $\delta_1$ and $\delta_2$ can be determined by comparing eq.~\eqref{eq:path1} and~\eqref{eq:path2}. We find,
\beq
\delta_1(\zb) -\delta_2(z) = \overline{F}_2(z,\zb) - F_1(z,\zb) = -G_{-1,1}(\zb)+\log{2}\,G_{-1}(\zb)-\log{2}\,G_{1}(\zb)\,,
\eeq
and so
\beq\bsp
\delta_1(\zb) &\,=  -G_{-1,1}(\zb)+\log{2}\,G_{-1}(\zb)-\log{2}\,G_{1}(\zb) + a\,,\\
\delta_2(z) &\, = a\,,
\esp\eeq
for some constant $a\in\mathbb{C}$. The value of $a$ is fixed by requiring $F$ to vanish at the origin, and we find $a=0$. Hence, 
\beq
F(z,\zb) = \xG_{-1/\zb,1}(z) = G_1(\zb)\,G_{-1/\zb}(z)+G_{-1/\zb,1}(z)-G_{-1,1}(\zb)+\log{2}\,G_{-1}(\zb)-\log{2}\,G_{1}(\zb)\,.
\eeq
We have applied the previous algorithm to the construction of all gSVMPLs with $a_i\in \{-1,0,1,-1/\zb\}$ up to weight five. The results are shown in Appendix~\ref{app:SVMPLs} up to weight three.

\subsubsection{The contribution from $C_k^{(2)}$}
Let us now return to the computation of the coefficients $C_k^{(2)}$. We start by writing the two-loop coefficient of eq.~\eqref{eq:C221n2} in terms of gSVMPLs,
\beq
C_{k}^{(2)}(z)  = C_{k}^{(2,1)}(z) + C_{k}^{(2,2)}(z)\,,
\eeq
with
\beq\bsp
C_2^{(2,1)}(z) =&  \frac{|z|\,(z-\zb)}{2\pi\,|1+z|^2|1-z|^2}\left[\xG_{1,0}(z)-\xG_{0,1}(z)\right]\,,\\
C_2^{(2,2)}(z)=& \frac{|z|\,(1-|z|^2)}{2\pi\,|1+z|^2|1-z|^2}\left[\xG_{0,1}(z)+\xG_{1,0}(z)- 2\xG_{-1/z^*,0}(z) -\zeta_2 \right]\,.
\esp\eeq
 Higher loop results can then be obtained from the recursion in the number of loops,
 \beq\label{eq:C2_loop_recursion}
C_{k+1}^{(2)}(z) = \left(\cX\ast C_k^{(2)}\right)(z)\,.
\eeq
Since $1+|z|^2$ is never zero, all singularities of the integrand in the convolution integral are isolated, and so we can use Stokes theorem to reduce the convolution integral to a sum over residues of the single-valued antiholomorphic primitive. The single-valued antiholomorphic primitive can be computed using the algorithm outlined in the previous section. We thus obtain an effective way to obtain higher loop results in terms of gSVMPLs. Results through weight five are collected in Appendix~\ref{app:C2results}. We observe that we can write the results in the form,
\beq\bsp\label{eq:C2k_to_cC}
C_{k}^{(2)}(z) =& \; \frac{|z|\,(z-\zb)}{2\pi\,|1+z|^2|1-z|^2}\cC^{(2,1)}_k + \frac{|z|\,(1-|z|^2)}{2\pi\,|1+z|^2|1-z|^2}\cC^{(2,2)}_k\,,
\esp\eeq
where the functions $\cC^{(2,i)}_k$ have uniform weight $k$. It follows from the argument in Appendix~\ref{app:proof} that this structure holds at any loop order. Finally, we note that at two and three loops, the results can be expressed in terms of SVHPLs and ordinary HPLs evaluated at $|z|^2$. Starting from four loops, we obtain genuine gSVMPLs that can no longer be expressed in terms of HPLs.

\section{Transcendental weight properties of the BFKL ladder at NLLA}
\label{sec:weight}
\subsection{Transcendental weight of the BFKL ladder in QCD}

In the previous section we have determined the BFKL ladder at NLLA in QCD through five loops in momentum space. We have observed that the full result involves polylogarithms of different weight. More precisely, the BFKL ladder at NLLA in QCD at $k$ loops in momentum space involves functions of weights up to $k$, and so the QCD result is not a maximal weight function, as expected. This matches the corresponding analysis in moment space~\cite{Kotikov:2000pm,Kotikov:2002ab,Kotikov:2003fb}, where it was observed that the anomalous dimensions of the leading-twist operators which control the Bjorken scaling violations in ${\cal N}=4$ SYM have a uniform and maximal transcendental weight in moment space which matches the maximal weight part of the corresponding anomalous dimensions in QCD. 
The goal of this section is to extend the analysis of the transcendental weight from moment to momentum space. We start by analysing the BFKL ladder at NLLA in QCD and we identify the terms which contribute to the maximal weight part, before extending the analysis to more general $SU(N_c)$ gauge theories in subsequent sections.

In order to understand the transcendental weight properties of the QCD result, we start from eq.~\eqref{eq:NLL_breakdown} and we classify the contributions which give rise to functions of weight $k$ in momentum space. Using the results of Appendix~\ref{app:proof}, we can show that the functions $f^{{LL}}_{k}$ and $C^{(i)}_{k}$ for $ i \in \{ 1,2 \}$ have uniform weight $k-1$ and $k$ respectively. Hence, we see that terms of weight strictly less than $k$ arise in eq.~\eqref{eq:NLL_breakdown} only in a limited number of places:
\begin{enumerate}
\item The term $\beta_0\, f^{{LL}}_{k}(z)$ involving the beta function has weight $k-1$, and so it is always of lower weight.
\item The contribution from the cusp anomalous dimension in eq.~\eqref{eq:NLL_breakdown}, $\gamma_K^{(2)}\,f^{{LL}}_{k-1}(z)$, involves a mixture of weights $k-2\le w \le k$. Lower weight terms arise entirely because the cusp anomalous dimension in QCD is not of maximal weight.
\item Using the explicit results through five loops of Section~\ref{sec:C3}, we see that in QCD the functions $C^{(3)}_{k}$ involve terms of weight $0\le w\le k$.
\end{enumerate}

Let us compare our analysis in momentum space to the corresponding analysis in moment space of ref.~\cite{Kotikov:2000pm,Kotikov:2002ab,Kotikov:2003fb}, and let us compare the QCD result to the corresponding result in $\cN=4$ SYM. The analysis of the terms involving the beta function and the cusp anomalous dimension is identical in moment and momentum space. The contribution from $C^{(3)}_{k}$, however, is substantially different in QCD and $\cN=4$ SYM, because it vanishes in the maximally supersymmetric Yang-Mills theory~\cite{Kotikov:2000pm} (cf. eq.~\eqref{nloeigenv} and~\eqref{eq:delta_N=4}). Since $C^{(3)}_{k}$ contains terms of weight $k$, we conclude that, unlike for the analysis of anomalous dimensions in moment space, the BFKL ladder in momentum space at NLLA in $\cN=4$ SYM is not equal to the maximal weight terms in QCD. This prompts the question if there is any other theory which agrees with the maximal weight part of the BFKL ladder in QCD order by order in the perturbative expansion. This question will be analysed in the next section.

\subsection{The BFKL ladder in generic gauge theories}
\label{sec:generic}

We study the transcendental weight properties of the BFKL ladder at NLLA in a generic $SU(N_c)$ gauge theory with scalar or fermionic matter in arbitrary representations. Our starting point is the BFKL eigenvalue at NLO in a generic theory~\cite{Kotikov:2000pm}. Inspired by eq.~\eqref{eq:NLL_breakdown}, we write the BFKL eigenvalue in a generic theory as
\beq
{\Delta}_{\nu n}=\frac{1}{4}\delta^{(1)}_{\nu n} +\frac{1}{4}\delta^{(2)}_{\nu n}+\frac{1}{4}\delta^{(3)}_{\nu n}(\tilde{N}_f,\tilde{N}_s)+\frac{3}{2}\zeta_3 + \gamma^{(2)}(\tilde n_f,\tilde n_s)\,\chi_{\nu n} - \frac{1}{8} \beta_0(\tilde n_f,\tilde n_s)\, \chi^2_{\nu n}\,.
\eeq
The quantities $\delta^{(1)}_{\nu n}$ and $\delta^{(2)}_{\nu n}$ are independent of the theory under consideration, and so they are the same as in QCD~\cite{Kotikov:2000pm}, i.e., they are given by eq.~\eqref{eq:DELTAS} and eq.~\eqref{eq:DELTAS_2}. The one-loop beta function and the two-loop cusp anomalous dimension depend on the details of the theory only through one-loop corrections to the gluon propagator, and so they are independent of the details of the theory (e.g., Yukawa couplings or the scalar potential). In DRED, they are given by
\beq\bsp
\beta_0(\tilde n_f,\tilde n_s) &\,= \frac{11}{3}- \frac{2\tilde{n}_f}{3 N_c} - \frac{\tilde{n}_s}{6 N_c}\,,\\
\gamma^{(2)}(\tilde n_f,\tilde n_s) &\, = \frac{1}{4}\left( \frac{64}{9} - \frac{10 \, \tilde{n}_f}{9 N_c}- \frac{4 \, \tilde{n}_s}{9 N_c}\right) - \frac{\zeta_2}{2}\,,
\esp\eeq
where we defined
\beq
\tilde{n}_f = \sum_R n_f^R T_R {\rm~~and~~}\tilde{n}_s = \sum_R n_s^R T_R\,,
\eeq
where the sum runs over all irreducible representations $R$ of $SU(N_c)$, and $n_f^R$ and $n_s^R$ denote the number of Weyl fermions and real scalars transforming in the representation $R$. The index $T_R$ of the representation is defined through $\textrm{Tr}(T_R^aT_R^b) = T_R\,\delta^{ab}$, with $T_R^a$ the infinitesimal generators of the representation $R$. We fix the normalisation of the structure constants of $SU(N_c)$ such that for the fundamental representation $T_F=1/2$. The contribution from $\delta_{\nu n}^{(3)}$ comes from (scalar) QED-type diagrams~\cite{Kotikov:2000pm}, and it is determined entirely by the matter content of the theory. We find
\beq
\delta^{(3)}_{\nu n}(\tilde{N}_f,\tilde{N}_s) = \delta^{(3,1)}_{\nu n}(\tilde{N}_f,\tilde{N}_s) + \delta^{(3,2)}_{\nu n}(\tilde{N}_f,\tilde{N}_s)\,,
\eeq
with
\beq
\tilde{N}_x = \frac{1}{2}\sum_R n^R_x \,T_R\,(2C_R-N_c)\,,\qquad x=f,s\,,
\eeq
and $C_R$ is the quadratic Casimir of the representation $R$. The functions $\delta^{(3,i)}_{\nu n}$ are given by
\beq\bsp
\delta^{(3,1)}_{\nu n}(\tilde{N}_f,\tilde{N}_s)&\,=\frac{f(\gamma)}{8}\left[\delta_{n0}\left(2\tilde{N}_s+12\tilde{N}_f -30 N_c^2\right)+ \delta_{|n|2}\left(N_c^2 -2\tilde{N}_f +\tilde{N}_s\right)\right]\,, \\
\delta^{(3,2)}_{\nu n}(\tilde{N}_f,\tilde{N}_s)&\,=\frac{f(\gamma)}{8}\left[\frac{(3\delta_{|n|2}-2\delta_{n0})(2\gamma-1)}{2(2\gamma-3)(2\gamma+1)}\left(N_c^2 -2\tilde{N}_f +\tilde{N}_s\right)\right]\,,
\esp\eeq
with $\gamma=\frac{1}{2}+i\nu$ and
\beq
f(\gamma)=\frac{1}{4\pi^2(1-2\gamma)} \Gamma(1-\gamma)\Gamma(\gamma) \bigg[ \psi(1-\gamma)-\psi(\gamma) \bigg]\,.
\eeq
We have checked by explicit computations through five loops that the Fourier-Mellin transforms of the type $\cF\left[\delta^{(3,1)}_{\nu n}\,\chi_{\nu n}^{k}\right]$ give rise to functions of uniform weight $k+2$, while the remaining contributions from $\delta^{(3,2)}_{\nu n}$ only produce lower weight terms. While we currently have no proof that this statement holds at arbitrary loop orders, we believe that our explicit results through five loops provide compelling evidence that this is indeed the case.

In a theory where the gauge group is minimally coupled to matter, the BFKL eigenvalue at NLLA is determined entirely by the gauge group and matter content of the theory~\cite{Kotikov:2000pm}, but it is independent of the details of the other interactions in the theory (e.g., the Yukawa couplings between the fermions and the scalar). As a consequence, we can repeat the analysis of the transcendental weight properties for generic gauge theories as a function of the fermionic and scalar matter content of the theory. Following our analysis in QCD in the previous section, a set of necessary and sufficient conditions for a theory to have a BFKL ladder at NLLA of uniform transcendental weight in momentum space are:
\begin{enumerate}
\item The one-loop beta function vanishes, i.e., we have
\beq\label{eq:beta_constraint}
\frac{11}{3}- \frac{2\tilde{n}_f}{3 N_c} - \frac{\tilde{n}_s}{6 N_c} = 0\,.
\eeq
\item The two-loop cusp anomalous dimension is proportional to $\zeta_2$. In DRED, this implies the following constraint on the matter content,
\beq\label{eq:gamma_constraint}
\frac{16}{9} - \frac{5 \, \tilde{n}_f}{18 N_c}- \frac{\, \tilde{n}_s}{9 N_c}=0\,.
\eeq
\item The contribution from $\delta^{(3,2)}_{\nu n}$ vanishes, which implies
\beq\label{eq:delta_constraint}
2\tilde{N}_f = N_c^2 +\tilde{N}_s\,.
\eeq
\end{enumerate}
We interpret eq.~\eqref{eq:beta_constraint}, \eqref{eq:gamma_constraint} and~\eqref{eq:delta_constraint} as a set of conditions on the matter content of a theory for the BFKL ladder at NLLA to have maximal weight. Before we solve the constraints in the next section in the case of adjoint and fundamental matter, we make the following observation which is independent of the representations of the matter fields. If eq.~\eqref{eq:delta_constraint} is satisfied, then the term proportional to $\delta_{|n|2}$ is absent from $\delta^{(3,1)}_{\nu n}$,
\beq
\delta^{(3,1)}_{\nu n}(\tilde{N}_f,2\tilde{N}_f-N_c^2) =  2f(\gamma) \,\delta_{n0}\,\left(\tilde{N}_f -2 N_c^2\right)\,.
\eeq
As the missing terms evaluate to terms of maximal weight, we are led to conclude that there is no theory such that the BFKL ladder at NLLA has uniform and maximal weight and agrees with the maximal weight terms in QCD.

\subsection{Theories with adjoint and fundamental matter}
\label{sec:fund_adj}
In this section we study the conditions in eq.~\eqref{eq:beta_constraint}, \eqref{eq:gamma_constraint} and~\eqref{eq:delta_constraint} in the case of theories with matter only in the fundamental and adjoint representations. The indices and Casimir operators of the adjoint and fundamental representations are
\beq
T_A = C_A = N_c {\rm~~and~~} C_F = T_F\,\frac{N_c^2-1}{N_c}\,.
\eeq
We can then write
\beq\bsp\label{eq:Ns_F_A}
\tilde{n}_x &\,= T_A n^{A}_{x} + T_F\,n^{F}_{x} {\rm~~and~~} \widetilde{N}_{x} = \frac{1}{2}T_A\,(2C_A-N_c)n^{A}_{x}+\frac{1}{2}\,T_F\,(2C_F-N_c) n_x^F\,, \qquad x=s,f\,.
\esp\eeq

In the following we only look for solutions to the constraints in eq.~\eqref{eq:beta_constraint}, \eqref{eq:gamma_constraint} and~\eqref{eq:delta_constraint} that are valid for an arbitrary number $N_c$ of colours\footnote{We have also searched for other solutions by varying the parameters $N_c$, $n_f^F$, $n_s^F$, $n_f^{A}$ and $n_s^{A}$ independently between 0 and 35, and we have not found any other solutions than those described here. We therefore  conjecture that these are the only solutions.}. Inserting eq.~\eqref{eq:Ns_F_A} into eq.~\eqref{eq:delta_constraint}, we find
\beq
T_F\,(2C_F-N_c)\,n_f^F + N_c^2\,n_f^{A} = 
N_c^2 + \frac{1}{2}N_c^2\,n_s^{A} +\frac{1}{2}\,T_F\,(2C_F-N_c)\,n_s^{F}\,.
\eeq
Since we are looking for solutions that are valid for an arbitrary number of colours, we find
a relation between the number of fermions and scalars in the adjoint and fundamental representations,
\beq\bsp\label{eq:2nff=nsf}
2\,n_f^F&\, = n_s^F {\rm~~and~~}2\,n_f^{A} = 2 +n_s^{A}\,.
\esp\eeq
Inserting this solution into the constraints ~\eqref{eq:beta_constraint} and ~\eqref{eq:gamma_constraint}, we see that the number of fermions in the adjoint and fundamental representations must be related by
\beq\label{eq:zerobeta}
4-n_f^A -T_F\frac{n_f^F}{N_c}= 0\,.
\eeq
The relations~\eqref{eq:2nff=nsf} and~\eqref{eq:zerobeta} are necessary conditions for a gauge theory to have a BFKL ladder at NLLA of uniform and maximal transcendental weight.

At this point we observe that eq.~\eqref{eq:2nff=nsf} describes the spectrum of a gauge theory with $\cN$ supersymmetries and $n_F\equiv n_f^F$ chiral multiplets\footnote{We consider chiral multiplets in $\cN=1$ supersymmetry, consisting of a Weyl fermion and two real scalar on-shell degrees of freedom.} in the fundamental representation and $n_A\equiv n_f^{A} - \cN$ chiral multiplets in the adjoint representation. Indeed, in terms of the parameters $\cN$, $n_F$ and $n_A$, eq.~\eqref{eq:2nff=nsf} can be cast in the form,
\beq\bsp\label{eq:susy_constraints}
n_f^{A} &\,= \cN+n_A\,,\\
 n_f^{F}&\, = n_{F}\,,\\
n_s^{A} &\,= 2(\cN-1)+2n_A\,,\\
 n_s^{F} &\,= 2 n_{F}\,.
\esp\eeq
We stress that our analysis can only constrain the scalar and fermionic matter of the theory, and it is insensitive to other aspects like supersymmetry. As a consequence, classifying the matter content in terms of supersymmetric multiplets is at this point merely a matter of convenience, and any theory with the same fermionic and scalar matter content would equally well solve the constraints, independently of supersymmetry.

We can insert eq.~\eqref{eq:susy_constraints} into eq.~\eqref{eq:zerobeta}, and obtain the equation,
\beq\label{susyconstr}
n_A+\frac{n_F}{2N_c}+\cN=4\,.
\eeq 
This equation has only four positive integer solutions which are shown in Table~\ref{tab:solutions}. For these theories, the BFKL eigenvalue at NLO takes the form,
\beq
\Delta_{\nu n} = \frac{1}{4}\delta^{(1)}_{\nu n} + \frac{1}{4}\delta^{(2)}_{\nu n} +\frac{3}{2}\zeta_3 + \frac{\zeta_2}{2}\chi_{\nu n}+f(\gamma)\,(N_c^2+1)\,(n_f^{A}-4)\,\delta_{n 0}\,.
\eeq
It is easy to see that theories with $(\cN,n_A,n_F) = (4,0,0)$ have the same field content as $\cN=4$ SYM, and so $\cN=4$ SYM satisfies all the constraints, as expected.
Similarly, theories with $(\cN,n_A,n_F) = (2,0,4N_c)$ have the same field content as $\cN=2$ superconformal QCD with $N_f=2N_c$ hypermultiplets~\cite{Gadde:2009dj}. The remaining theories have the field content of an $\cN=1$ theory. We observe that $(\cN,n_A,n_F) = (1,0,6N_c)$ corresponds to the matter content of $\cN=1$ super-QCD. Moreover, the matter content is such that the theory is at the upper end of the conformal window~\cite{Seiberg:1994pq}. Note that the lower end of the conformal window, corresponding to $(\cN,n_A,n_F) = (1,0,3N_c)$, does not solve our constraints for uniform and maximal transcendental weight. We currently do not have any interpretation of the second $\cN=1$ solution as a superconformal theory.

\begin{center}
\begin{table}[!t]
\begin{center}
\begin{tabular}{c|cccc}
\hline\hline
$\cN$ & 4 & 2 & 1 &1\\
$n_A$ & 0 & 0&0 &2\\
$n_F$ & 0 & $4N_c$&$6N_c$ &$2N_c$\\
\hline\hline
\end{tabular}
\caption{\label{tab:solutions}The four solutions to the constraints in eq.~\eqref{eq:susy_constraints} into eq.~\eqref{eq:beta_constraint} and~\eqref{eq:gamma_constraint}.}
\end{center}
\end{table}
\end{center}

\subsection{Discussion}
In the previous sections we have derived a set of conditions on the matter content of a gauge theory so that the BFKL ladder at NLLA is a function of uniform transcendental weight in momentum space. In this section we discuss some implications of our analysis.

Our analysis is valid only for a specific gauge-theory correlator, namely the BFKL ladder at NLLA, and so we can strictly speaking not make any statement about generic scattering amplitudes or correlations functions. In other words, we cannot exclude that there are theories that do not fulfil the criteria of the previous sections, but where all the scattering amplitudes or other correlatiors are functions of uniform and maximal weight. However, we deem it unnatural that the same mechanism which would make scattering amplitudes and correlations functions have uniform weight would fail for the BFKL ladder. We therefore expect that the conditions derived in the previous section are more generally necessary conditions for a theory to have the property of maximal transcendental weight similar to the $\cN=4$ SYM theory.

The analysis of the previous section shows that the theories with the maximal weight property have a highly constrained matter content. We find it intriguing that in all cases the field content can be arranged into supersymmetric multiplets, although supersymmetry was not an input to our analysis. {The different choices for the matter content that solve all the constraints, arranged into supersymmetric multiplets, is shown in Table~\ref{tab:solutions}. Note that we cannot distinguish, e.g., an $\cN=2$ theory from an $\cN=1$ theory with additional chiral multiplets in the adjoint representation.} Moreover, we find that the vanishing of the beta function is not independent from the other constraints, but eq.~\eqref{eq:gamma_constraint} and~\eqref{eq:delta_constraint} alone are sufficient to find all the solutions in Table~\ref{tab:solutions}, and all of these solutions have a vanishing one-loop beta function. The $\cN=4$ and $\cN=2$ solutions are known to be superconformal, while one of the $\cN=1$ solutions in Table~\ref{tab:solutions} has a matter content that puts it right at the edge of the conformal window of $\cN=1$ super-QCD. We currently ignore if there is a superconformal theory that matches the second $\cN=1$ solution, but if there is, this may point to the intriguing possibility of a connection between the maximal weight property and superconformal symmetry.

We stress that our results are a necessary condition for a theory to have the property of maximal weight. Indeed, even if the BFKL ladder is of uniform and maximal weight, this may not be the case for other quantities in the same theory. It is therefore interesting to analyse the theories that we have identified further.
We have checked that all theories with a matter content as in Table~\ref{tab:solutions} give rise to one-loop amplitudes with maximal transcendental weight~\cite{Bern:1993mq,Bern:1993tz,Bern:1994zx,Dixon:1996wi}. The same conclusion holds for the single-emission soft gluon current through two loops~\cite{Catani:2000pi,Duhr:2013msa,Li:2013lsa} in all of these theories. The cusp anomalous dimension and the beta function are known to three~\cite{Korchemsky:1987wg,Moch:2004pa,Grozin:2014hna,Grozin:2015kna} and five loops~\cite{Caswell:1974gg,Jones:1974mm,Tarasov:1980au,Larin:1993tp,vanRitbergen:1997va,Czakon:2004bu,Herzog:2017ohr} respectively. These quantities, however, depend on the details of the theory under consideration (e.g., Yukawa and scalar couplings), and so it is hard to make generic statements only based on the matter content of the theory. 
It would be interesting to confront the theories described in Table~\ref{tab:solutions} to explicit computations, in order to further constrain the space of possible maximal weight theories. 
For example, it is known that the four-point two-loop amplitude in $\cN=2$ superconformal QCD (SCQCD) does not have maximal weight~\cite{Andree:2010na,Leoni:2014fja,Leoni:2015zxa}, which rules out $\cN=2$ SCQCD as a candidate for a maximal weight theory like $\cN=4$ SYM. 

We conclude this section by commenting on possible shortcomings of our analysis. First, our results only hold for $SU(N_c)$ gauge theories in four dimensions with additional scalar or fermionic matter in the adjoint or fundamental representations. It would be interesting to repeat the analysis of Sections~\ref{sec:generic} and~\ref{sec:fund_adj} for matter transforming in other irreducible representations of $SU(N_c)$. Second, since our analysis relies on the BFKL ladder, the conclusions are only valid for theories where the gauge bosons are minimally coupled to the matter. Indeed, if additional (higher-dimensional) operators are present in the theory, they may alter the high-energy behaviour of the theory, and therefore the BFKL ladder, which would invalidate our analysis.

\section{Concluding remarks}
\label{sec:concl}

In this paper, we have addressed three questions related to the NLO corrections to the eigenvalue of the BFKL equation.
Firstly, we have noted that the NLO corrections to the eigenfunctions computed by Chirilli and Kovchegov
can be made to vanish by taking the scale of the coupling to be the geometric mean of the transverse momenta at the ends of the BFKL ladder. Secondly, we have found the functions which describe the analytic structure of the BFKL ladder at NLLA. These are the gSVMPLs
recently introduced by Schnetz~\cite{Schnetz:2016fhy}, and we have developed techniques to evaluate the BFKL ladder at NLLA to high loop order. Finally, using the freedom in defining the matter content of the NLO BFKL eigenvalue, we have proven that there is no gauge theory of uniform and maximal transcendental weight such that in momentum space it matches the maximal weight part of QCD. However, we have identified a set of conditions which allow us to constrain the field content of theories for which the BFKL ladder has maximal weight.

Let us comment on potential ramifications of our work. First, we have only been concerned with the study of formal analytic properties of the BFKL ladder at NLLA. It would be interesting to investigate potential phenomenological applications of our results. Second, it is natural to ask if or how some of the properties of the BFKL ladder that we have studied manifest themselves also beyond NLLA. A study of the BFKL ladder beyond NLLA, however, is yet to be undertaken. The keystone of the BFKL theory is the
gluon Reggeisation at leading logarithmic~\cite{Lipatov:1976zz,Balitsky:1979ap} and NLL~\cite{Fadin:2006bj,Kozlov:2011zza,Kozlov:2012zza,Kozlov:2012zz,Fadin:2015zea} accuracy. This entails the multi-Regge pole structure of (the real part of) the QCD amplitudes in multi-Regge kinematics and the
Regge factorisation of those amplitudes. We already know that the Regge factorisation is broken at NNLO accuracy~\cite{DelDuca:2001gu}.
The violation can be explained through the infrared structure of massless gauge theories by showing that the real part of the amplitudes becomes non-diagonal in the $t$-channel-exchange basis~\cite{Bret:2011xm,DelDuca:2011ae}. Accordingly, it can be predicted how the violation propagates to higher loops, and the three-loop prediction for the violation~\cite{DelDuca:2013ara,DelDuca:2014cya} has been confirmed by the explicit computation of the three-loop four-point function of ${\cal N}=4$ SYM~\cite{Henn:2016jdu}. In the Regge theory, that violation is due to the contribution of the three-Reggeised-gluon exchange~\cite{Fadin:2016wso,Caron-Huot:2017fxr}. Thus, it is conceivable that the violations of the BFKL-ladder structure can be computed and kept under control, allowing a study of the BFKL ladder beyond NLLA, which would likely provide more analytic tools to examine the behaviour of QCD and of massless gauge theories, and thereby help to unravel even more the fascinating mathematical structure underlying gauge theory amplitudes.

\section*{Acknowledgements}
The authors are grateful to Lance Dixon and Falko Dulat for inspiring discussions and a critical reading of the manuscript, and in particular to Lance Dixon for collaboration during the early stages of this project.
CD, RM and BV acknowledge the hospitality of the ETH Zurich at various stages of this project and RM and BV also acknowledge the hospitality of the Theory Department at CERN. This work is supported by the European Research Council
(ERC) under the Horizon 2020 Research and Innovation Programme through the grant 637019 (MathAm).


\appendix

\section{Pure functions from convolutions}\label{app:proof}

In this appendix we present the proof that the functions $f_k^{LL}$, $C_k^{(1)}$ and $C_k^{(2)}$ have uniform weight at any loop order.
We first prove the following result: assume that we have a function $a_{\nu n}$ in moment space whose Fourier-Mellin transform evaluates to an expression of the form,
\beq\label{eq:proof_start}
\cF[a_{\nu n}] = \frac{|z|{A}(z) }{(z-b)(\zb - c)}\,,
\eeq
where $A(z)$ is a single-valued pure function of weight $k$, and $b$ and $c$ are complex numbers. We wish to show that $\cF[a_{\nu n}\chi_{\nu n}]$ consists of a single-valued pure function of weight $k+1$ multiplied by the same rational prefactor. Using the convolution product, we find
\beq\bsp
\cF[a_{\nu n} \chi_{\nu n}]&=\cF[a_{\nu n}]*\cF[\chi_{\nu n}] \\
&= \int d^2 \omega \frac{|z|{A}(\omega) }{(\omega-b)(\omegab - c)(\omega-z)(\omegab-\zb)}\\
&=\frac{|z|}{(z-b)(\zb-c)}\int d^2 \omega \left( \frac{1}{\omega-z}-\frac{1}{\omega-b} \right)\left( \frac{1}{\omegab-\zb} - \frac{1}{\omegab-c} \right)A(\omega)\,.
\esp\eeq
We can solve this integral in terms of residues as described in Section~\ref{sec:C1result}. We start by computing the single-valued primitive,
\beq
\int_{SV} d \omegab \left( \frac{1}{\omegab-\zb} - \frac{1}{\omegab-c} \right)A(\omega) \equiv \widetilde{A}(\omega)\,.
\eeq
Since $A(\omega)$ is assumed to be a pure function of weight $k$, $\widetilde{A}(\omega)$ is a pure function of weight $k+1$.
The holomorphic residues are 
\beq\bsp
\cF[a_{\nu n} \chi_{\nu n}] &= \frac{|z|}{(z-b)(\zb-c)}\left( \text{Res}_{\omega = b}\frac{\widetilde{A}(\omega)}{\omega-b} - \text{Res}_{\omega = z}\frac{\widetilde{A}(\omega)}{\omega-z} \right)\\
    &= \frac{|z|}{(z-b)(\zb-c)}\left(\textrm{Reg}_{\omega=b}\widetilde{A}(\omega)-\textrm{Reg}_{\omega=z}\widetilde{A}(\omega)\right)\,,
\esp\eeq
where $\textrm{Reg}_{\omega=a}\widetilde{A}(\omega)$ denotes the shuffle-regulated value of $\widetilde{A}(\omega)$ at the point $\omega=a$, defined as follows: since $\widetilde{A}$ has weight $k+1$, close to every point $\omega=a$ it admits an expansion of the type,
\beq
\widetilde{A}(\omega) = \sum_{i=0}^{k+1}\log^i\left|1-\frac{\omega}{a}\right|^2\,\widetilde{A}^{(i)}(\omega)\,,
\eeq
where the functions $\widetilde{A}^{(i)}$ are analytic at $\omega=a$, i.e., they admit a Taylor expansion in a neighbourhood of $\omega=a$. The shuffle-regulated value of $\widetilde{A}$ at $\omega=a$ is then defined as
\beq
\textrm{Reg}_{\omega=a}\widetilde{A}(\omega) \equiv \widetilde{A}^{(0)}(a)\,.
\eeq
Since $\widetilde{A}(\omega)$ is a pure function of weight $k+1$, its shuffle-regulated values remain pure and have the same weight. We have thus shown that
\beq
\cF[a_{\nu n} \chi_{\nu n}](z) = \frac{|z|}{(z-b)(\zb-c)}\left[\widetilde{A}^{(0)}(b)-\widetilde{A}^{(0)}(z)\right]\,,
\eeq
where the right-hand side is a pure function of weight $k+1$, which completes the proof.

We can use the previous result to prove that the functions $f_k^{LL}$, $C_k^{(1)}$ and $C_k^{(2)}$ have uniform weight at any loop order. We know by explicit computation that this statement is true for small numbers of loops. Using partial fractioning, the loop recursion in eq.~\eqref{eq:C1_loop_recursion} and~\eqref{eq:C2_loop_recursion} can be written in a form that matches eq.~\eqref{eq:proof_start}. Hence, the loop recursion will increase the weight of the functions by precisely one unit.

\section{Generalised single-valued multiple polylogarithms}
\label{app:SVMPLs}
In this section we present the generalised single-valued polylogarithms that cannot be expressed in terms of ordinary SVMPLs,
\enlargethispage{1cm}
\begin{align}
\xG_{-1/\zb}(z)=&\,G_{-1/\zb}(z) \nonumber \\
\xG_{-1,-1/\zb}(z)=&\,G_{-1,-1/\zb}(z)+G_{1,-1}(\zb)+\log 2\,G_{-1}(\zb)-\log 2\,G_1(\zb)\nonumber \\ 
\xG_{0,-1/\zb}(z)=&\,G_{0,-1/\zb}(z)\nonumber \\ 
\xG_{1,-1/\zb}(z)=&\,G_{-1,1}(\zb)+G_{1,-1/\zb}(z)-\log 2\,G_{-1}(\zb)+\log 2\,G_1(\zb)\nonumber \\ 
\xG_{-1/\zb,-1}(z)=&\,-G_{1,-1}(\zb)+G_{-1/\zb,-1}(z)+G_{-1/\zb}(z)G_{-1}(\zb)-\log 2\,G_{-1}(\zb)+\log 2\,G_1(\zb)\nonumber \\ 
\xG_{-1/\zb,0}(z)=&\,G_{-1/\zb,0}(z)+G_0(\zb)G_{-1/\zb}(z)\nonumber \\ 
\xG_{-1/\zb,1}(z)=&\,-G_{-1,1}(\zb)+G_{-1/\zb,1}(z)+G_1(\zb)G_{-1/\zb}(z)+\log 2\,G_{-1}(\zb)-\log 2\,G_1(\zb)\nonumber \\ 
\xG_{-1/\zb,-1/\zb}(z)=&\,G_{-1/\zb,-1/\zb}(z) \nonumber \\
%
\xG_{-1,-1,-1/\zb}(z)=&\,G_{-1}(z)G_{1,-1}(\zb)+G_{-1,-1,-1/\zb}(z)+G_{0,1,-1}(\zb)+G_{1,-1,-1}(\zb)-G_{1,1,-1}(\zb) \nonumber \\
& + \log 2\,G_{-1,-1}(\zb)+\log 2\,G_{0,-1}(\zb)-\log 2\,G_{0,1}(\zb)-2\log 2\,G_{1,-1}(\zb) \nonumber \\
& +\log 2\,G_{1,1}(\zb)+\frac{1}{2}\zeta_2G_1(\zb)-\log^22G_{-1}(\zb)+\log^22G_1(\zb) \nonumber \\
& +\log 2\,G_{-1}(z)G_{-1}(\zb)-\log 2\,G_{-1}(z)G_1(\zb)\nonumber \\ 
\xG_{-1,0,-1/\zb}(z)=&\,G_{-1,0,-1/\zb}(z)+G_{0,1,-1}(\zb)+\log 2\,G_{0,-1}(\zb)-\log 2\,G_{0,1}(\zb)+\frac{1}{2}\zeta_2G_{-1}(\zb)\nonumber \\ 
\xG_{-1,1,-1/\zb}(z)=&\,G_{-1}(z)G_{-1,1}(\zb)+G_{-1,1,-1/\zb}(z)+G_{0,1,-1}(\zb)-\log 2\,G_{-1,1}(\zb)+\log 2\,G_{0,-1}(\zb) \nonumber \\
& -\log 2\,G_{0,1}(\zb)+\log 2\,G_{1,-1}(\zb)+\frac{1}{2}\zeta_2G_{-1}(\zb)+2\log^22G_{-1}(\zb) \nonumber \\
& -2\log^22G_1(\zb)-\log 2\,G_{-1}(z)G_{-1}(\zb)+\log 2\,G_{-1}(z)G_1(\zb)\nonumber \\ 
\xG_{-1,-1/\zb,-1}(z)=&\,G_{-1}(\zb)G_{-1,-1/\zb}(z)-G_{-1}(z)G_{1,-1}(\zb)+G_{-1,1,-1}(\zb)+G_{-1,-1/\zb,-1}(z) \nonumber \\
& -2G_{0,1,-1}(\zb)+2G_{1,1,-1}(\zb)-\log 2\,G_{-1,1}(\zb)-2\log 2\,G_{0,-1}(\zb) \nonumber \\
& +2\log 2\,G_{0,1}(\zb)+3\log 2\,G_{1,-1}(\zb)-2\log 2\,G_{1,1}(\zb)-\zeta_2G_1(\zb) \nonumber \\
& +2\log^22G_{-1}(\zb)-2\log^22G_1(\zb)-\log 2\,G_{-1}(z)G_{-1}(\zb) \nonumber \\
& +\log 2\,G_{-1}(z)G_1(\zb)\nonumber \\ 
\xG_{-1,-1/\zb,0}(z)=&\,G_0(\zb)G_{-1,-1/\zb}(z)+G_{-1,-1/\zb,0}(z)+G_{1,0,-1}(\zb)-\frac{1}{2}\zeta_2G_{-1}(\zb)-\frac{1}{2}\zeta_2G_1(\zb)\nonumber \\ 
\xG_{-1,-1/\zb,1}(z)=&\,-G_{-1}(z)G_{-1,1}(\zb)+G_1(\zb)G_{-1,-1/\zb}(z)+G_{-1,-1/\zb,1}(z)-G_{0,-1,1}(\zb) \nonumber \\
& -G_{0,1,-1}(\zb)+G_{1,-1,1}(\zb)+2G_{1,1,-1}(\zb)+\log 2\,G_{-1,1}(\zb)+\log 2\,G_{1,-1}(\zb) \nonumber \\
& -2\log 2\,G_{1,1}(\zb)-\frac{1}{2}\zeta_2G_{-1}(\zb)-\frac{1}{2}\zeta_2G_1(\zb)+\log 2\,G_{-1}(z)G_{-1}(\zb) \nonumber \\
& -\log 2\,G_{-1}(z)G_1(\zb)\nonumber \\ 
\xG_{-1,-1/\zb,-1/\zb}(z)=&\,G_{-1,-1/\zb,-1/\zb}(z)+G_{1,1,-1}(\zb)+\log 2\,G_{1,-1}(\zb)-\log 2\,G_{1,1}(\zb) \nonumber \\
& -\frac{1}{2}\zeta_2G_1(\zb)+\frac{1}{2}\log^22G_{-1}(\zb)-\frac{1}{2}\log^22G_1(\zb)\nonumber \\ 
\xG_{0,-1,-1/\zb}(z)=&\,G_0(z)G_{1,-1}(\zb)+G_{0,-1,-1/\zb}(z)+G_{1,-1,0}(\zb)+\log 2\,G_{-1,0}(\zb)-\log 2\,G_{1,0}(\zb) \nonumber \\
& +\frac{1}{2}\zeta_2G_1(\zb)+\log 2\,G_{-1}(\zb)G_0(z)-\log 2\,G_1(\zb)G_0(z)\nonumber 
\end{align}
\begin{align}
\xG_{0,0,-1/\zb}(z)=&\,G_{0,0,-1/\zb}(z)\nonumber \\ 
\xG_{0,1,-1/\zb}(z)=&\,G_0(z)G_{-1,1}(\zb)+G_{-1,1,0}(\zb)+G_{0,1,-1/\zb}(z)-\log 2\,G_{-1,0}(\zb)+\log 2\,G_{1,0}(\zb) \nonumber \\
& +\frac{1}{2}\zeta_2G_{-1}(\zb)-\log 2\,G_{-1}(\zb)G_0(z)+\log 2\,G_1(\zb)G_0(z)\nonumber \\
\xG_{0,-1/\zb,-1}(z)=&\,G_0(z)(-G_{1,-1}(\zb))+G_{-1}(\zb)G_{0,-1/\zb}(z)-G_{0,1,-1}(\zb)+G_{0,-1/\zb,-1}(z) \nonumber \\
& -G_{1,-1,0}(\zb)-\log 2\,G_{-1,0}(\zb)-\log 2\,G_{0,-1}(\zb)+\log 2\,G_{0,1}(\zb)+\log 2\,G_{1,0}(\zb) \nonumber \\
& -\frac{1}{2}\zeta_2G_{-1}(\zb)-\frac{1}{2}\zeta_2G_1(\zb)-\log 2\,G_{-1}(\zb)G_0(z)+\log 2\,G_1(\zb)G_0(z)\nonumber \\ 
\xG_{0,-1/\zb,0}(z)=&\,G_0(\zb)G_{0,-1/\zb}(z)+G_{0,-1/\zb,0}(z)\nonumber \\ 
\xG_{0,-1/\zb,1}(z)=&\,G_0(z)(-G_{-1,1}(\zb))+G_1(\zb)G_{0,-1/\zb}(z)-G_{-1,1,0}(\zb)-G_{0,-1,1}(\zb)+G_{0,-1/\zb,1}(z) \nonumber \\
& +\log 2\,G_{-1,0}(\zb)+\log 2\,G_{0,-1}(\zb)-\log 2\,G_{0,1}(\zb)-\log 2\,G_{1,0}(\zb) \nonumber \\
& -\frac{1}{2}\zeta_2G_{-1}(\zb)-\frac{1}{2}\zeta_2G_1(\zb)+\log 2\,G_{-1}(\zb)G_0(z)-\log 2\,G_1(\zb)G_0(z)\nonumber \\ 
\xG_{0,-1/\zb,-1/\zb}(z)=&\,G_{0,-1/\zb,-1/\zb}(z)\nonumber \\
%
%
\xG_{1,-1,-1/\zb}(z)=&\,G_1(z)G_{1,-1}(\zb)+G_{0,-1,1}(\zb)+G_{1,-1,-1/\zb}(z)+\log 2\,G_{-1,1}(\zb)-\log 2\,G_{0,-1}(\zb) \nonumber \\
& +\log 2\,G_{0,1}(\zb)-\log 2\,G_{1,-1}(\zb)+\frac{1}{2}\zeta_2G_1(\zb)-2\log^22G_{-1}(\zb) \nonumber \\
& +2\log^22G_1(\zb)+\log 2\,G_1(z)G_{-1}(\zb)-\log 2\,G_1(z)G_1(\zb)\nonumber \\ 
\xG_{1,0,-1/\zb}(z)=&\,G_{0,-1,1}(\zb)+G_{1,0,-1/\zb}(z)-\log 2\,G_{0,-1}(\zb)+\log 2\,G_{0,1}(\zb)+\frac{1}{2}\zeta_2G_1(\zb)\nonumber \\ 
\xG_{1,1,-1/\zb}(z)=&\,G_1(z)G_{-1,1}(\zb)-G_{-1,-1,1}(\zb)+G_{-1,1,1}(\zb)+G_{0,-1,1}(\zb)+G_{1,1,-1/\zb}(z) \nonumber \\
& +\log 2\,G_{-1,-1}(\zb)-2\log 2\,G_{-1,1}(\zb)-\log 2\,G_{0,-1}(\zb)+\log 2\,G_{0,1}(\zb) \nonumber \\
& +\log 2\,G_{1,1}(\zb)+\frac{1}{2}\zeta_2G_{-1}(\zb)+\log^22G_{-1}(\zb)-\log^22G_1(\zb) \nonumber \\
& -\log 2\,G_1(z)G_{-1}(\zb)+\log 2\,G_1(z)G_1(\zb)\nonumber \\ 
\xG_{1,-1/\zb,-1}(z)=&\,G_1(z)(-G_{1,-1}(\zb))+G_{-1}(\zb)G_{1,-1/\zb}(z)+2G_{-1,-1,1}(\zb)+G_{-1,1,-1}(\zb) \nonumber \\
& -G_{0,-1,1}(\zb)-G_{0,1,-1}(\zb)+G_{1,-1/\zb,-1}(z)-2\log 2\,G_{-1,-1}(\zb) \nonumber \\
& +\log 2\,G_{-1,1}(\zb)+\log 2\,G_{1,-1}(\zb)-\frac{1}{2}\zeta_2G_{-1}(\zb)-\frac{1}{2}\zeta_2G_1(\zb) \nonumber \\
& -\log 2\,G_{-1}(\zb)G_1(z)+\log 2\,G_1(\zb)G_1(z)\nonumber \\ 
\xG_{1,-1/\zb,0}(z)=&\,G_0(\zb)G_{1,-1/\zb}(z)+G_{-1,0,1}(\zb)+G_{1,-1/\zb,0}(z)-\frac{1}{2}\zeta_2G_{-1}(\zb)-\frac{1}{2}\zeta_2G_1(\zb)\nonumber \\ 
\xG_{1,-1/\zb,1}(z)=&\,-G_1(z)G_{-1,1}(\zb)+G_1(\zb)G_{1,-1/\zb}(z)+2G_{-1,-1,1}(\zb)-2G_{0,-1,1}(\zb)+G_{1,-1,1}(\zb) \nonumber \\
& +G_{1,-1/\zb,1}(z)-2\log 2\,G_{-1,-1}(\zb)+3\log 2\,G_{-1,1}(\zb)+2\log 2\,G_{0,-1}(\zb) \nonumber \\
& -2\log 2\,G_{0,1}(\zb)-\log 2\,G_{1,-1}(\zb)-\zeta_2G_{-1}(\zb)-2\log^22G_{-1}(\zb) \nonumber \\
& +2\log^22G_1(\zb)+\log 2\,G_1(z)G_{-1}(\zb)-\log 2\,G_1(z)G_1(\zb)\nonumber \\ 
\xG_{1,-1/\zb,-1/\zb}(z)=&\,G_{-1,-1,1}(\zb)+G_{1,-1/\zb,-1/\zb}(z)-\log 2\,G_{-1,-1}(\zb)+\log 2\,G_{-1,1}(\zb) \nonumber \\
& -\frac{1}{2}\zeta_2G_{-1}(\zb)-\frac{1}{2}\log^22G_{-1}(\zb)+\frac{1}{2}\log^22G_1(\zb)\nonumber \\ 
\xG_{-1/\zb,-1,-1}(z)=&\,G_{-1}(\zb)G_{-1/\zb,-1}(z)+G_{-1/\zb}(z)G_{-1,-1}(\zb)-G_{-1,1,-1}(\zb)+G_{0,1,-1}(\zb) \nonumber \\
& -G_{1,-1,-1}(\zb)-G_{1,1,-1}(\zb)+G_{-1/\zb,-1,-1}(z)-\log 2\,G_{-1,-1}(\zb) \nonumber \\
& +\log 2\,G_{-1,1}(\zb)+\log 2\,G_{0,-1}(\zb)-\log 2\,G_{0,1}(\zb)-\log 2\,G_{1,-1}(\zb) \nonumber \\
& +\log 2\,G_{1,1}(\zb)+\frac{1}{2}\zeta_2G_1(\zb)-\log^22G_{-1}(\zb)+\log^22G_1(\zb)\nonumber 
\end{align}
\begin{align}
\xG_{-1/\zb,-1,0}(z)=&\,G_{-1/\zb}(z)G_{0,-1}(\zb)+G_0(\zb)G_{-1/\zb,-1}(z)-G_{0,1,-1}(\zb)-G_{1,0,-1}(\zb) \nonumber \\
& +G_{-1/\zb,-1,0}(z)-\log 2\,G_{0,-1}(\zb)+\log 2\,G_{0,1}(\zb)+\frac{1}{2}\zeta_2G_1(\zb)\nonumber \\
\xG_{-1/\zb,-1,1}(z)=&\,G_{-1/\zb}(z)G_{1,-1}(\zb)+G_1(\zb)G_{-1/\zb,-1}(z)+G_{0,-1,1}(\zb)-G_{1,-1,1}(\zb)-2G_{1,1,-1}(\zb) \nonumber \\
& +G_{-1/\zb,-1,1}(z)-\log 2\,G_{0,-1}(\zb)+\log 2\,G_{0,1}(\zb)-2\log 2\,G_{1,-1}(\zb) \nonumber \\
& +2\log 2\,G_{1,1}(\zb)+\frac{1}{2}\zeta_2G_1(\zb)-2\log^22G_{-1}(\zb)+2\log^22G_1(\zb) \nonumber \\
& +2\log 2\,G_{-1/\zb}(z)G_{-1}(\zb)-2\log 2\,G_1(\zb)G_{-1/\zb}(z)\nonumber \\ 
\xG_{-1/\zb,-1,-1/\zb}(z)=&\,G_{-1/\zb}(z)G_{1,-1}(\zb)-2G_{1,1,-1}(\zb)+G_{-1/\zb,-1,-1/\zb}(z)-2\log 2\,G_{1,-1}(\zb) \nonumber \\
& +2\log 2\,G_{1,1}(\zb)+\zeta_2G_1(\zb)-\log^22G_{-1}(\zb)+\log^22G_1(\zb) \nonumber \\
& +\log 2\,G_{-1/\zb}(z)G_{-1}(\zb)-\log 2\,G_1(\zb)G_{-1/\zb}(z)\nonumber \\ 
%
\xG_{-1/\zb,0,-1}(z)=&\,G_{-1/\zb}(z)G_{-1,0}(\zb)+G_{-1}(\zb)G_{-1/\zb,0}(z)+G_{0,1,-1}(\zb)+G_{-1/\zb,0,-1}(z) \nonumber \\
& +\log 2\,G_{0,-1}(\zb)-\log 2\,G_{0,1}(\zb)+\frac{1}{2}\zeta_2G_{-1}(\zb)\nonumber \\ 
\xG_{-1/\zb,0,0}(z)=&\,G_{-1/\zb}(z)G_{0,0}(\zb)+G_0(\zb)G_{-1/\zb,0}(z)+G_{-1/\zb,0,0}(z)\nonumber \\ 
\xG_{-1/\zb,0,1}(z)=&\,G_{-1/\zb}(z)G_{1,0}(\zb)+G_1(\zb)G_{-1/\zb,0}(z)+G_{0,-1,1}(\zb)+G_{-1/\zb,0,1}(z) \nonumber \\
& -\log 2\,G_{0,-1}(\zb)+\log 2\,G_{0,1}(\zb)+\frac{1}{2}\zeta_2G_1(\zb)\nonumber \\ 
\xG_{-1/\zb,0,-1/\zb}(z)=&\,G_{-1/\zb,0,-1/\zb}(z)\nonumber \\ 
\xG_{-1/\zb,1,-1}(z)=&\,G_{-1}(\zb)G_{-1/\zb,1}(z)+G_{-1/\zb}(z)G_{-1,1}(\zb)-2G_{-1,-1,1}(\zb)-G_{-1,1,-1}(\zb)\nonumber \\
& +G_{0,1,-1}(\zb) +G_{-1/\zb,1,-1}(z)+2\log 2\,G_{-1,-1}(\zb)-2\log 2\,G_{-1,1}(\zb) \nonumber \\
& +\log 2\,G_{0,-1}(\zb)-\log 2\,G_{0,1}(\zb)+\frac{1}{2}\zeta_2G_{-1}(\zb)+2\log^22G_{-1}(\zb) \nonumber \\
& -2\log^22G_1(\zb)-2\log 2\,G_{-1/\zb}(z)G_{-1}(\zb)+2\log 2\,G_1(\zb)G_{-1/\zb}(z)\nonumber \\ 
\xG_{-1/\zb,1,0}(z)=&\,G_{-1/\zb}(z)G_{0,1}(\zb)+G_0(\zb)G_{-1/\zb,1}(z)-G_{-1,0,1}(\zb)-G_{0,-1,1}(\zb)+G_{-1/\zb,1,0}(z) \nonumber \\
& +\log 2\,G_{0,-1}(\zb)-\log 2\,G_{0,1}(\zb)+\frac{1}{2}\zeta_2G_{-1}(\zb)\nonumber \\ 
\xG_{-1/\zb,1,1}(z)=&\,G_{-1/\zb}(z)G_{1,1}(\zb)+G_1(\zb)G_{-1/\zb,1}(z)-G_{-1,-1,1}(\zb)-G_{-1,1,1}(\zb)+G_{0,-1,1}(\zb) \nonumber \\
& -G_{1,-1,1}(\zb)+G_{-1/\zb,1,1}(z)+\log 2\,G_{-1,-1}(\zb)-\log 2\,G_{-1,1}(\zb) \nonumber \\
& -\log 2\,G_{0,-1}(\zb)+\log 2\,G_{0,1}(\zb)+\log 2\,G_{1,-1}(\zb)-\log 2\,G_{1,1}(\zb) \nonumber \\
& +\frac{1}{2}\zeta_2G_{-1}(\zb)+\log^22G_{-1}(\zb)-\log^22G_1(\zb)\nonumber \\ 
\xG_{-1/\zb,1,-1/\zb}(z)=&\,G_{-1/\zb}(z)G_{-1,1}(\zb)-2G_{-1,-1,1}(\zb)+G_{-1/\zb,1,-1/\zb}(z)+2\log 2\,G_{-1,-1}(\zb) \nonumber \\
& -2\log 2\,G_{-1,1}(\zb)+\zeta_2G_{-1}(\zb)+\log^22G_{-1}(\zb)-\log^22G_1(\zb) \nonumber \\
& -\log 2\,G_{-1/\zb}(z)G_{-1}(\zb)+\log 2\,G_1(\zb)G_{-1/\zb}(z)\nonumber \\ 
\xG_{-1/\zb,-1/\zb,-1}(z)=&\,G_{-1}(\zb)G_{-1/\zb,-1/\zb}(z)-G_{-1/\zb}(z)G_{1,-1}(\zb)+G_{1,1,-1}(\zb)+G_{-1/\zb,-1/\zb,-1}(z) \nonumber \\
& +\log 2\,G_{1,-1}(\zb)-\log 2\,G_{1,1}(\zb)-\frac{1}{2}\zeta_2G_1(\zb)+\frac{1}{2}\log^22G_{-1}(\zb) \nonumber \\
& -\frac{1}{2}\log^22G_1(\zb)-\log 2\,G_{-1/\zb}(z)G_{-1}(\zb)+\log 2\,G_1(\zb)G_{-1/\zb}(z)\nonumber \\ 
\xG_{-1/\zb,-1/\zb,0}(z)=&\,G_0(\zb)G_{-1/\zb,-1/\zb}(z)+G_{-1/\zb,-1/\zb,0}(z)\nonumber \\
\xG_{-1/\zb,-1/\zb,1}(z)=&-G_{-1/\zb}(z)G_{-1,1}(\zb)+G_1(\zb)G_{-1/\zb,-1/\zb}(z)+G_{-1,-1,1}(\zb)+G_{-1/\zb,-1/\zb,1}(z) \nonumber 
\end{align}
\begin{align}
& -\log 2\,G_{-1,-1}(\zb)+\log 2\,G_{-1,1}(\zb)-\frac{1}{2}\zeta_2G_{-1}(\zb)-\frac{1}{2}\log^22G_{-1}(\zb) \nonumber \\
& +\frac{1}{2}\log^22G_1(\zb)+\log 2\,G_{-1/\zb}(z)G_{-1}(\zb)-\log 2\,G_1(\zb)G_{-1/\zb}(z)\nonumber \\ 
\xG_{-1/\zb,-1/\zb,-1/\zb}(z)=&\,G_{-1/\zb,-1/\zb,-1/\zb}(z)\nonumber
\end{align}

\section{Analytic results for the BFKL ladder through five loops}
\subsection{Analytic results for the functions $C_k^{(1)}$}
\label{app:C1results}
In this appendix we present the analytic results for the functions $C_k^{(1)}$through five loops.
Writing
\beq
C_k^{(1)}(z) = \frac{|z|}{2\pi\,|1-z|^2}\,\cC_k^{(1)}(z)\,,
\eeq
we find for the first few orders 
\beq\bsp\label{eq:C1}
\cC_2^{(1)}(z) \,=\, &-\cG_{0,0}(z)\,,\\
\cC_3^{(1)}(z) \,=\, &2\cG_{0,0,0}(z)-2\cG_{0,0,1}(z)-2\cG_{1,0,0}(z)-8\zeta_3\,, \\
\cC_4^{(1)}(z) \,=\, &-2\cG_{0,0,0,0}(z)+4\cG_{0,0,0,1}(z)+2\cG_{0,0,1,0}(z)+4\cG_{0,0,1,1}(z)+2\cG_{0,1,0,0}(z)\\
&+4\cG_{1,0,0,0}(z)-4\cG_{1,0,0,1}(z)-\cG_{1,1,0,0}(z)+12\zeta_3 \cG_0(z)-16\zeta_3\cG_1(z)\,,\\
\cC_5^{(1)}(z) \,=\, &2\cG_{0,0,0,0,0}(z)-6\cG_{0,0,0,0,1}(z)-6\cG_{0,0,0,1,0}(z)+12\cG_{0,0,0,1,1}(z)-4\cG_{0,0,1,0,0}(z)\\
&+6\cG_{0,0,1,0,1}(z)+6\cG_{0,0,1,1,0}(z)-12\cG_{0,0,1,1,1}(z)-6\cG_{0,1,0,0,0}(z)+6\cG_{0,1,0,0,1}(z)\\
&+6\cG_{0,1,1,0,0}(z)-6\cG_{1,0,0,0,0}(z)+12\cG_{1,0,0,0,1}(z)+6\cG_{1,0,0,1,0}(z)-12\cG_{1,0,0,1,1}(z)\\
&+6\cG_{1,0,1,0,0}(z)+12\cG_{1,1,0,0,0}(z)-12\cG_{1,1,0,0,1}(z)-12\cG_{1,1,1,0,0}(z)\\
&-16\zeta_3\cG_{0,0}(z)+24\zeta_3\cG_{0,1}(z)+36\zeta_3\cG_{1,0}(z)-48\zeta_3\cG_{1,1}(z)-36\zeta_5\,.
\esp\eeq

\subsection{Analytic results for the functions $C_k^{(2)}$}
\label{app:C2results}
In this section we present the results through weight five for the functions $\cC^{(2,i)}_k$ defined in eq.~\eqref{eq:C2k_to_cC}.
\beq\bsp
\cC^{(2,1)}_2 =& \; \xG_{1,0}(z)-\xG_{0,1}(z)\,,\\
\cC^{(2,2)}_2 =& \; -2\xG_{-\frac{1}{\bar{z}},0}(z)+\xG_{0,1}(z)+\xG_{1,0}(z)-\zeta_2\,,\\
\nonumber\\
\cC^{(2,1)}_3 =& \; \xG_{0,0,1}(z)-2\xG_{0,1,1}(z)-\xG_{1,0,0}(z)+2\xG_{1,1,0}(z)\,,\\
\cC^{(2,2)}_3 =& \; 2\xG_{0,-\frac{1}{\bar{z}},0}(z)+2\xG_{-\frac{1}{\bar{z}},0,0}(z)-4\xG_{-\frac{1}{\bar{z}},-\frac{1}{\bar{z}},0}(z)-\xG_{0,0,1}(z)-2\xG_{0,1,0}(z)\\&+2\xG_{0,1,1}(z)-\xG_{1,0,0}(z)+2\xG_{1,1,0}(z)+\zeta_2\xG_0(z)-2\zeta_2\xG_{-\frac{1}{\bar{z}}}(z)-2\zeta_3\,,\\
\nonumber\\
\cC^{(2,1)}_4 =& \; -2\xG_{-1,-1,-\frac{1}{\bar{z}},0}(z)+\xG_{-1,0,-\frac{1}{\bar{z}},0}(z)+\xG_{0,-1,-\frac{1}{\bar{z}},0}(z)-\xG_{0,1,-\frac{1}{\bar{z}},0}(z)-\xG_{1,0,-\frac{1}{\bar{z}},0}(z)\\&+2\xG_{1,1,-\frac{1}{\bar{z}},0}(z)+2\xG_{-1,-1,0,1}(z)+\xG_{-1,0,1,0}(z)-2\xG_{-1,0,1,1}(z)-\xG_{0,-1,0,1}(z)\\&-\xG_{0,0,0,1}(z)-\xG_{0,0,1,0}(z)+4\xG_{0,0,1,1}(z)+\xG_{0,1,0,0}(z)+3\xG_{0,1,0,1}(z)-6\xG_{0,1,1,1}(z)\\&+\xG_{1,0,0,0}(z)-3\xG_{1,0,1,0}(z)-4\xG_{1,1,0,0}(z)+6\xG_{1,1,1,0}(z)-\zeta_2\xG_{-1,-1}(z)-\frac{1}{2}\zeta_2\xG_{-1,0}(z)\\&+\frac{1}{2}\zeta_2\xG_{0,-1}(z)-\frac{1}{2}\zeta_2\xG_{0,1}(z)+\frac{1}{2}\zeta_2\xG_{1,0}(z)+\zeta_2\xG_{1,1}(z)+\frac{1}{2}\zeta_3\xG_{-1}(z)-\frac{15}{2}\zeta_3\xG_1(z)\,,
\esp\eeq
\beq\bsp
\cC^{(2,2)}_4 =& \; 2\xG_{-1,-1,-\frac{1}{\bar{z}},0}(z)-\xG_{-1,0,-\frac{1}{\bar{z}},0}(z)-\xG_{0,-1,-\frac{1}{\bar{z}},0}(z)-2\xG_{0,0,-\frac{1}{\bar{z}},0}(z)-\xG_{0,1,-\frac{1}{\bar{z}},0}(z)\\&-4\xG_{0,-\frac{1}{\bar{z}},0,0}(z)+8\xG_{0,-\frac{1}{\bar{z}},-\frac{1}{\bar{z}},0}(z)-\xG_{1,0,-\frac{1}{\bar{z}},0}(z)+2\xG_{1,1,-\frac{1}{\bar{z}},0}(z)-2\xG_{-\frac{1}{\bar{z}},0,0,0}(z)\\&+8\xG_{-\frac{1}{\bar{z}},0,-\frac{1}{\bar{z}},0}(z)+8\xG_{-\frac{1}{\bar{z}},-\frac{1}{\bar{z}},0,0}(z)-16\xG_{-\frac{1}{\bar{z}},-\frac{1}{\bar{z}},-\frac{1}{\bar{z}},0}(z)-2\xG_{-1,-1,0,1}(z)\\&-\xG_{-1,0,1,0}(z)+2\xG_{-1,0,1,1}(z)+\xG_{0,-1,0,1}(z)+\xG_{0,0,0,1}(z)+3\xG_{0,0,1,0}(z)\\&-4\xG_{0,0,1,1}(z)+3\xG_{0,1,0,0}(z)-3\xG_{0,1,0,1}(z)-6\xG_{0,1,1,0}(z)+6\xG_{0,1,1,1}(z)\\&+\xG_{1,0,0,0}(z)-3\xG_{1,0,1,0}(z)-4\xG_{1,1,0,0}(z)+6\xG_{1,1,1,0}(z)+8\zeta_3\xG_{-\frac{1}{\bar{z}}}(z)\\&+4\zeta_2\xG_{0,-\frac{1}{\bar{z}}}(z)+2\zeta_2\xG_{-\frac{1}{\bar{z}},0}(z)-8\zeta_2\xG_{-\frac{1}{\bar{z}},-\frac{1}{\bar{z}}}(z)+\zeta_2\xG_{-1,-1}(z)\\&+\frac{1}{2}\zeta_2\xG_{-1,0}(z)-\frac{1}{2}\zeta_2\xG_{0,-1}(z)-\zeta_2\xG_{0,0}(z)-\frac{1}{2}\zeta_2\xG_{0,1}(z)+\frac{1}{2}\zeta_2\xG_{1,0}(z)\\&+\zeta_2\xG_{1,1}(z)-\frac{1}{2}\zeta_3\xG_{-1}(z)+2\zeta_3\xG_0(z)-\frac{15}{2}\zeta_3\xG_1(z)-\frac{5\zeta_4}{4}\,,\\
\esp\eeq

\beq\bsp
\cC^{(2,1)}_5 =& \;12\xG_{-1,-1,-1,0,1}(z)-12\xG_{-1,-1,-1,-\frac{1}{\bar{z}},0}(z)-6\xG_{-1,-1,0,0,1}(z)+12\xG_{-1,-1,0,-\frac{1}{\bar{z}},0}(z)\\
&+6\xG_{-1,-1,-\frac{1}{\bar{z}},0,0}(z)-12\xG_{-1,-1,-\frac{1}{\bar{z}},-\frac{1}{\bar{z}},0}(z)-6\xG_{-1,0,-1,0,1}(z)+6\xG_{-1,0,-1,-\frac{1}{\bar{z}},0}(z)\\
&-3\xG_{-1,0,0,1,0}(z)+6\xG_{-1,0,0,1,1}(z)-4\xG_{-1,0,0,-\frac{1}{\bar{z}},0}(z)-2\xG_{-1,0,1,0,0}(z)+6\xG_{-1,0,1,0,1}(z)\\
&+6\xG_{-1,0,1,1,0}(z)-12\xG_{-1,0,1,1,1}(z)-3\xG_{-1,0,-\frac{1}{\bar{z}},0,0}(z)+6\xG_{-1,0,-\frac{1}{\bar{z}},-\frac{1}{\bar{z}},0}(z)\\
&-6\xG_{0,-1,-1,0,1}(z)+6\xG_{0,-1,-1,-\frac{1}{\bar{z}},0}(z)+3\xG_{0,-1,0,0,1}(z)-6\xG_{0,-1,0,-\frac{1}{\bar{z}},0}(z)\\
&-3\xG_{0,-1,-\frac{1}{\bar{z}},0,0}(z)+6\xG_{0,-1,-\frac{1}{\bar{z}},-\frac{1}{\bar{z}},0}(z)+2\xG_{0,0,-1,0,1}(z)-2\xG_{0,0,-1,-\frac{1}{\bar{z}},0}(z)\\
&+\xG_{0,0,0,0,1}(z)+2\xG_{0,0,0,1,0}(z)-6\xG_{0,0,0,1,1}(z)-9\xG_{0,0,1,0,1}(z)-5\xG_{0,0,1,1,0}(z)\\
&+18\xG_{0,0,1,1,1}(z)+2\xG_{0,0,1,-\frac{1}{\bar{z}},0}(z)-2\xG_{0,1,0,0,0}(z)-4\xG_{0,1,0,0,1}(z)+12\xG_{0,1,0,1,1}(z)\\
&+6\xG_{0,1,0,-\frac{1}{\bar{z}},0}(z)+5\xG_{0,1,1,0,0}(z)+12\xG_{0,1,1,0,1}(z)-24\xG_{0,1,1,1,1}(z)-6\xG_{0,1,1,-\frac{1}{\bar{z}},0}(z)\\
&+3\xG_{0,1,-\frac{1}{\bar{z}},0,0}(z)-6\xG_{0,1,-\frac{1}{\bar{z}},-\frac{1}{\bar{z}},0}(z)-\xG_{1,0,0,0,0}(z)+4\xG_{1,0,0,1,0}(z)+4\xG_{1,0,0,-\frac{1}{\bar{z}},0}(z)\\
&+9\xG_{1,0,1,0,0}(z)-12\xG_{1,0,1,1,0}(z)-6\xG_{1,0,1,-\frac{1}{\bar{z}},0}(z)+3\xG_{1,0,-\frac{1}{\bar{z}},0,0}(z)-6\xG_{1,0,-\frac{1}{\bar{z}},-\frac{1}{\bar{z}},0}(z)\\
&+6\xG_{1,1,0,0,0}(z)-12\xG_{1,1,0,1,0}(z)-12\xG_{1,1,0,-\frac{1}{\bar{z}},0}(z)-18\xG_{1,1,1,0,0}(z)+24\xG_{1,1,1,1,0}(z)\\
&+12\xG_{1,1,1,-\frac{1}{\bar{z}},0}(z)-6\xG_{1,1,-\frac{1}{\bar{z}},0,0}(z)+12\xG_{1,1,-\frac{1}{\bar{z}},-\frac{1}{\bar{z}},0}(z)-6\xG_{-1,-1,-1}(z)\zeta_2\\
&-6\xG_{-1,-1,-\frac{1}{\bar{z}}}(z)\zeta_2+3\xG_{-1,0,-1}(z)\zeta_2+\xG_{-1,0,0}(z)\zeta_2+3\xG_{-1,0,-\frac{1}{\bar{z}}}(z)\zeta_2+3\xG_{0,-1,-1}(z)\zeta_2\\
&+3\xG_{0,-1,-\frac{1}{\bar{z}}}(z)\zeta_2-\xG_{0,0,-1}(z)\zeta_2+\xG_{0,0,1}(z)\zeta_2-3\xG_{0,1,1}(z)\zeta_2-3\xG_{0,1,-\frac{1}{\bar{z}}}(z)\zeta_2\\
&-\xG_{1,0,0}(z)\zeta_2-3\xG_{1,0,1}(z)\zeta_2-3\xG_{1,0,-\frac{1}{\bar{z}}}(z)\zeta_2+6\xG_{1,1,1}(z)\zeta_2+6\xG_{1,1,-\frac{1}{\bar{z}}}(z)\zeta_2\\
&-3\xG_{-1,-1}(z)\zeta_3-\frac{3}{2}\xG_{-1,0}(z)\zeta_3+\frac{3}{2}\xG_{0,-1}(z)\zeta_3+\frac{23}{2}\xG_{0,1}(z)\zeta_3+\frac{21}{2}\xG_{1,0}(z)\zeta_3\\
&-39\xG_{1,1}(z)\zeta_3-\frac{3}{4}\xG_{-1}(z)\zeta_4+\frac{3}{4}\xG_1(z)\zeta_4\,,
\esp\eeq
\beq\bsp
\cC^{(2,2)}_5 =& \; -12\xG_{-1,-1,-1,0,1}(z)+12\xG_{-1,-1,-1,-\frac{1}{\bar{z}},0}(z)+6\xG_{-1,-1,0,0,1}(z)-12\xG_{-1,-1,0,-\frac{1}{\bar{z}},0}(z)\\
&-6\xG_{-1,-1,-\frac{1}{\bar{z}},0,0}(z)+12\xG_{-1,-1,-\frac{1}{\bar{z}},-\frac{1}{\bar{z}},0}(z)+6\xG_{-1,0,-1,0,1}(z)-6\xG_{-1,0,-1,-\frac{1}{\bar{z}},0}(z)\\
&+3\xG_{-1,0,0,1,0}(z)-6\xG_{-1,0,0,1,1}(z)+4\xG_{-1,0,0,-\frac{1}{\bar{z}},0}(z)+2\xG_{-1,0,1,0,0}(z)-6\xG_{-1,0,1,0,1}(z)\\
&-6\xG_{-1,0,1,1,0}(z)+12\xG_{-1,0,1,1,1}(z)+3\xG_{-1,0,-\frac{1}{\bar{z}},0,0}(z)-6\xG_{-1,0,-\frac{1}{\bar{z}},-\frac{1}{\bar{z}},0}(z)\\
&+6\xG_{0,-1,-1,0,1}(z)-6\xG_{0,-1,-1,-\frac{1}{\bar{z}},0}(z)-3\xG_{0,-1,0,0,1}(z)+6\xG_{0,-1,0,-\frac{1}{\bar{z}},0}(z)\\
&+3\xG_{0,-1,-\frac{1}{\bar{z}},0,0}(z)-6\xG_{0,-1,-\frac{1}{\bar{z}},-\frac{1}{\bar{z}},0}(z)-2\xG_{0,0,-1,0,1}(z)+2\xG_{0,0,-1,-\frac{1}{\bar{z}},0}(z)\\
&-\xG_{0,0,0,0,1}(z)-4\xG_{0,0,0,1,0}(z)+6\xG_{0,0,0,1,1}(z)+2\xG_{0,0,0,-\frac{1}{\bar{z}},0}(z)-6\xG_{0,0,1,0,0}(z)\\
&+9\xG_{0,0,1,0,1}(z)+13\xG_{0,0,1,1,0}(z)-18\xG_{0,0,1,1,1}(z)+2\xG_{0,0,1,-\frac{1}{\bar{z}},0}(z)+6\xG_{0,0,-\frac{1}{\bar{z}},0,0}(z)\\
&-12\xG_{0,0,-\frac{1}{\bar{z}},-\frac{1}{\bar{z}},0}(z)-4\xG_{0,1,0,0,0}(z)+4\xG_{0,1,0,0,1}(z)+12\xG_{0,1,0,1,0}(z)-12\xG_{0,1,0,1,1}(z)\\
&+6\xG_{0,1,0,-\frac{1}{\bar{z}},0}(z)+13\xG_{0,1,1,0,0}(z)-12\xG_{0,1,1,0,1}(z)-24\xG_{0,1,1,1,0}(z)+24\xG_{0,1,1,1,1}(z)\\
&-6\xG_{0,1,1,-\frac{1}{\bar{z}},0}(z)+3\xG_{0,1,-\frac{1}{\bar{z}},0,0}(z)-6\xG_{0,1,-\frac{1}{\bar{z}},-\frac{1}{\bar{z}},0}(z)+6\xG_{0,-\frac{1}{\bar{z}},0,0,0}(z)\\
&-24\xG_{0,-\frac{1}{\bar{z}},0,-\frac{1}{\bar{z}},0}(z)-24\xG_{0,-\frac{1}{\bar{z}},-\frac{1}{\bar{z}},0,0}(z)+48\xG_{0,-\frac{1}{\bar{z}},-\frac{1}{\bar{z}},-\frac{1}{\bar{z}},0}(z)-\xG_{1,0,0,0,0}(z)\\
&+4\xG_{1,0,0,1,0}(z)+4\xG_{1,0,0,-\frac{1}{\bar{z}},0}(z)+9\xG_{1,0,1,0,0}(z)-12\xG_{1,0,1,1,0}(z)-6\xG_{1,0,1,-\frac{1}{\bar{z}},0}(z)\\
&+3\xG_{1,0,-\frac{1}{\bar{z}},0,0}(z)-6\xG_{1,0,-\frac{1}{\bar{z}},-\frac{1}{\bar{z}},0}(z)+6\xG_{1,1,0,0,0}(z)-12\xG_{1,1,0,1,0}(z)-12\xG_{1,1,0,-\frac{1}{\bar{z}},0}(z)\\
&-18\xG_{1,1,1,0,0}(z)+24\xG_{1,1,1,1,0}(z)+12\xG_{1,1,1,-\frac{1}{\bar{z}},0}(z)-6\xG_{1,1,-\frac{1}{\bar{z}},0,0}(z)+12\xG_{1,1,-\frac{1}{\bar{z}},-\frac{1}{\bar{z}},0}(z)\\
&+2\xG_{-\frac{1}{\bar{z}},0,0,0,0}(z)-16\xG_{-\frac{1}{\bar{z}},0,0,-\frac{1}{\bar{z}},0}(z)-24\xG_{-\frac{1}{\bar{z}},0,-\frac{1}{\bar{z}},0,0}(z)+48\xG_{-\frac{1}{\bar{z}},0,-\frac{1}{\bar{z}},-\frac{1}{\bar{z}},0}(z)\\
&-12\xG_{-\frac{1}{\bar{z}},-\frac{1}{\bar{z}},0,0,0}(z)+48\xG_{-\frac{1}{\bar{z}},-\frac{1}{\bar{z}},0,-\frac{1}{\bar{z}},0}(z)+48\xG_{-\frac{1}{\bar{z}},-\frac{1}{\bar{z}},-\frac{1}{\bar{z}},0,0}(z)-96\xG_{-\frac{1}{\bar{z}},-\frac{1}{\bar{z}},-\frac{1}{\bar{z}},-\frac{1}{\bar{z}},0}(z)\\
&+6\xG_{-1,-1,-1}(z)\zeta_2+6\xG_{-1,-1,-\frac{1}{\bar{z}}}(z)\zeta_2-3\xG_{-1,0,-1}(z)\zeta_2-\xG_{-1,0,0}(z)\zeta_2-3\xG_{-1,0,-\frac{1}{\bar{z}}}(z)\zeta_2\\
&-3\xG_{0,-1,-1}(z)\zeta_2-3\xG_{0,-1,-\frac{1}{\bar{z}}}(z)\zeta_2+\xG_{0,0,-1}(z)\zeta_2+\xG_{0,0,0}(z)\zeta_2+\xG_{0,0,1}(z)\zeta_2\\
&-6\xG_{0,0,-\frac{1}{\bar{z}}}(z)\zeta_2-3\xG_{0,1,1}(z)\zeta_2-3\xG_{0,1,-\frac{1}{\bar{z}}}(z)\zeta_2-6\xG_{0,-\frac{1}{\bar{z}},0}(z)\zeta_2+24\xG_{0,-\frac{1}{\bar{z}},-\frac{1}{\bar{z}}}(z)\zeta_2\\
&-\xG_{1,0,0}(z)\zeta_2-3\xG_{1,0,1}(z)\zeta_2-3\xG_{1,0,-\frac{1}{\bar{z}}}(z)\zeta_2+6\xG_{1,1,1}(z)\zeta_2+6\xG_{1,1,-\frac{1}{\bar{z}}}(z)\zeta_2\\
&-2\xG_{-\frac{1}{\bar{z}},0,0}(z)\zeta_2+24\xG_{-\frac{1}{\bar{z}},0,-\frac{1}{\bar{z}}}(z)\zeta_2+12\xG_{-\frac{1}{\bar{z}},-\frac{1}{\bar{z}},0}(z)\zeta_2-48\xG_{-\frac{1}{\bar{z}},-\frac{1}{\bar{z}},-\frac{1}{\bar{z}}}(z)\zeta_2\\
&+\frac{4}{3}\log^3(2)\zeta_2+3\xG_{-1,-1}(z)\zeta_3+\frac{3}{2}\xG_{-1,0}(z)\zeta_3-\frac{3}{2}\xG_{0,-1}(z)\zeta_3-2\xG_{0,0}(z)\zeta_3+\frac{55}{2}\xG_{0,1}(z)\zeta_3\\
&-24\xG_{0,-\frac{1}{\bar{z}}}(z)\zeta_3+\frac{21}{2}\xG_{1,0}(z)\zeta_3-39\xG_{1,1}(z)\zeta_3-12\xG_{-\frac{1}{\bar{z}},0}(z)\zeta_3+48\xG_{-\frac{1}{\bar{z}},-\frac{1}{\bar{z}}}(z)\zeta_3-6\zeta_2\zeta_3\\
&+\frac{3}{4}\xG_{-1}(z)\zeta_4+\frac{5}{4}\xG_0(z)\zeta_4+\frac{3}{4}\xG_1(z)\zeta_4-\frac{21}{2}\xG_{-\frac{1}{\bar{z}}}(z)\zeta_4-2\zeta_5-\frac{4}{3}\zeta_2\log^32\,.
\esp\eeq

\subsection{Analytic results for the functions $C_k^{(3)}$}
\label{app:C3results}
The functions $C_k^{(3,0)}(z)$ and $C_k^{(3,2)}(z)$ can through five loops be written in the form
\beq\bsp
C_k^{(3,0)}(z) =&\, R^{(0)}_{\text{max}}\mathfrak{a}^{(0)}_{k} + A\frac{1+|z|^2}{|z|} \mathfrak{b}^{(0)}_{k,\text{sym}}+ A\frac{1-|z|^2}{|z|} \mathfrak{b}^{(0)}_{k,\text{asym}}  + A\frac{1+|z|^4}{|z|^2} \mathfrak{c}^{(0)}_{k,\text{sym}}\nonumber\\
C_k^{(3,2)}(z) =&\, R^{(2)}_{\text{max}}\mathfrak{a}^{(2)}_{k} + A\frac{1+|z|^2}{|z|}\left(\frac{z}{\bar{z}}+\frac{\bar{z}}{z}\right) \mathfrak{b}^{(2)}_{k,\text{sym}}+ A\frac{1-|z|^2}{|z|}\left(\frac{z}{\bar{z}}+\frac{\bar{z}}{z}\right) \mathfrak{b}^{(2)}_{k,\text{asym}}  \\
&+ A\frac{1+|z|^4}{|z|^2}\left(\frac{z}{\bar{z}}+\frac{\bar{z}}{z}\right) \mathfrak{c}^{(2)}_{k,\text{sym}} + A\left(\frac{z}{\bar{z}}+\frac{\bar{z}}{z}\right) \fd^{(2)}_k,\nonumber
\esp\eeq
where the functions $R^{(i)}_{\text{max}}$ are the rational functions multiplying the highest weight terms at each order,
\beq\bsp
R^{(0)}_{\text{max}} &= A\,\frac{{|z|}^4-22 \,{|z|}^2+1}{{|z|}^2}-96 \\
R^{(2)}_{\text{max}} &= A\,\frac{3 \,{|z|}^4-2 \,{|z|}^2+3 }{{|z|}^2}\,\left[\left(\frac{z}{\bar{z}}\right)^2+\left(\frac{\bar z}{{z}}\right)^2\right]\,.
\esp\eeq
The transcendental coefficients appearing in the expansion of $C_k^{(3,0)}$ are
\begin{align}
\fa^{(0)}_{2}=&\,\frac{1}{32}G_0(|z|^2)\text{Ti}_1(|z|)-\frac{1}{16}\text{Ti}_2(|z|)\,,\nonumber \\ 
\fb^{(0)}_{2,\text{sym}}=&\,\frac{1}{16}\,,\nonumber \\ 
\fb^{(0)}_{2,\text{asym}}=&\,-\frac{1}{32}G_0(|z|^2)\,,\nonumber \\ 
\fc^{(0)}_{2,\text{sym}}=&\,0\,,\nonumber \\ 
&\nonumber \\ 
\fa^{(0)}_{3}=&\,-\frac{1}{32}\text{Ti}_1(|z|)G_{0,0}(|z|^2)+\frac{1}{16}G_0(|z|^2)\Big(-\text{Ti}_{-1,1}(|z|)-\text{Ti}_{1,1}(|z|)+\text{Ti}_2(|z|)\Big) \nonumber \\
&+\frac{1}{8}\text{Ti}_{-2,1}(|z|)+\frac{1}{8}\text{Ti}_{-1,2}(|z|)+\frac{1}{8}\text{Ti}_{1,2}(|z|)+\frac{1}{8}\text{Ti}_{2,1}(|z|)+\frac{1}{32}\zeta_2\text{Ti}_1(|z|)-\frac{1}{8}\text{Ti}_3(|z|)\,,\nonumber \\ 
\fb^{(0)}_{3,\text{sym}}=&\,-\frac{1}{8}\,,\nonumber \\ 
\fb^{(0)}_{3,\text{asym}}=&\,\frac{1}{32}G_{0,0}(|z|^2)+\frac{1}{16}G_0(|z|^2)-\frac{1}{16}G_{-1,0}(|z|^2)-\frac{1}{32}\,\zeta_2\,,\nonumber \\ 
\fc^{(0)}_{3,\text{sym}}=&\,-\frac{1}{8}G_0(|z|^2)\text{Ti}_1(|z|)+\frac{1}{4}\text{Ti}_2(|z|)\,,\nonumber \\ 
&\nonumber \\ 
\fa^{(0)}_{4}=&\,+\frac{1}{32}\text{Ti}_1(|z|)G_{0,0,0}(|z|^2)+\frac{1}{16}G_{0,0}(|z|^2)\Big(2\text{Ti}_{-1,1}(|z|)+2\text{Ti}_{1,1}(|z|)-\text{Ti}_2(|z|)\Big) \nonumber \\
&+\frac{1}{32}G_0(|z|^2)\Big(-8\text{Ti}_{-1,2}(|z|)-8\text{Ti}_{1,2}(|z|)+8\text{Ti}_{-1,-1,1}(|z|)+8\text{Ti}_{-1,1,1}(|z|)+8\text{Ti}_{1,-1,1}(|z|) \nonumber \\
&+8\text{Ti}_{1,1,1}(|z|)-\zeta_2\text{Ti}_1(|z|)+4\text{Ti}_3(|z|)\Big)-\frac{1}{8}\zeta_2\text{Ti}_{-1,1}(|z|)-\frac{1}{8}\zeta_2\text{Ti}_{1,1}(|z|)-\frac{1}{2}\text{Ti}_{-3,1}(|z|) \nonumber \\
&+\frac{1}{2}\text{Ti}_{-1,3}(|z|)+\frac{1}{2}\text{Ti}_{1,3}(|z|)-\frac{1}{2}\text{Ti}_{3,1}(|z|)-\frac{1}{2}\text{Ti}_{-2,-1,1}(|z|)-\frac{1}{2}\text{Ti}_{-2,1,1}(|z|)\nonumber \\
&-\frac{1}{2}\text{Ti}_{-1,-2,1}(|z|)-\frac{1}{2}\text{Ti}_{-1,-1,2}(|z|)-\frac{1}{2}\text{Ti}_{-1,1,2}(|z|)-\frac{1}{2}\text{Ti}_{-1,2,1}(|z|)-\frac{1}{2}\text{Ti}_{1,-2,1}(|z|)\nonumber \\
&-\frac{1}{2}\text{Ti}_{1,-1,2}(|z|)-\frac{1}{2}\text{Ti}_{1,1,2}(|z|)-\frac{1}{2}\text{Ti}_{1,2,1}(|z|)-\frac{1}{2}\text{Ti}_{2,-1,1}(|z|)-\frac{1}{2}\text{Ti}_{2,1,1}(|z|)-\frac{1}{8}\zeta_3\text{Ti}_1(|z|) \nonumber \\
&+\frac{1}{16}\zeta_2\text{Ti}_2(|z|)-\frac{\text{Ti}_4(|z|)}{4},\nonumber \\ 
\fb^{(0)}_{4,\text{sym}}=&-\frac{1}{16}G_{0,0}(|z|^2)+\frac{\zeta_2}{16}+\frac{1}{4},\nonumber \\ 
\fb^{(0)}_{4,\text{asym}}=&-\frac{1}{32}G_{0,0,0}(|z|^2)-\frac{1}{8}G_{0,0}(|z|^2)+\frac{1}{32}G_0(|z|^2)(\zeta_2-4)+\frac{1}{4}G_{-1,0}(|z|^2)-\frac{1}{4}G_{-1,-1,0}(|z|^2),\nonumber \\
&+\frac{1}{8}G_{-1,0,0}(|z|^2) +\frac{1}{8}G_{0,-1,0}(|z|^2)+\frac{\zeta_2}{8}+\frac{\zeta_3}{8}-\frac{1}{8}\zeta_2G_{-1}(|z|^2)\nonumber 
\end{align}
\begin{align}
\fc^{(0)}_{4,\text{sym}}=&\frac{1}{4}\text{Ti}_1(|z|)G_{0,0}(|z|^2)+\frac{1}{2}G_0(|z|^2)\Big(\text{Ti}_{-1,1}(|z|)+\text{Ti}_{1,1}(|z|)+\text{Ti}_1(|z|)-\text{Ti}_2(|z|)\Big)-\text{Ti}_{-2,1}(|z|) \nonumber \\
&-\text{Ti}_{-1,2}(|z|)-\text{Ti}_{1,2}(|z|)-\text{Ti}_{2,1}(|z|)-\frac{1}{4}\zeta_2\text{Ti}_1(|z|)-\text{Ti}_2(|z|)+\text{Ti}_3(|z|),\nonumber \\
\nonumber\\
\fa^{(0)}_{5}=&-\frac{1}{32}G_{0,0,0,0}(|z|^2)\text{Ti}_1(|z|)+\frac{1}{16}G_{0,0,0}(|z|^2)\Big(\text{Ti}_2(|z|)-3\text{Ti}_{-1,1}(|z|)-3\text{Ti}_{1,1}(|z|)\Big) \nonumber \\
&+\frac{1}{32}G_{0,0}(|z|^2)\Big(-4\text{Ti}_3(|z|)-12\text{Ti}_{-2,1}(|z|)+12\text{Ti}_{-1,2}(|z|)+12\text{Ti}_{1,2}(|z|)-12\text{Ti}_{2,1}(|z|) \nonumber \\
&-24\text{Ti}_{-1,-1,1}(|z|)-24\text{Ti}_{-1,1,1}(|z|)-24\text{Ti}_{1,-1,1}(|z|)-24\text{Ti}_{1,1,1}(|z|)+\text{Ti}_1(|z|)\zeta_2\Big) \nonumber \\
&+\frac{1}{16}G_0(|z|^2)\Big(4\text{Ti}_4(|z|)+20\text{Ti}_{-3,1}(|z|)+12\text{Ti}_{-2,2}(|z|)-12\text{Ti}_{-1,3}(|z|)-12\text{Ti}_{1,3}(|z|) \nonumber \\
&+12\text{Ti}_{2,2}(|z|)+20\text{Ti}_{3,1}(|z|)+24\text{Ti}_{-1,-1,2}(|z|)+24\text{Ti}_{-1,1,2}(|z|)+24\text{Ti}_{1,-1,2}(|z|) \nonumber \\
&+24\text{Ti}_{1,1,2}(|z|)-24\text{Ti}_{-1,-1,-1,1}(|z|)-24\text{Ti}_{-1,-1,1,1}(|z|)-24\text{Ti}_{-1,1,-1,1}(|z|) \nonumber \\
&-24\text{Ti}_{-1,1,1,1}(|z|)-24\text{Ti}_{1,-1,-1,1}(|z|)-24\text{Ti}_{1,-1,1,1}(|z|)-24\text{Ti}_{1,1,-1,1}(|z|)\nonumber \\
&-24\text{Ti}_{1,1,1,1}(|z|)-\text{Ti}_2(|z|)\zeta_2+3\text{Ti}_{-1,1}(|z|)\zeta_2+3\text{Ti}_{1,1}(|z|)\zeta_2+3\text{Ti}_1(|z|)\zeta_3\Big) \nonumber \\
&+\frac{21}{128}\zeta_4\text{Ti}_1(|z|)-\frac{\text{Ti}_5(|z|)}{2}-\frac{3}{2}\text{Ti}_{-4,1}(|z|)-\frac{5}{2}\text{Ti}_{-3,2}(|z|)-\frac{3}{2}\text{Ti}_{-2,3}(|z|)+\frac{3}{2}\text{Ti}_{-1,4}(|z|) \nonumber \\
&+\frac{3}{2}\text{Ti}_{1,4}(|z|)-\frac{3}{2}\text{Ti}_{2,3}(|z|)-\frac{5}{2}\text{Ti}_{3,2}(|z|)-\frac{3}{2}\text{Ti}_{4,1}(|z|)+3\text{Ti}_{-3,-1,1}(|z|)+3\text{Ti}_{-3,1,1}(|z|) \nonumber \\
&+3\text{Ti}_{-2,-2,1}(|z|)+3\text{Ti}_{-2,2,1}(|z|)+3\text{Ti}_{-1,-3,1}(|z|)-3\text{Ti}_{-1,-1,3}(|z|)-3\text{Ti}_{-1,1,3}(|z|) \nonumber \\
&+3\text{Ti}_{-1,3,1}(|z|)+3\text{Ti}_{1,-3,1}(|z|)-3\text{Ti}_{1,-1,3}(|z|)-3\text{Ti}_{1,1,3}(|z|)+3\text{Ti}_{1,3,1}(|z|) \nonumber \\
&+3\text{Ti}_{2,-2,1}(|z|)+3\text{Ti}_{2,2,1}(|z|)+3\text{Ti}_{3,-1,1}(|z|)+3\text{Ti}_{3,1,1}(|z|)+3\text{Ti}_{-2,-1,-1,1}(|z|) \nonumber \\
&+3\text{Ti}_{-2,-1,1,1}(|z|)+3\text{Ti}_{-2,1,-1,1}(|z|)+3\text{Ti}_{-2,1,1,1}(|z|)+3\text{Ti}_{-1,-2,-1,1}(|z|) \nonumber \\
&+3\text{Ti}_{-1,-2,1,1}(|z|)+3\text{Ti}_{-1,-1,-2,1}(|z|)+3\text{Ti}_{-1,-1,-1,2}(|z|)+3\text{Ti}_{-1,-1,1,2}(|z|) \nonumber \\
&+3\text{Ti}_{-1,-1,2,1}(|z|)+3\text{Ti}_{-1,1,-2,1}(|z|)+3\text{Ti}_{-1,1,-1,2}(|z|)+3\text{Ti}_{-1,1,1,2}(|z|)+3\text{Ti}_{-1,1,2,1}(|z|) \nonumber \\
&+3\text{Ti}_{-1,2,-1,1}(|z|)+3\text{Ti}_{-1,2,1,1}(|z|)+3\text{Ti}_{1,-2,-1,1}(|z|)+3\text{Ti}_{1,-2,1,1}(|z|)+3\text{Ti}_{1,-1,-2,1}(|z|) \nonumber \\
&+3\text{Ti}_{1,-1,-1,2}(|z|)+3\text{Ti}_{1,-1,1,2}(|z|)+3\text{Ti}_{1,-1,2,1}(|z|)+3\text{Ti}_{1,1,-2,1}(|z|)+3\text{Ti}_{1,1,-1,2}(|z|) \nonumber \\
&+3\text{Ti}_{1,1,1,2}(|z|)+3\text{Ti}_{1,1,2,1}(|z|)+3\text{Ti}_{1,2,-1,1}(|z|)+3\text{Ti}_{1,2,1,1}(|z|)+3\text{Ti}_{2,-1,-1,1}(|z|) \nonumber \\
&+3\text{Ti}_{2,-1,1,1}(|z|)+3\text{Ti}_{2,1,-1,1}(|z|)+3\text{Ti}_{2,1,1,1}(|z|)+\frac{1}{8}\text{Ti}_3(|z|)\zeta_2+\frac{3}{8}\text{Ti}_{-2,1}(|z|)\zeta_2 \nonumber \\
&-\frac{3}{8}\text{Ti}_{-1,2}(|z|)\zeta_2-\frac{3}{8}\text{Ti}_{1,2}(|z|)\zeta_2+\frac{3}{8}\text{Ti}_{2,1}(|z|)\zeta_2+\frac{3}{4}\text{Ti}_{-1,-1,1}(|z|)\zeta_2+\frac{3}{4}\text{Ti}_{-1,1,1}(|z|)\zeta_2 \nonumber \\
&+\frac{3}{4}\text{Ti}_{1,-1,1}(|z|)\zeta_2+\frac{3}{4}\text{Ti}_{1,1,1}(|z|)\zeta_2-\frac{3}{8}\text{Ti}_2(|z|)\zeta_3+\frac{3}{4}\text{Ti}_{-1,1}(|z|)\zeta_3+\frac{3}{4}\text{Ti}_{1,1}(|z|)\zeta_3,\nonumber \\ 
\fb^{(0)}_{5,\text{sym}}=&\frac{1}{8}G_{0,0,0}(|z|^2)+\frac{3}{8}G_{0,0}(|z|^2)-\frac{1}{8}\zeta_2G_0(|z|^2)-\frac{1}{4}G_{0,-1,0}(|z|^2)-\frac{3}{8}\zeta_3-\frac{3}{8}\zeta_2-\frac{1}{2},\nonumber \\ 
\fb^{(0)}_{5,\text{asym}}=&+\frac{1}{32}G_{0,0,0,0}(|z|^2)+\frac{3}{16}G_{0,0,0}(|z|^2)+\frac{1}{32}G_{0,0}(|z|^2)(16-\zeta_2)+\frac{1}{16}G_0(|z|^2)(-3\zeta_2-3\zeta_3+4) \nonumber \\
&+\frac{3}{16}\zeta_2G_{-1,0}(|z|^2)-\frac{3}{4}\zeta_2G_{-1,-1}(|z|^2)+\frac{3}{8}\zeta_2G_{0,-1}(|z|^2)-G_{-1,0}(|z|^2)+\frac{3}{2}G_{-1,-1,0}(|z|^2) \nonumber \\
&-\frac{3}{4}G_{-1,0,0}(|z|^2)-\frac{3}{4}G_{0,-1,0}(|z|^2)-\frac{3}{2}G_{-1,-1,-1,0}(|z|^2)+\frac{3}{4}G_{-1,-1,0,0}(|z|^2) \nonumber \\
&+\frac{3}{4}G_{-1,0,-1,0}(|z|^2)-\frac{3}{16}G_{-1,0,0,0}(|z|^2)+\frac{3}{4}G_{0,-1,-1,0}(|z|^2)-\frac{3}{8}G_{0,-1,0,0}(|z|^2) \nonumber \\
&-\frac{1}{4}G_{0,0,-1,0}(|z|^2)-\frac{\zeta_2}{2}-\frac{3\zeta_3}{4}-\frac{21\zeta_4}{128}+\frac{3}{4}\zeta_2G_{-1}(|z|^2)+\frac{3}{4}\zeta_3G_{-1}(|z|^2),\nonumber 
\end{align}

\begin{align}
\fc^{(0)}_{5,\text{sym}}=&-\frac{3}{8}\text{Ti}_1(|z|)G_{0,0,0}(|z|^2)-\frac{3}{4}G_{0,0}(|z|^2)\Big(2\text{Ti}_{-1,1}(|z|)+2\text{Ti}_{1,1}(|z|)+2\text{Ti}_1(|z|)-\text{Ti}_2(|z|)\Big) \nonumber \\
&+\frac{1}{8}G_0(|z|^2)\Big(-24\text{Ti}_{-1,1}(|z|)+24\text{Ti}_{-1,2}(|z|)-24\text{Ti}_{1,1}(|z|)+24\text{Ti}_{1,2}(|z|)-24\text{Ti}_{-1,-1,1}(|z|) \nonumber \\
&-24\text{Ti}_{-1,1,1}(|z|)-24\text{Ti}_{1,-1,1}(|z|)-24\text{Ti}_{1,1,1}(|z|)+3\zeta_2\text{Ti}_1(|z|)-16\text{Ti}_1(|z|)+24\text{Ti}_2(|z|) \nonumber \\
&-12\text{Ti}_3(|z|)\Big)+\frac{3}{2}\zeta_2\text{Ti}_{-1,1}(|z|)+\frac{3}{2}\zeta_2\text{Ti}_{1,1}(|z|)+6\text{Ti}_{-3,1}(|z|)+6\text{Ti}_{-2,1}(|z|)+6\text{Ti}_{-1,2}(|z|) \nonumber \\
&-6\text{Ti}_{-1,3}(|z|)+6\text{Ti}_{1,2}(|z|)-6\text{Ti}_{1,3}(|z|)+6\text{Ti}_{2,1}(|z|)+6\text{Ti}_{3,1}(|z|)+6\text{Ti}_{-2,-1,1}(|z|) \nonumber \\
&+6\text{Ti}_{-2,1,1}(|z|)+6\text{Ti}_{-1,-2,1}(|z|)+6\text{Ti}_{-1,-1,2}(|z|)+6\text{Ti}_{-1,1,2}(|z|)+6\text{Ti}_{-1,2,1}(|z|) \nonumber \\
&+6\text{Ti}_{1,-2,1}(|z|)+6\text{Ti}_{1,-1,2}(|z|)+6\text{Ti}_{1,1,2}(|z|)+6\text{Ti}_{1,2,1}(|z|)+6\text{Ti}_{2,-1,1}(|z|)+6\text{Ti}_{2,1,1}(|z|) \nonumber \\
&+\frac{3}{2}\zeta_2\text{Ti}_1(|z|)+\frac{3}{2}\zeta_3\text{Ti}_1(|z|)-\frac{3}{4}\zeta_2\text{Ti}_2(|z|)+4\text{Ti}_2(|z|)-6\text{Ti}_3(|z|)+3\text{Ti}_4(|z|), \nonumber
\end{align}
\enlargethispage{10pt}
The transcendental coefficients appearing in the expansion of $C_k^{(3,2)}$ are
\begin{align}
\fa^{(2)}_{2}=&\frac{1}{64}G_0(|z|^2)\text{Ti}_1(|z|)-\frac{1}{32}\text{Ti}_2(|z|),\nonumber \\ 
\fb^{(2)}_{2,\text{sym}}=&\frac{3}{32},\nonumber \\ 
\fb^{(2)}_{2,\text{asym}}=&-\frac{3}{64}G_0(|z|^2),\nonumber \\ 
\fc^{(2)}_{2,\text{sym}}=&\fd^{(2)}_{2}=0, \nonumber \\
\nonumber\\
\fa^{(2)}_{3}=&-\frac{1}{64}\text{Ti}_1(|z|)G_{0,0}(|z|^2)+\frac{1}{32}G_0(|z|^2)\Big(-\text{Ti}_{-1,1}(|z|)-\text{Ti}_{1,1}(|z|)+\text{Ti}_2(|z|)\Big)+\frac{1}{16}\text{Ti}_{-2,1}(|z|) \nonumber \\
&+\frac{1}{16}\text{Ti}_{-1,2}(|z|)+\frac{1}{16}\text{Ti}_{1,2}(|z|)+\frac{1}{16}\text{Ti}_{2,1}(|z|)+\frac{1}{64}\zeta_2\text{Ti}_1(|z|)-\frac{\text{Ti}_3(|z|)}{16},\nonumber \\ 
\fb^{(2)}_{3,\text{sym}}=&-\frac{1}{16},\nonumber \\ 
\fb^{(2)}_{3,\text{asym}}=&\frac{3}{64}G_{0,0}(|z|^2)+\frac{1}{32}G_0(|z|^2)-\frac{3}{32}G_{-1,0}(|z|^2)-\frac{3\zeta_2}{64},\nonumber \\ 
\fc^{(2)}_{3,\text{sym}}=&\fd^{(2)}_{3}=-\frac{1}{8}G_0(|z|^2)\text{Ti}_1(|z|)+\frac{1}{4}\text{Ti}_2(|z|) ,\nonumber \\
\nonumber\\
\fa^{(2)}_{4}=&\frac{1}{64}\text{Ti}_1(|z|)G_{0,0,0}(|z|^2)+\frac{1}{32}G_{0,0}(|z|^2)\Big(2\text{Ti}_{-1,1}(|z|)+2\text{Ti}_{1,1}(|z|)-\text{Ti}_2(|z|)\Big) \nonumber \\
&+\frac{1}{64}G_0(|z|^2)\Big(-8\text{Ti}_{-1,2}(|z|)-8\text{Ti}_{1,2}(|z|)+8\text{Ti}_{-1,-1,1}(|z|)+8\text{Ti}_{-1,1,1}(|z|)+8\text{Ti}_{1,-1,1}(|z|) \nonumber \\
&+8\text{Ti}_{1,1,1}(|z|)-\zeta_2\text{Ti}_1(|z|)+4\text{Ti}_3(|z|)\Big)-\frac{1}{16}\zeta_2\text{Ti}_{-1,1}(|z|)-\frac{1}{16}\zeta_2\text{Ti}_{1,1}(|z|)-\frac{1}{4}\text{Ti}_{-3,1}(|z|) \nonumber \\
&+\frac{1}{4}\text{Ti}_{-1,3}(|z|)+\frac{1}{4}\text{Ti}_{1,3}(|z|)-\frac{1}{4}\text{Ti}_{3,1}(|z|)-\frac{1}{4}\text{Ti}_{-2,-1,1}(|z|)-\frac{1}{4}\text{Ti}_{-2,1,1}(|z|) \nonumber \\
&-\frac{1}{4}\text{Ti}_{-1,-2,1}(|z|)-\frac{1}{4}\text{Ti}_{-1,-1,2}(|z|)-\frac{1}{4}\text{Ti}_{-1,1,2}(|z|)-\frac{1}{4}\text{Ti}_{-1,2,1}(|z|)-\frac{1}{4}\text{Ti}_{1,-2,1}(|z|) \nonumber \\
&-\frac{1}{4}\text{Ti}_{1,-1,2}(|z|)-\frac{1}{4}\text{Ti}_{1,1,2}(|z|)-\frac{1}{4}\text{Ti}_{1,2,1}(|z|)-\frac{1}{4}\text{Ti}_{2,-1,1}(|z|)-\frac{1}{4}\text{Ti}_{2,1,1}(|z|) \nonumber \\
&-\frac{1}{16}\zeta_3\text{Ti}_1(|z|)+\frac{1}{32}\zeta_2\text{Ti}_2(|z|)-\frac{\text{Ti}_4(|z|)}{8},\nonumber 
\end{align}
\begin{align}
\fb^{(2)}_{4,\text{sym}}=&-\frac{1}{96}G_{0,0}(|z|^2)+\frac{\zeta_2}{96}+\frac{1}{24},\nonumber \\ 
\fb^{(2)}_{4,\text{asym}}=&-\frac{3}{64}G_{0,0,0}(|z|^2)-\frac{7}{48}G_{0,0}(|z|^2)+\frac{1}{192}(9\zeta_2-4)G_0(|z|^2)+\frac{7}{24}G_{-1,0}(|z|^2)-\frac{3}{8}G_{-1,-1,0}(|z|^2) \nonumber \\
&+\frac{3}{16}G_{-1,0,0}(|z|^2)+\frac{3}{16}G_{0,-1,0}(|z|^2)+\frac{7\zeta_2}{48}+\frac{3\zeta_3}{16}-\frac{3}{16}\zeta_2G_{-1}(|z|^2),\nonumber \\
\fc^{(2)}_{4,\text{sym}}=&+\frac{1}{4}\text{Ti}_1(|z|)G_{0,0}(|z|^2)+\frac{1}{6}G_0(|z|^2)\Big(3\text{Ti}_{-1,1}(|z|)+3\text{Ti}_{1,1}(|z|)+2\text{Ti}_1(|z|)-3\text{Ti}_2(|z|)\Big) \nonumber \\
&-\text{Ti}_{-2,1}(|z|)-\text{Ti}_{-1,2}(|z|)-\text{Ti}_{1,2}(|z|)-\text{Ti}_{2,1}(|z|)-\frac{1}{4}\zeta_2\text{Ti}_1(|z|)-\frac{2\text{Ti}_2(|z|)}{3}+\text{Ti}_3(|z|),\nonumber \\ 
\fd^{(2)}_{4}=&+\frac{1}{4}\text{Ti}_1(|z|)G_{0,0}(|z|^2)+\frac{1}{2}G_0(|z|^2)\Big(\text{Ti}_{-1,1}(|z|)+\text{Ti}_{1,1}(|z|)+\text{Ti}_1(|z|)-\text{Ti}_2(|z|)\Big) \nonumber \\
&-\text{Ti}_{-2,1}(|z|)-\text{Ti}_{-1,2}(|z|)-\text{Ti}_{1,2}(|z|)-\text{Ti}_{2,1}(|z|)-\frac{1}{4}\zeta_2\text{Ti}_1(|z|)-\text{Ti}_2(|z|)+\text{Ti}_3(|z|), \nonumber\\
\fa^{(2)}_{5}=&-\frac{1}{64}G_{0,0,0,0}(|z|^2)\text{Ti}_1(|z|)+\frac{1}{32}G_{0,0,0}(|z|^2)\Big(\text{Ti}_2(|z|)-3\text{Ti}_{-1,1}(|z|)-3\text{Ti}_{1,1}(|z|)\Big) \nonumber \\
&+\frac{1}{64}G_{0,0}(|z|^2)\Big(-4\text{Ti}_3(|z|)-12\text{Ti}_{-2,1}(|z|)+12\text{Ti}_{-1,2}(|z|)+12\text{Ti}_{1,2}(|z|)-12\text{Ti}_{2,1}(|z|) \nonumber \\
&-24\text{Ti}_{-1,-1,1}(|z|)-24\text{Ti}_{-1,1,1}(|z|)-24\text{Ti}_{1,-1,1}(|z|)-24\text{Ti}_{1,1,1}(|z|)+\text{Ti}_1(|z|)\zeta_2\Big) \nonumber \\
&+\frac{1}{32}G_0(|z|^2)\Big(4\text{Ti}_4(|z|)+20\text{Ti}_{-3,1}(|z|)+12\text{Ti}_{-2,2}(|z|)-12\text{Ti}_{-1,3}(|z|)-12\text{Ti}_{1,3}(|z|) \nonumber \\
&+12\text{Ti}_{2,2}(|z|)+20\text{Ti}_{3,1}(|z|)+24\text{Ti}_{-1,-1,2}(|z|)+24\text{Ti}_{-1,1,2}(|z|)+24\text{Ti}_{1,-1,2}(|z|) \nonumber \\
&+24\text{Ti}_{1,1,2}(|z|)-24\text{Ti}_{-1,-1,-1,1}(|z|)-24\text{Ti}_{-1,-1,1,1}(|z|)-24\text{Ti}_{-1,1,-1,1}(|z|) \nonumber \\
&-24\text{Ti}_{-1,1,1,1}(|z|)-24\text{Ti}_{1,-1,-1,1}(|z|)-24\text{Ti}_{1,-1,1,1}(|z|)-24\text{Ti}_{1,1,-1,1}(|z|) \nonumber \\
&-24\text{Ti}_{1,1,1,1}(|z|)-\text{Ti}_2(|z|)\zeta_2+3\text{Ti}_{-1,1}(|z|)\zeta_2+3\text{Ti}_{1,1}(|z|)\zeta_2+3\text{Ti}_1(|z|)\zeta_3\Big) \nonumber \\
&+\frac{21}{256}\zeta_4\text{Ti}_1(|z|)-\frac{\text{Ti}_5(|z|)}{4}-\frac{3}{4}\text{Ti}_{-4,1}(|z|)-\frac{5}{4}\text{Ti}_{-3,2}(|z|)-\frac{3}{4}\text{Ti}_{-2,3}(|z|)+\frac{3}{4}\text{Ti}_{-1,4}(|z|) \nonumber \\
&+\frac{3}{4}\text{Ti}_{1,4}(|z|)-\frac{3}{4}\text{Ti}_{2,3}(|z|)-\frac{5}{4}\text{Ti}_{3,2}(|z|)-\frac{3}{4}\text{Ti}_{4,1}(|z|)+\frac{3}{2}\text{Ti}_{-3,-1,1}(|z|)+\frac{3}{2}\text{Ti}_{-3,1,1}(|z|) \nonumber \\
&+\frac{3}{2}\text{Ti}_{-2,-2,1}(|z|)+\frac{3}{2}\text{Ti}_{-2,2,1}(|z|)+\frac{3}{2}\text{Ti}_{-1,-3,1}(|z|)-\frac{3}{2}\text{Ti}_{-1,-1,3}(|z|)-\frac{3}{2}\text{Ti}_{-1,1,3}(|z|) \nonumber \\
&+\frac{3}{2}\text{Ti}_{-1,3,1}(|z|)+\frac{3}{2}\text{Ti}_{1,-3,1}(|z|)-\frac{3}{2}\text{Ti}_{1,-1,3}(|z|)-\frac{3}{2}\text{Ti}_{1,1,3}(|z|)+\frac{3}{2}\text{Ti}_{1,3,1}(|z|) \nonumber \\
&+\frac{3}{2}\text{Ti}_{2,-2,1}(|z|)+\frac{3}{2}\text{Ti}_{2,2,1}(|z|)+\frac{3}{2}\text{Ti}_{3,-1,1}(|z|)+\frac{3}{2}\text{Ti}_{3,1,1}(|z|)+\frac{3}{2}\text{Ti}_{-2,-1,-1,1}(|z|) \nonumber \\
&+\frac{3}{2}\text{Ti}_{-2,-1,1,1}(|z|)+\frac{3}{2}\text{Ti}_{-2,1,-1,1}(|z|)+\frac{3}{2}\text{Ti}_{-2,1,1,1}(|z|)+\frac{3}{2}\text{Ti}_{-1,-2,-1,1}(|z|) \nonumber \\
&+\frac{3}{2}\text{Ti}_{-1,-2,1,1}(|z|)+\frac{3}{2}\text{Ti}_{-1,-1,-2,1}(|z|)+\frac{3}{2}\text{Ti}_{-1,-1,-1,2}(|z|)+\frac{3}{2}\text{Ti}_{-1,-1,1,2}(|z|) \nonumber \\
&+\frac{3}{2}\text{Ti}_{-1,-1,2,1}(|z|)+\frac{3}{2}\text{Ti}_{-1,1,-2,1}(|z|)+\frac{3}{2}\text{Ti}_{-1,1,-1,2}(|z|)+\frac{3}{2}\text{Ti}_{-1,1,1,2}(|z|) \nonumber \\
&+\frac{3}{2}\text{Ti}_{-1,1,2,1}(|z|)+\frac{3}{2}\text{Ti}_{-1,2,-1,1}(|z|)+\frac{3}{2}\text{Ti}_{-1,2,1,1}(|z|)+\frac{3}{2}\text{Ti}_{1,-2,-1,1}(|z|) \nonumber \\
&+\frac{3}{2}\text{Ti}_{1,-2,1,1}(|z|)+\frac{3}{2}\text{Ti}_{1,-1,-2,1}(|z|)+\frac{3}{2}\text{Ti}_{1,-1,-1,2}(|z|)+\frac{3}{2}\text{Ti}_{1,-1,1,2}(|z|) \nonumber \\
&+\frac{3}{2}\text{Ti}_{1,-1,2,1}(|z|)+\frac{3}{2}\text{Ti}_{1,1,-2,1}(|z|)+\frac{3}{2}\text{Ti}_{1,1,-1,2}(|z|)+\frac{3}{2}\text{Ti}_{1,1,1,2}(|z|)+\frac{3}{2}\text{Ti}_{1,1,2,1}(|z|) \nonumber \\
&+\frac{3}{2}\text{Ti}_{1,2,-1,1}(|z|)+\frac{3}{2}\text{Ti}_{1,2,1,1}(|z|)+\frac{3}{2}\text{Ti}_{2,-1,-1,1}(|z|)+\frac{3}{2}\text{Ti}_{2,-1,1,1}(|z|)+\frac{3}{2}\text{Ti}_{2,1,-1,1}(|z|) \nonumber 
\end{align}
\begin{align}
&+\frac{3}{2}\text{Ti}_{2,1,1,1}(|z|)+\frac{1}{16}\text{Ti}_3(|z|)\zeta_2+\frac{3}{16}\text{Ti}_{-2,1}(|z|)\zeta_2-\frac{3}{16}\text{Ti}_{-1,2}(|z|)\zeta_2-\frac{3}{16}\text{Ti}_{1,2}(|z|)\zeta_2 \nonumber \\
&+\frac{3}{16}\text{Ti}_{2,1}(|z|)\zeta_2+\frac{3}{8}\text{Ti}_{-1,-1,1}(|z|)\zeta_2+\frac{3}{8}\text{Ti}_{-1,1,1}(|z|)\zeta_2+\frac{3}{8}\text{Ti}_{1,-1,1}(|z|)\zeta_2 \nonumber \\
&+\frac{3}{8}\text{Ti}_{1,1,1}(|z|)\zeta_2-\frac{3}{16}\text{Ti}_2(|z|)\zeta_3+\frac{3}{8}\text{Ti}_{-1,1}(|z|)\zeta_3+\frac{3}{8}\text{Ti}_{1,1}(|z|)\zeta_3,\nonumber \\ 
\fb^{(2)}_{5,\text{sym}}=&\frac{1}{48}G_{0,0,0}(|z|^2)-\frac{5}{144}G_{0,0}(|z|^2)-\frac{1}{48}\zeta_2G_0(|z|^2)-\frac{1}{24}G_{0,-1,0}(|z|^2)+\frac{5\zeta_2}{144}-\frac{\zeta_3}{16}-\frac{1}{36},\nonumber \\ 
\fb^{(2)}_{5,\text{asym}}=&+\frac{3}{64}G_{0,0,0,0}(|z|^2)+\frac{25}{96}G_{0,0,0}(|z|^2)+\frac{1}{576}G_{0,0}(|z|^2)\Big(272-27\zeta_2\Big) \nonumber \\
&+\frac{1}{288}G_0(|z|^2)\Big(-75\zeta_2-81\zeta_3+4\Big)+\frac{9}{32}\zeta_2G_{-1,0}(|z|^2)-\frac{9}{8}\zeta_2G_{-1,-1}(|z|^2) \nonumber \\
&+\frac{9}{16}\zeta_2G_{0,-1}(|z|^2)-\frac{17}{18}G_{-1,0}(|z|^2)+\frac{7}{4}G_{-1,-1,0}(|z|^2)-\frac{7}{8}G_{-1,0,0}(|z|^2)-\frac{7}{8}G_{0,-1,0}(|z|^2) \nonumber \\
&-\frac{9}{4}G_{-1,-1,-1,0}(|z|^2)+\frac{9}{8}G_{-1,-1,0,0}(|z|^2)+\frac{9}{8}G_{-1,0,-1,0}(|z|^2)-\frac{9}{32}G_{-1,0,0,0}(|z|^2) \nonumber \\
&+\frac{9}{8}G_{0,-1,-1,0}(|z|^2)-\frac{9}{16}G_{0,-1,0,0}(|z|^2)-\frac{3}{8}G_{0,0,-1,0}(|z|^2)-\frac{17\zeta_2}{36}-\frac{7\zeta_3}{8}-\frac{63\zeta_4}{256} \nonumber \\
&+\frac{7}{8}\zeta_2G_{-1}(|z|^2)+\frac{9}{8}\zeta_3G_{-1}(|z|^2),\nonumber \\ 
\fc^{(2)}_{5,\text{sym}}=&-\frac{3}{8}\text{Ti}_1(|z|)G_{0,0,0}(|z|^2)+\frac{1}{4}G_{0,0}(|z|^2)\Big(-6\text{Ti}_{-1,1}(|z|)-6\text{Ti}_{1,1}(|z|)-4\text{Ti}_1(|z|)+3\text{Ti}_2(|z|)\Big) \nonumber \\
&+\frac{1}{72}G_0(|z|^2)\Big(-144\text{Ti}_{-1,1}(|z|)+216\text{Ti}_{-1,2}(|z|)-144\text{Ti}_{1,1}(|z|)+216\text{Ti}_{1,2}(|z|) \nonumber \\
&-216\text{Ti}_{-1,-1,1}(|z|)-216\text{Ti}_{-1,1,1}(|z|)-216\text{Ti}_{1,-1,1}(|z|)-216\text{Ti}_{1,1,1}(|z|)+27\zeta_2\text{Ti}_1(|z|) \nonumber \\
&-64\text{Ti}_1(|z|)+144\text{Ti}_2(|z|)-108\text{Ti}_3(|z|)\Big)+\frac{3}{2}\zeta_2\text{Ti}_{-1,1}(|z|)+\frac{3}{2}\zeta_2\text{Ti}_{1,1}(|z|)+6\text{Ti}_{-3,1}(|z|) \nonumber \\
&+4\text{Ti}_{-2,1}(|z|)+4\text{Ti}_{-1,2}(|z|)-6\text{Ti}_{-1,3}(|z|)+4\text{Ti}_{1,2}(|z|)-6\text{Ti}_{1,3}(|z|)+4\text{Ti}_{2,1}(|z|) \nonumber \\
&+6\text{Ti}_{3,1}(|z|)+6\text{Ti}_{-2,-1,1}(|z|)+6\text{Ti}_{-2,1,1}(|z|)+6\text{Ti}_{-1,-2,1}(|z|)+6\text{Ti}_{-1,-1,2}(|z|) \nonumber \\
&+6\text{Ti}_{-1,1,2}(|z|)+6\text{Ti}_{-1,2,1}(|z|)+6\text{Ti}_{1,-2,1}(|z|)+6\text{Ti}_{1,-1,2}(|z|)+6\text{Ti}_{1,1,2}(|z|) \nonumber \\
&+6\text{Ti}_{1,2,1}(|z|)+6\text{Ti}_{2,-1,1}(|z|)+6\text{Ti}_{2,1,1}(|z|)+\zeta_2\text{Ti}_1(|z|)+\frac{3}{2}\zeta_3\text{Ti}_1(|z|)-\frac{3}{4}\zeta_2\text{Ti}_2(|z|) \nonumber \\
&+\frac{16}{9}\text{Ti}_2(|z|)-4\text{Ti}_3(|z|)+3\text{Ti}_4(|z|),\nonumber \\ 
\fd^{(2)}_{5}=&-\frac{3}{8}\text{Ti}_1(|z|)G_{0,0,0}(|z|^2)-\frac{3}{4}G_{0,0}(|z|^2)\Big(2\text{Ti}_{-1,1}(|z|)+2\text{Ti}_{1,1}(|z|)+2\text{Ti}_1(|z|)-\text{Ti}_2(|z|)\Big) \nonumber \\
&+\frac{1}{8}G_0(|z|^2)\Big(-24\text{Ti}_{-1,1}(|z|)+24\text{Ti}_{-1,2}(|z|)-24\text{Ti}_{1,1}(|z|)+24\text{Ti}_{1,2}(|z|)-24\text{Ti}_{-1,-1,1}(|z|) \nonumber \\
&-24\text{Ti}_{-1,1,1}(|z|)-24\text{Ti}_{1,-1,1}(|z|)-24\text{Ti}_{1,1,1}(|z|)+3\zeta_2\text{Ti}_1(|z|)-16\text{Ti}_1(|z|)+24\text{Ti}_2(|z|) \nonumber \\
&-12\text{Ti}_3(|z|)\Big)+\frac{3}{2}\zeta_2\text{Ti}_{-1,1}(|z|)+\frac{3}{2}\zeta_2\text{Ti}_{1,1}(|z|)+6\text{Ti}_{-3,1}(|z|)+6\text{Ti}_{-2,1}(|z|)+6\text{Ti}_{-1,2}(|z|) \nonumber \\
&-6\text{Ti}_{-1,3}(|z|)+6\text{Ti}_{1,2}(|z|)-6\text{Ti}_{1,3}(|z|)+6\text{Ti}_{2,1}(|z|)+6\text{Ti}_{3,1}(|z|)+6\text{Ti}_{-2,-1,1}(|z|) \nonumber \\
&+6\text{Ti}_{-2,1,1}(|z|)+6\text{Ti}_{-1,-2,1}(|z|)+6\text{Ti}_{-1,-1,2}(|z|)+6\text{Ti}_{-1,1,2}(|z|)+6\text{Ti}_{-1,2,1}(|z|) \nonumber \\
&+6\text{Ti}_{1,-2,1}(|z|)+6\text{Ti}_{1,-1,2}(|z|)+6\text{Ti}_{1,1,2}(|z|)+6\text{Ti}_{1,2,1}(|z|)+6\text{Ti}_{2,-1,1}(|z|)+6\text{Ti}_{2,1,1}(|z|) \nonumber \\
&+\frac{3}{2}\zeta_2\text{Ti}_1(|z|)+\frac{3}{2}\zeta_3\text{Ti}_1(|z|)-\frac{3}{4}\zeta_2\text{Ti}_2(|z|)+4\text{Ti}_2(|z|)-6\text{Ti}_3(|z|)+3\text{Ti}_4(|z|). \nonumber
\end{align}

\bibliography{refs}

\providecommand{\href}[2]{#2}\begingroup\raggedright\begin{thebibliography}{10}

\bibitem{Fadin:1975cb}
V.~S. Fadin, E.~A. Kuraev, and L.~N. Lipatov, {\it {On the Pomeranchuk
  Singularity in Asymptotically Free Theories}},  {\em Phys. Lett.} {\bf B60}
  (1975) 50--52.

\bibitem{Kuraev:1976ge}
E.~A. Kuraev, L.~N. Lipatov, and V.~S. Fadin, {\it {Multi - Reggeon Processes
  in the Yang-Mills Theory}},  {\em Sov. Phys. JETP} {\bf 44} (1976) 443--450.
  [Zh. Eksp. Teor. Fiz.71,840(1976)].

\bibitem{Kuraev:1977fs}
E.~A. Kuraev, L.~N. Lipatov, and V.~S. Fadin, {\it {The Pomeranchuk Singularity
  in Nonabelian Gauge Theories}},  {\em Sov. Phys. JETP} {\bf 45} (1977)
  199--204. [Zh. Eksp. Teor. Fiz.72,377(1977)].

\bibitem{Balitsky:1978ic}
I.~I. Balitsky and L.~N. Lipatov, {\it {The Pomeranchuk Singularity in Quantum
  Chromodynamics}},  {\em Sov. J. Nucl. Phys.} {\bf 28} (1978) 822--829. [Yad.
  Fiz.28,1597(1978)].

\bibitem{Fadin:1998py}
V.~S. Fadin and L.~N. Lipatov, {\it {BFKL pomeron in the next-to-leading
  approximation}},  {\em Phys. Lett.} {\bf B429} (1998) 127--134,
  [\href{http://xxx.lanl.gov/abs/hep-ph/9802290}{{\tt hep-ph/9802290}}].

\bibitem{Ciafaloni:1998gs}
M.~Ciafaloni and G.~Camici, {\it {Energy scale(s) and next-to-leading BFKL
  equation}},  {\em Phys. Lett.} {\bf B430} (1998) 349--354,
  [\href{http://xxx.lanl.gov/abs/hep-ph/9803389}{{\tt hep-ph/9803389}}].

\bibitem{Chirilli:2013kca}
G.~A. Chirilli and Y.~V. Kovchegov, {\it {Solution of the NLO BFKL Equation and
  a Strategy for Solving the All-Order BFKL Equation}},  {\em JHEP} {\bf 06}
  (2013) 055, [\href{http://xxx.lanl.gov/abs/1305.1924}{{\tt 1305.1924}}].

\bibitem{Chirilli:2014dcb}
G.~A. Chirilli and Y.~V. Kovchegov, {\it {$\gamma^* \gamma^*$ Cross Section at
  NLO and Properties of the BFKL Evolution at Higher Orders}},  {\em JHEP} {\bf
  05} (2014) 099, [\href{http://xxx.lanl.gov/abs/1403.3384}{{\tt 1403.3384}}].
  [Erratum: JHEP08,075(2015)].

\bibitem{DelDuca:2013lma}
V.~Del~Duca, L.~J. Dixon, C.~Duhr, and J.~Pennington, {\it {The BFKL equation,
  Mueller-Navelet jets and single-valued harmonic polylogarithms}},  {\em JHEP}
  {\bf 02} (2014) 086, [\href{http://xxx.lanl.gov/abs/1309.6647}{{\tt
  1309.6647}}].

\bibitem{BrownSVHPLs}
F.~C.~S. Brown, {\it Single-valued multiple polylogarithms in one variable},
  {\em C. R. Acad. Sci. Paris, Ser. I} {\bf 338} (2004) 527.

\bibitem{Schnetz:2016fhy}
O.~Schnetz, {\it {Numbers and Functions in Quantum Field Theory}},
  \href{http://xxx.lanl.gov/abs/1606.08598}{{\tt 1606.08598}}.

\bibitem{Kotikov:2000pm}
A.~V. Kotikov and L.~N. Lipatov, {\it {NLO corrections to the BFKL equation in
  QCD and in supersymmetric gauge theories}},  {\em Nucl. Phys.} {\bf B582}
  (2000) 19--43, [\href{http://xxx.lanl.gov/abs/hep-ph/0004008}{{\tt
  hep-ph/0004008}}].

\bibitem{Kotikov:2002ab}
A.~V. Kotikov and L.~N. Lipatov, {\it {DGLAP and BFKL equations in the $N=4$
  supersymmetric gauge theory}},  {\em Nucl. Phys.} {\bf B661} (2003) 19--61,
  [\href{http://xxx.lanl.gov/abs/hep-ph/0208220}{{\tt hep-ph/0208220}}].
  [Erratum: Nucl. Phys.B685,405(2004)].

\bibitem{Kotikov:2003fb}
A.~V. Kotikov, L.~N. Lipatov, and V.~N. Velizhanin, {\it {Anomalous dimensions
  of Wilson operators in N=4 SYM theory}},  {\em Phys. Lett.} {\bf B557} (2003)
  114--120, [\href{http://xxx.lanl.gov/abs/hep-ph/0301021}{{\tt
  hep-ph/0301021}}].

\bibitem{Kotikov:2004er}
A.~V. Kotikov, L.~N. Lipatov, A.~I. Onishchenko, and V.~N. Velizhanin, {\it
  {Three loop universal anomalous dimension of the Wilson operators in $N=4$
  SUSY Yang-Mills model}},  {\em Phys. Lett.} {\bf B595} (2004) 521--529,
  [\href{http://xxx.lanl.gov/abs/hep-th/0404092}{{\tt hep-th/0404092}}].
  [Erratum: Phys. Lett.B632,754(2006)].

\bibitem{Moch:2004pa}
S.~Moch, J.~A.~M. Vermaseren, and A.~Vogt, {\it {The Three loop splitting
  functions in QCD: The Nonsinglet case}},  {\em Nucl. Phys.} {\bf B688} (2004)
  101--134, [\href{http://xxx.lanl.gov/abs/hep-ph/0403192}{{\tt
  hep-ph/0403192}}].

\bibitem{Lipatov:1985uk}
L.~N. Lipatov, {\it {The Bare Pomeron in Quantum Chromodynamics}},  {\em Sov.
  Phys. JETP} {\bf 63} (1986) 904--912. [Zh. Eksp. Teor. Fiz.90,1536(1986)].

\bibitem{Dixon:2012yy}
L.~J. Dixon, C.~Duhr, and J.~Pennington, {\it {Single-valued harmonic
  polylogarithms and the multi-Regge limit}},  {\em JHEP} {\bf 10} (2012) 074,
  [\href{http://xxx.lanl.gov/abs/1207.0186}{{\tt 1207.0186}}].

\bibitem{Remiddi:1999ew}
E.~Remiddi and J.~A.~M. Vermaseren, {\it {Harmonic polylogarithms}},  {\em Int.
  J. Mod. Phys.} {\bf A15} (2000) 725--754,
  [\href{http://xxx.lanl.gov/abs/hep-ph/9905237}{{\tt hep-ph/9905237}}].

\bibitem{Goncharov:1998}
A.~B. Goncharov, {\it {Multiple polylogarithms, cyclotomy and modular
  complexes}},  {\em Math. Research Letters} {\bf 5} (1998), no.~4 497.

\bibitem{Goncharov:2001}
A.~B. Goncharov, {\it {Multiple polylogarithms and mixed Tate motives}},
  \href{http://xxx.lanl.gov/abs/math/0103059v4}{{\tt math/0103059v4}}.

\bibitem{BrownSVMPLs}
F.~C.~S. Brown, {\it {Single-valued hyperlogarithms and unipotent differential
  equations}},  {\em \verb+http://www.ihes.fr/~brown/RHpaper5.pdf+}.

\bibitem{Brown:2013gia}
F.~Brown, {\it {Single-valued Motivic Periods and Multiple Zeta Values}},  {\em
  SIGMA} {\bf 2} (2014) e25, [\href{http://xxx.lanl.gov/abs/1309.5309}{{\tt
  1309.5309}}].

\bibitem{Brown_Notes}
F.~C.~S. Brown, {\it {Notes on motivic periods}},  {\em {}} (2015)
  [\href{http://xxx.lanl.gov/abs/1512.06410}{{\tt 1512.06410}}].

\bibitem{DelDuca:2016lad}
V.~Del~Duca, S.~Druc, J.~Drummond, C.~Duhr, F.~Dulat, R.~Marzucca,
  G.~Papathanasiou, and B.~Verbeek, {\it {Multi-Regge kinematics and the moduli
  space of Riemann spheres with marked points}},  {\em JHEP} {\bf 08} (2016)
  152, [\href{http://xxx.lanl.gov/abs/1606.08807}{{\tt 1606.08807}}].

\bibitem{ArkaniHamed:2010gh}
N.~Arkani-Hamed, J.~L. Bourjaily, F.~Cachazo, and J.~Trnka, {\it {Local
  Integrals for Planar Scattering Amplitudes}},  {\em JHEP} {\bf 06} (2012)
  125, [\href{http://xxx.lanl.gov/abs/1012.6032}{{\tt 1012.6032}}].

\bibitem{Vermaseren:1998uu}
J.~A.~M. Vermaseren, {\it {Harmonic sums, Mellin transforms and integrals}},
  {\em Int. J. Mod. Phys.} {\bf A14} (1999) 2037--2076,
  [\href{http://xxx.lanl.gov/abs/hep-ph/9806280}{{\tt hep-ph/9806280}}].

\bibitem{Moch:2001zr}
S.~Moch, P.~Uwer, and S.~Weinzierl, {\it {Nested sums, expansion of
  transcendental functions and multiscale multiloop integrals}},  {\em J. Math.
  Phys.} {\bf 43} (2002) 3363--3386,
  [\href{http://xxx.lanl.gov/abs/hep-ph/0110083}{{\tt hep-ph/0110083}}].

\bibitem{Weinzierl:2004bn}
S.~Weinzierl, {\it {Expansion around half-integer values, binomial sums and
  inverse binomial sums}},  {\em J. Math. Phys.} {\bf 45} (2004) 2656--2673,
  [\href{http://xxx.lanl.gov/abs/hep-ph/0402131}{{\tt hep-ph/0402131}}].

\bibitem{Schnetz:2013hqa}
O.~Schnetz, {\it {Graphical functions and single-valued multiple
  polylogarithms}},  {\em Commun. Num. Theor. Phys.} {\bf 08} (2014) 589--675,
  [\href{http://xxx.lanl.gov/abs/1302.6445}{{\tt 1302.6445}}].

\bibitem{Gadde:2009dj}
A.~Gadde, E.~Pomoni, and L.~Rastelli, {\it {The Veneziano Limit of N = 2
  Superconformal QCD: Towards the String Dual of N = 2 SU(N(c)) SYM with N(f) =
  2 N(c)}},  \href{http://xxx.lanl.gov/abs/0912.4918}{{\tt 0912.4918}}.

\bibitem{Seiberg:1994pq}
N.~Seiberg, {\it {Electric - magnetic duality in supersymmetric nonAbelian
  gauge theories}},  {\em Nucl. Phys.} {\bf B435} (1995) 129--146,
  [\href{http://xxx.lanl.gov/abs/hep-th/9411149}{{\tt hep-th/9411149}}].

\bibitem{Bern:1993mq}
Z.~Bern, L.~J. Dixon, and D.~A. Kosower, {\it {One loop corrections to five
  gluon amplitudes}},  {\em Phys. Rev. Lett.} {\bf 70} (1993) 2677--2680,
  [\href{http://xxx.lanl.gov/abs/hep-ph/9302280}{{\tt hep-ph/9302280}}].

\bibitem{Bern:1993tz}
Z.~Bern and A.~G. Morgan, {\it {Supersymmetry relations between contributions
  to one loop gauge boson amplitudes}},  {\em Phys. Rev.} {\bf D49} (1994)
  6155--6163, [\href{http://xxx.lanl.gov/abs/hep-ph/9312218}{{\tt
  hep-ph/9312218}}].

\bibitem{Bern:1994zx}
Z.~Bern, L.~J. Dixon, D.~C. Dunbar, and D.~A. Kosower, {\it {One loop n point
  gauge theory amplitudes, unitarity and collinear limits}},  {\em Nucl. Phys.}
  {\bf B425} (1994) 217--260,
  [\href{http://xxx.lanl.gov/abs/hep-ph/9403226}{{\tt hep-ph/9403226}}].

\bibitem{Dixon:1996wi}
L.~J. Dixon, {\it {Calculating scattering amplitudes efficiently}},  in {\em
  {QCD and beyond. Proceedings, Theoretical Advanced Study Institute in
  Elementary Particle Physics, TASI-95, Boulder, USA, June 4-30, 1995}},
  pp.~539--584, 1996.
\newblock \href{http://xxx.lanl.gov/abs/hep-ph/9601359}{{\tt hep-ph/9601359}}.

\bibitem{Catani:2000pi}
S.~Catani and M.~Grazzini, {\it {The soft gluon current at one loop order}},
  {\em Nucl. Phys.} {\bf B591} (2000) 435--454,
  [\href{http://xxx.lanl.gov/abs/hep-ph/0007142}{{\tt hep-ph/0007142}}].

\bibitem{Duhr:2013msa}
C.~Duhr and T.~Gehrmann, {\it {The two-loop soft current in dimensional
  regularization}},  {\em Phys. Lett.} {\bf B727} (2013) 452--455,
  [\href{http://xxx.lanl.gov/abs/1309.4393}{{\tt 1309.4393}}].

\bibitem{Li:2013lsa}
Y.~Li and H.~X. Zhu, {\it {Single soft gluon emission at two loops}},  {\em
  JHEP} {\bf 11} (2013) 080, [\href{http://xxx.lanl.gov/abs/1309.4391}{{\tt
  1309.4391}}].

\bibitem{Korchemsky:1987wg}
G.~P. Korchemsky and A.~V. Radyushkin, {\it {Renormalization of the Wilson
  Loops Beyond the Leading Order}},  {\em Nucl. Phys.} {\bf B283} (1987)
  342--364.

\bibitem{Grozin:2014hna}
A.~Grozin, J.~M. Henn, G.~P. Korchemsky, and P.~Marquard, {\it {Three Loop Cusp
  Anomalous Dimension in QCD}},  {\em Phys. Rev. Lett.} {\bf 114} (2015), no.~6
  062006, [\href{http://xxx.lanl.gov/abs/1409.0023}{{\tt 1409.0023}}].

\bibitem{Grozin:2015kna}
A.~Grozin, J.~M. Henn, G.~P. Korchemsky, and P.~Marquard, {\it {The three-loop
  cusp anomalous dimension in QCD and its supersymmetric extensions}},  {\em
  JHEP} {\bf 01} (2016) 140, [\href{http://xxx.lanl.gov/abs/1510.07803}{{\tt
  1510.07803}}].

\bibitem{Caswell:1974gg}
W.~E. Caswell, {\it {Asymptotic Behavior of Nonabelian Gauge Theories to Two
  Loop Order}},  {\em Phys. Rev. Lett.} {\bf 33} (1974) 244.

\bibitem{Jones:1974mm}
D.~R.~T. Jones, {\it {Two Loop Diagrams in Yang-Mills Theory}},  {\em Nucl.
  Phys.} {\bf B75} (1974) 531.

\bibitem{Tarasov:1980au}
O.~V. Tarasov, A.~A. Vladimirov, and A.~{\relax Yu}. Zharkov, {\it {The
  Gell-Mann-Low Function of QCD in the Three Loop Approximation}},  {\em Phys.
  Lett.} {\bf B93} (1980) 429--432.

\bibitem{Larin:1993tp}
S.~A. Larin and J.~A.~M. Vermaseren, {\it {The Three loop QCD Beta function and
  anomalous dimensions}},  {\em Phys. Lett.} {\bf B303} (1993) 334--336,
  [\href{http://xxx.lanl.gov/abs/hep-ph/9302208}{{\tt hep-ph/9302208}}].

\bibitem{vanRitbergen:1997va}
T.~van Ritbergen, J.~A.~M. Vermaseren, and S.~A. Larin, {\it {The Four loop
  beta function in quantum chromodynamics}},  {\em Phys. Lett.} {\bf B400}
  (1997) 379--384, [\href{http://xxx.lanl.gov/abs/hep-ph/9701390}{{\tt
  hep-ph/9701390}}].

\bibitem{Czakon:2004bu}
M.~Czakon, {\it {The Four-loop QCD beta-function and anomalous dimensions}},
  {\em Nucl. Phys.} {\bf B710} (2005) 485--498,
  [\href{http://xxx.lanl.gov/abs/hep-ph/0411261}{{\tt hep-ph/0411261}}].

\bibitem{Herzog:2017ohr}
F.~Herzog, B.~Ruijl, T.~Ueda, J.~A.~M. Vermaseren, and A.~Vogt, {\it {The
  five-loop beta function of Yang-Mills theory with fermions}},  {\em JHEP}
  {\bf 02} (2017) 090, [\href{http://xxx.lanl.gov/abs/1701.01404}{{\tt
  1701.01404}}].

\bibitem{Andree:2010na}
R.~Andree and D.~Young, {\it {Wilson Loops in N=2 Superconformal Yang-Mills
  Theory}},  {\em JHEP} {\bf 09} (2010) 095,
  [\href{http://xxx.lanl.gov/abs/1007.4923}{{\tt 1007.4923}}].

\bibitem{Leoni:2014fja}
M.~Leoni, A.~Mauri, and A.~Santambrogio, {\it {Four-point amplitudes in
  $\mathcal{N}=2$ SCQCD}},  {\em JHEP} {\bf 09} (2014) 017,
  [\href{http://xxx.lanl.gov/abs/1406.7283}{{\tt 1406.7283}}]. [Erratum:
  JHEP02,022(2015)].

\bibitem{Leoni:2015zxa}
M.~Leoni, A.~Mauri, and A.~Santambrogio, {\it {On the amplitude/Wilson loop
  duality in N=2 SCQCD}},  {\em Phys. Lett.} {\bf B747} (2015) 325--330,
  [\href{http://xxx.lanl.gov/abs/1502.07614}{{\tt 1502.07614}}].

\bibitem{Lipatov:1976zz}
L.~N. Lipatov, {\it {Reggeization of the Vector Meson and the Vacuum
  Singularity in Nonabelian Gauge Theories}},  {\em Sov. J. Nucl. Phys.} {\bf
  23} (1976) 338--345. [Yad. Fiz.23,642(1976)].

\bibitem{Balitsky:1979ap}
I.~I. Balitsky, L.~N. Lipatov, and V.~S. Fadin, {\it {REGGE PROCESSES IN
  NONABELIAN GAUGE THEORIES. (IN RUSSIAN)}}, .

\bibitem{Fadin:2006bj}
V.~S. Fadin, R.~Fiore, M.~G. Kozlov, and A.~V. Reznichenko, {\it {Proof of the
  multi-Regge form of QCD amplitudes with gluon exchanges in the NLA}},  {\em
  Phys. Lett.} {\bf B639} (2006) 74--81,
  [\href{http://xxx.lanl.gov/abs/hep-ph/0602006}{{\tt hep-ph/0602006}}].

\bibitem{Kozlov:2011zza}
M.~G. Kozlov, A.~V. Reznichenko, and V.~S. Fadin, {\it {Check of the
  gluon-reggeization condition in the next-to-leading order: Quark part}},
  {\em Phys. Atom. Nucl.} {\bf 74} (2011) 758--770. [Yad. Fiz.74,784(2011)].

\bibitem{Kozlov:2012zza}
M.~G. Kozlov, A.~V. Reznichenko, and V.~S. Fadin, {\it {Check of the
  gluon-Reggeization condition in the next-to-leading order: Gluon part}},
  {\em Phys. Atom. Nucl.} {\bf 75} (2012) 493--506.

\bibitem{Kozlov:2012zz}
M.~G. Kozlov, A.~V. Reznichenko, and V.~S. Fadin, {\it {Impact factor for gluon
  production in multi-Regge kinematics in the next-to-leading order}},  {\em
  Phys. Atom. Nucl.} {\bf 75} (2012) 850--865.

\bibitem{Fadin:2015zea}
V.~S. Fadin, M.~G. Kozlov, and A.~V. Reznichenko, {\it {Gluon Reggeization in
  Yang-Mills Theories}},  {\em Phys. Rev.} {\bf D92} (2015), no.~8 085044,
  [\href{http://xxx.lanl.gov/abs/1507.00823}{{\tt 1507.00823}}].

\bibitem{DelDuca:2001gu}
V.~Del~Duca and E.~W.~N. Glover, {\it {The High-energy limit of QCD at two
  loops}},  {\em JHEP} {\bf 10} (2001) 035,
  [\href{http://xxx.lanl.gov/abs/hep-ph/0109028}{{\tt hep-ph/0109028}}].

\bibitem{Bret:2011xm}
V.~Del~Duca, C.~Duhr, E.~Gardi, L.~Magnea, and C.~D. White, {\it {An infrared
  approach to Reggeization}},  {\em Phys. Rev.} {\bf D85} (2012) 071104,
  [\href{http://xxx.lanl.gov/abs/1108.5947}{{\tt 1108.5947}}].

\bibitem{DelDuca:2011ae}
V.~Del~Duca, C.~Duhr, E.~Gardi, L.~Magnea, and C.~D. White, {\it {The Infrared
  structure of gauge theory amplitudes in the high-energy limit}},  {\em JHEP}
  {\bf 12} (2011) 021, [\href{http://xxx.lanl.gov/abs/1109.3581}{{\tt
  1109.3581}}].

\bibitem{DelDuca:2013ara}
V.~Del~Duca, G.~Falcioni, L.~Magnea, and L.~Vernazza, {\it {High-energy QCD
  amplitudes at two loops and beyond}},  {\em Phys. Lett.} {\bf B732} (2014)
  233--240, [\href{http://xxx.lanl.gov/abs/1311.0304}{{\tt 1311.0304}}].

\bibitem{DelDuca:2014cya}
V.~Del~Duca, G.~Falcioni, L.~Magnea, and L.~Vernazza, {\it {Analyzing
  high-energy factorization beyond next-to-leading logarithmic accuracy}},
  {\em JHEP} {\bf 02} (2015) 029,
  [\href{http://xxx.lanl.gov/abs/1409.8330}{{\tt 1409.8330}}].

\bibitem{Henn:2016jdu}
J.~M. Henn and B.~Mistlberger, {\it {Four-Gluon Scattering at Three Loops,
  Infrared Structure, and the Regge Limit}},  {\em Phys. Rev. Lett.} {\bf 117}
  (2016), no.~17 171601, [\href{http://xxx.lanl.gov/abs/1608.00850}{{\tt
  1608.00850}}].

\bibitem{Fadin:2016wso}
V.~S. Fadin, {\it {Particularities of the NNLLA BFKL}},  {\em AIP Conf. Proc.}
  {\bf 1819} (2017), no.~1 060003,
  [\href{http://xxx.lanl.gov/abs/1612.04481}{{\tt 1612.04481}}].

\bibitem{Caron-Huot:2017fxr}
S.~Caron-Huot, E.~Gardi, and L.~Vernazza, {\it {Two-parton scattering in the
  high-energy limit}},  \href{http://xxx.lanl.gov/abs/1701.05241}{{\tt
  1701.05241}}.

\end{thebibliography}\endgroup
\end{document}